\documentclass{aastex631}

\usepackage{xcolor}
\usepackage{appendix}
\usepackage{ulem}
\usepackage{comment}
\usepackage{natbib}
\usepackage{enumerate}   
\usepackage[colorinlistoftodos]{todonotes}
 
\definecolor{Mycolor2}{HTML}{FEAB1F}
\definecolor{Mycolor3}{HTML}{006A4E}

\shorttitle{Initial State Vectors Derivation Using PINN}
\shortauthors{S. Chatterjee and M. Dikpati}

\begin{document}

\title[]{A Physics Informed Neural Network For Deriving MHD State Vectors From Global Active Regions Observations}

\correspondingauthor{Subhamoy Chatterjee}
\email{subhamoy.chatterjee@swri.org}

\author[0000-0002-5014-7022]{Subhamoy Chatterjee} 
\affil{Southwest Research Institute, Boulder, CO 80302, USA}

\author[0000-0002-2227-0488]{Mausumi Dikpati} 
\affil{High Altitude Observatory, NSF-NCAR, 3080 Center Green Drive, Boulder, CO 80301, USA}

\begin{abstract} 
Solar active regions (ARs) do not appear randomly but cluster along longitudinally warped toroidal bands (‘toroids’) that encode information about magnetic structures in the tachocline, where global-scale organization likely originates. Global MagnetoHydroDynamic Shallow-Water Tachocline (MHD-SWT) models have shown potential to simulate such toroids, matching observations qualitatively. For week-scale early prediction of flare-producing AR emergence, forward-integration of these toroids is necessary. This requires model initialization with a dynamically self-consistent MHD state-vector that includes magnetic, flow fields, and shell-thickness variations. However, synoptic magnetograms provide only geometric shape of toroids, not the state-vector needed to initialize MHD-SWT models.  To address this challenging task, we develop PINNBARDS, a novel Physics-Informed Neural Network (PINN)-Based AR Distribution Simulator, that uses observational toroids and MHD-SWT equations to derive initial state-vector. Using Feb-14-2024 SDO/HMI synoptic map, we show that PINN converges to physically consistent, predominantly antisymmetric toroids, matching observed ones. Although surface data provides north and south toroids' central latitudes, and their latitudinal widths, they cannot determine tachocline field strengths, connected to AR emergence. We explore here solutions across a broad parameter range, finding hydrodynamically-dominated structures for weak fields ($\sim$2 kG) and overly rigid behavior for strong fields ($\sim$100 kG). We obtain best agreement with observations for 20–30 kG toroidal fields, and $\sim 10^{\circ}$ bandwidth, consistent with low-order longitudinal mode excitation. To our knowledge, this framework provides the first plausible method for reconstructing state-vectors for hidden tachocline magnetic structures from surface patterns; this could potentially lead to accurate prediction of flare-producing AR-emergence weeks ahead.

\end{abstract}


\section{Introduction} \label{sec:intro}

Active regions (ARs) are known to be the progenitors of solar energetic events, such as CMEs and flares, which often cause hazardous space weather and impact our technological society. ARs do not appear uniformly at all longitudes; neither do they appear at only one longitude all the time. In the former case, we would experience space weather all the time, and in the latter case, space weather events would occur in a perfectly periodic interval, and hence no challenge to simulate and predict that. Instead, ARs appear on the surface of the Sun with a significantly globally-organized spatiotemporal distribution pattern, which is more systematic than random. Previously, analyzing observations of AR distributions in synoptic magnetograms, \citet{dikpati2021deciphering} \citep[see 
also][]{norton2005recovering} reported that their global patterns can be represented by a superposition of Fourier modes with low longitudinal wavenumbers. These global wavy patterns evolve slowly with time, particularly during active phases of a solar cycle when a majority of energetic events like big CMEs and X-class flares occur \citep[see e.g., Figure 1 of][]{Raphaldini2024deciphering}.

A big AR often turns complex and becomes flare-prone, and in a few to several hours can cause space weather hazards \citep{Whitman2023}. Often activity nests form from multiple emergences within $\pm 7$ degrees of both latitude and longitude \citep[the term 'activity nests' was first introduced in][] {schrijver2000}. Also sometimes recurrent emergences occur at the same longitude where there are previously existing ARs \citep[see, e.g.,][] {raphaldini2023deciphering}; the biggest X-flare of cycle 24 during 2017 September storm was caused by such a recurrent emergence of a new region into the existing AR 12673. In both cases ARs can locally interact and develop complexities, and cause eruptions and associated energetic events. To protect our technological environment from hazardous space weather impacts well ahead in time, it is prudent to simulate the origin of these eruptions, namely where on the surface and at what time new, big ARs would emerge, whether it would emerge in a latitude-longitude location of an existing AR or within the vicinity of an AR, thereby forming activity nests.

If flare-prone ARs emerge at the front side of the Sun, only a few hours can be available to study their local dynamics for the development of flares, such as helicity imbalance, evolution of the AR morphology, percentage total area with shear angle greater than 45-degree, free magnetic energy density and several others \citep[see, e.g.,][]{dikpati2025}.
However, in the recent past, it has been extensively explored to demonstrate that combining global evolutionary patterns of toroids in which ARs are strung, particularly the latitude-longitude locations of magnetically complex ARs, and then, tracking their local dynamics to lead to their eruptive states, can significantly improve the lead time about upcoming big solar energetic events from hours to weeks. Proxies from local dynamics hinting towards eruptive events are many, and there isn't any single unique proxy that can be fully relied on to predict eruptions of ARs into big flares/CMEs. Furthermore these proxies indicate upcoming flares only with a few hours lead time, often not enough to protect our technological society from their hazardous impact.
Studying global evolutionary patterns of narrow, warped toroids, into which ARs are attached, it can be possible to increase the lead time from hours to days \citep{dikpati2025}.

The obvious question is: how can we simulate weeks ahead a future toroid pattern containing a new big AR emergence, which will be flare-prone? To seek answer to this question requires a physical model, which can be initialized by observations, and the physical equations can be integrated forward in time. 
A pluasible origin of systematic global distribution patterns of ARs in warped toroids is most likely the dynamics of magnetic fields in a subadiabatic layer at the bottom of the convection zone or at the tachocline, where the turbulence is much less and hence, global organization is possible. Over the past three decades, it has been shown by various authors \citet{Cally_Dikpati_Gilman2003, Miesch_Gilman_Dikpati2007, dikpati2017origin}) that global MHD models (3D thin-shell or shallow-water type quasi-3D models) operating in a relatively stably-stratified tachocline layer than the overlying turbulent convection zone, can produce certain outputs that could be identified as potential candidates for initiating the emergence of ARs in an ordered fashion. The physics is essentially based on the nonlinear interactions among global dynamo-generated toroidal magnetic fields, differential rotation and Rossby waves, which produce  the ``imprints" for the flux-emergence locations \citep[see, e.g.,][]{dikpati2017origin, dikpati2020spaceweather, dikpati2022space}. 

Such numerical models, in order to be integrated forward for simulating global toroid patterns and possible new big AR emergence weeks ahead, need to be initialized with observational data so that initial state-variables can be specified at $t=0$. On one hand, the model-state-variables in a quasi-3D shallow-water model are generally longitude and latitude components of magnetic and velocity fields ($a$, $b$, $u$, $v$), and the shell-thickness-variation ($h$). On the other hand, global synoptic maps give only a geometric pattern, namely the spatial distribution of ARs in warped toroid, as a combination of low-order longitudinal modes. Thus, deriving the model-required state-variables from the observed warped toroid pattern is a most challenging task. The derived state-variables will not only have to reproduce the warped toroid pattern, but also have to satisfy the shallow-water model equations. In contrast to purely data-driven machine learning models (\citep{chatterjee2022}), Physics Informed Neural Networks (PINNs) have emerged as a promising approach to perform data assimilation while solving Partial Differential Equations (PDEs) without requiring regular grid definition, creating surrogates of expensive numerical solvers and addressing inverse problems \citep{Raissi2019, Raissi2020}. PINNs have also been successfully used as solvers of hydrodynamic Shallow-Water equations on a sphere \citep{bihlo2022}. Therefore, PINN can be implemented as one of the best possible techniques to derive model-state-variables at $t=0$ from the observed AR distribution pattern. Application of machine learning models are, in general, increasingly showing potential for better use of big data, leading to improved predictions \citep[see, e.g., the application in for flare prediction by][]{inceoglu2018, korsos2021, liu2021}.

Where exactly the magnetic fluxtubes come from to manifest as emerged ARs at the surface is not fully settled yet; there exists a vast literature on this, suggesting fluxtubes come from somewhere in the convection zone \citep{nelson2013}, or at/near the base of the convection zone or tachocline \citep{hughes1997, manek2024}. However, based on the good correspondence between the global MHD model-generated warped toroid patterns and that observed at the surface \citep{dikpati2021deciphering}, we consider here that the fluxtubes rise from the tachocline and/or from the base of the convection zone, where global organization is feasible. 

Two key issues remain. The first concerns the time delay between flux-emergence “imprints” at the base and the warped toroid patterns observed at the surface. Models of flux emergence have heavily relied on thin flux tube approximation \citep{weber2014}. However, emerged ARs reveal a finite size of the fluxtubes, from which they originate. Only a few recent work address wide fluxtube simulations, and features of the surface emergences from wide toroidal band \citep{fournier2017}. Stronger fields rise faster, taking about 1-2 weeks, while weaker ones can take over a month. Thus, ARs visible at a given time may originate from different bottom toroidal bands. However, we focus on the strong fields, because strong, big ARs would emerge from the strong toroidal fields at the bottom. Focusing on the strong fields only helps achieve a clearer correspondence between the model-evolution of toroidal bands and surface warped toroid-patterns containing ARs. Selection of a high threshold in the magnetogram data for toroid derivation would ensure that we are dealing with the ARs emerging at the surface taking approximately the same time. Thus the observed surface distribution can be traced to a single toroid pattern formed about a week or two earlier.

The second issue is the peak toroidal field strength. Surface-derived toroid patterns reveal that the warping is a consequence of the combination of a few low-order longitudinal modes ($m>0$) along with the central latitude location for the $m=0$ component. But the only information about the field strength comes from the threshold value set to derive the warped toroids from synoptic magnetogram data. If the origin of the surface active regions is somewhere at the bottom of the convection zone or at the tachocline, a more subadiabatically stratified region than the turbulent bulk of the convection zone or the near-surface subphotospheric region, then flux tube simulations suggest that ARs originate from much stronger subsurface fields \citep{weber2014} than their surface signatures imply. While thin flux tube simulations \citep{weber2014, manek2018} indicate that the surface flux emergence occurs from strong toroidal fields of strength 50 kG or so, wide-band simulations indicate somewhat weaker fields can emerge coherently \citep{jouve2009}. By combining these results with shallow-water MHD models, a plausible range for the unperturbed field strengths for toroidal bands, which are unstable to low-order longitudinal modes, can be selected. Placing these bands in the tachocline at the central latitude, amplitude and phase of low-order longitudinal modes, derived from the surface observation, we can then implement a PINN to obtain the state variables that would satisfy the physical equations of the governing model, an MHD shallow-water model here, and the observed warped toroid patterns.

With the above reasoning, the obvious next question is how to implement the PINN for deriving the initial state variables from the observed warped toroid pattern at the surface. The main aim of this work is to build an inversion framework that uses a PINN to derive initial MHD state vectors by utilizing observational and physical constraints respectively coming from observed global organization of ARs and quasi-3D MHD shallow-water model equations. \S2 elaborates on them in the following two subsections (\S2.1 and \S2.2) and then describes the PINN architecture along with the training strategy in \S2.3. Results are described in \S3 and concluding remarks in \S4.

\section{Developing PINNBARDS -- the inversion framework using PINN}

As described earlier, deriving state vectors for the MHD shallow water model to forward integrate the model by data-driving, such derivation is a challenge, because the observationally-derived warped toroid pattern is a geometric pattern only.
We describe in the following several subsections the development of an inversion framework PINNBARDS that uses a Physics-Informed Neural Network (PINN) to derive initial MHD state vectors with both observational and physical constraints. One observational constraint is that that the derived latitude and longitude magnetic field components will have to resemble the warped toroid pattern in which the emerged Active Regions (ARs) are strung. Physical constraint is that and 3D MHD shallow-water equations, respectively. We elaborate on them in the following two subsections and then describe the PINN architecture along with the training strategy.

\subsection{Observational Constraint: Toroid Derivation}

Global evolutionary patterns of ARs are not totally random, they show a significant systematic organization. These organized patterns can be constructed through superposition of Fourier modes in longitude with low wavenumbers. The basic methodology for deriving global latitude-longitude distribution of ARs observed in synoptic magnetograms is based on an optimization technique \citep{norton2005recovering, dikpati2021deciphering}. HMI synoptic maps are constructed by combining near-central meridian data from full-disc magnetograms over a synodic solar rotation period, resulting in sine(latitude) $\times$ Carrington longitude charts. These maps provide a representation of the latitudinal distribution of active regions as well as their Carrington longitude positions. 

Given a synoptic chart at time $t$, global toroid patterns $P_c(\lambda,t)$ can be determined via a superposition of Fourier modes, following the expression: 

$$P_c(\lambda,t)=\sum_{m=0}^N q_m(t) sin(m\lambda + \zeta_m(t)), \quad\eqno(1)$$
in which, $q_m(t)$ is the Fourier amplitude of the m-th mode at time $t$ and $\zeta_m(t)$ the respective Fourier phase. $q_0$ represents the mean latitudinal position of the warped toroid containing ARs at time $t$. Both Fourier amplitude and phases are represented as a combination of mean ($\overline{q}_m, \overline{\zeta}_m$) and time-varying parts ($s_{i,m}$, $w_{i,m}$) respectively, at time $t_i$, $i$ being a discrete time index described as: 
$$q_m(t)=\overline{q}_m+\sum_i s_{i,m}(t_i), \quad\eqno(2)$$ 
and, $$\zeta_m(t)=\overline{\zeta}_m+\sum_i w_{i,m}(t_i). \quad\eqno(3)$$
Note that $t$ here denotes the time of the next toroid, i.e. in the next synoptic map in consideration. So $t$ can be a day, a week or our chosen cadence of the next toroid. In case of toroid derivation based on one synoptic map, derived toroid is not a function of $t$, which will be the consideration in the present calculation.

By following \citet{branch1999subspace}, ``Trust Region Reflective" (TRR) algorithm is employed to derive the resulting spatial distribution of active regions, in the form of two wavy-belts, one for each hemisphere. 

\begin{figure}[t!]
\hspace{0.0\textwidth}
\includegraphics[width=1.0\linewidth]%
{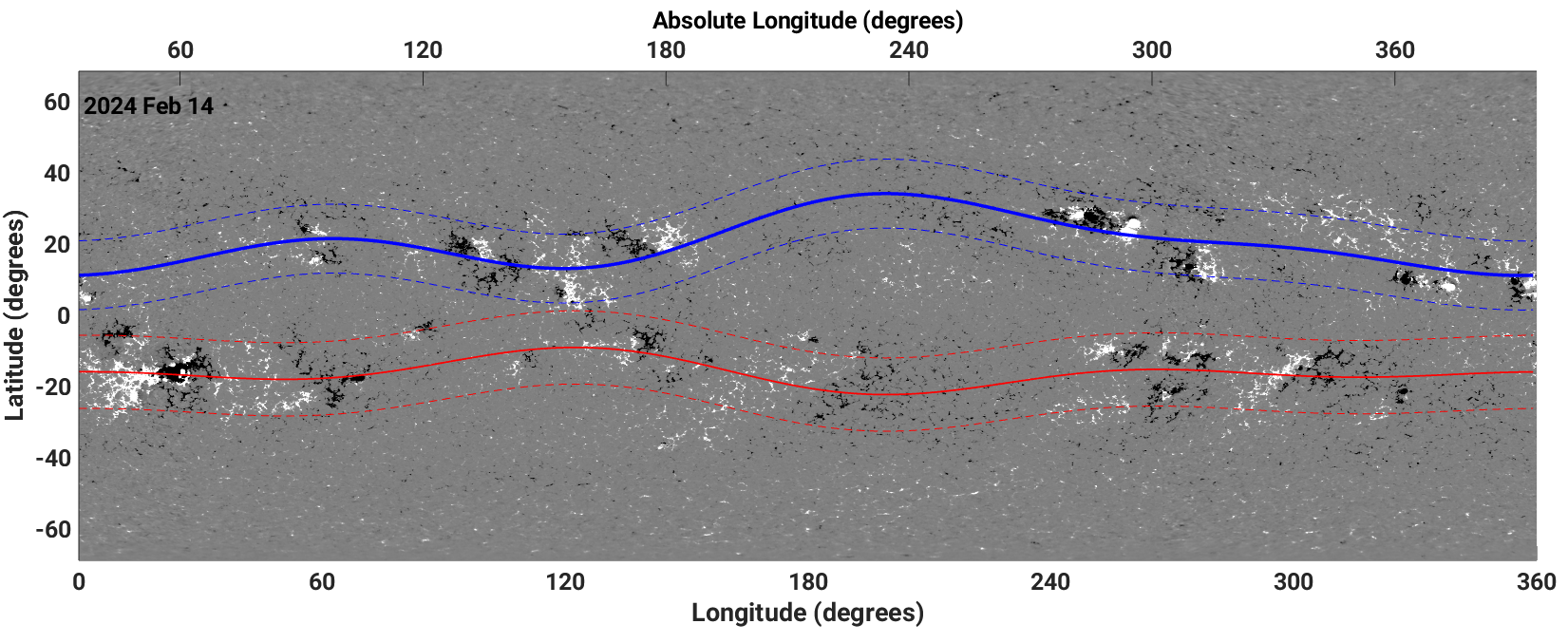}
\caption{Two warped toroid patterns displayed in North (blue) and South (red) hemispheres. Solid curves in blue and red respectively denote the central latitudes of the AR distributions in North and South; blue (red) dashed curves on both sides of solid blue (red) curve imply the width of the toroidal band in which ARs are strung. Absolute longitude (Carrington longitude) is shown on the top x-axis, while the bottom x-axis will be used in PINN model, and later on PINN-solutions will be compared with observations according to bottom x-axis. 
}
\label{fig:AR_toroid}
\end{figure}

Here we select an AR distribution on February 14, 2024, for our case study. Figure \ref{fig:AR_toroid} displays the derived toroids in the north and south; blue and red solid curves along the longitude respectively indicate the central latitudes for these two toroids, and the blue (red) dashed lines on both sides of the solid blue (red) curves reveal the latitudinal widths of the toroids. 

We set a threshold field of $\sim 0.2$ kG to derive the toroid-patterns. This is because our goal is to study the evolution of these toroids containing big, complex ARs, which are the flare-producing ones and hazardous for our technological society. This also ensures that these big, large ARs emerge approximately taking similar time from the tachocline to the surface, thus the coherence between the model-generated bottom toroidal bands and the warped toroids at the surface can be ensured. Note that the bottom horizontal-axis (longitude) is somewhat arbitrary, because the selection of zero-longitude does not matter, since these toroid-patterns are periodic in longitude. However, to compare with the observed magnetograms we have marked the absolute longitude at the top horizontal axis.

\subsection{Physical Constraint: MHD model}
The dynamics of global toroidal magnetic fields in a 3D MHD thin-shell model (shallow-water type) is governed by the following equations: (a) equations of motions for longitude and latitude velocity components (i.e., $u,v$; radial velocity, $w$ comes from mass continuity equation in this system of 3D MHD thin-shell model); (b) induction equations for longitude and latitude magnetic field components (i.e., $a,b$; radial magnetic field component $c$, comes from divergence-free condition); (c) equation for shell-thickness variation ($h$) in case of rigid bottom but deformable top, or hydrostatic pressure equation, in case of rigid top and bottom. By denoting the undisturbed shell-thickness as $H$, which is a fraction of solar radius ($\sim 0.03R_{\odot}$), the total thickness of the thin-shell in MHD shallow-water model can be written as $H(1+h)$. This includes the undisturbed as well as the deformable top. 
MHD shallow-water models are being used for the past 25 years, and the model equations along with formulation can be found in the literature\citep{Gilman_2000, Zaq2007, dikpati2018role}. However, we repeat them here briefly so that they can be correlated with the PINN flow-chart diagram depicted in Figure~\ref{fig:flow}. A single-layer shallow-water model has a rigid bottom and a deformable top. The layer has constant density. The main approximation used in a shallow-water model is that the horizontal extent and motions in it are much larger than the the shell-thickness and vertical motions, rendering the system to be in hydrostatic balance in vertical direction. The horizontal velocity components ($u, v$) and shell-thickness-variation ($h$) are functions of latitude ($\phi$), longitude ($\lambda$) and time ($t$) only, whereas the radial velocity is a linear function of depth. The nonlinear MHD shallow water equations can be written in dimensionless form in the rotating frame of reference as:

$${\partial u \over \partial t} = {v \over \cos\phi}
\left[{\partial v \over \partial\lambda} - {\partial \over
\partial\phi} (u \cos\phi)\right] - {1 \over \cos\phi} {\partial
\over \partial\lambda}\left( {u^2 + v^2 \over 2}\right)
-G {1 \over \cos\phi} {\partial h \over \partial\lambda}
+2\omega_c \,v\,\sin\phi $$
$$ -{b \over \cos\phi}
\left[{\partial b \over \partial\lambda} - {\partial \over
\partial\phi} (a \cos\phi)\right] + {1 \over \cos\phi} {\partial
\over \partial\lambda}\left( {a^2 + b^2 \over 2}\right), \eqno(4)
$$

$${\partial v \over \partial t} =
-{u \over \cos\phi}
\left[{\partial v \over \partial\lambda} - {\partial \over
\partial\phi} (u \cos\phi)\right] - {\partial
\over \partial\phi}\left( {u^2 + v^2 \over 2}\right)
-G {\partial h \over \partial\phi}
-2\omega_c \,u\,\sin\phi $$
$$+{a \over \cos\phi}
\left[{\partial b \over \partial\lambda} - {\partial \over
\partial\phi} (a \cos\phi)\right] + {\partial
\over \partial\phi}\left( {a^2 + b^2 \over 2}\right), \quad\eqno(5) $$

$${\partial \over \partial t}(1+h) = -{1 \over \cos\phi}
{\partial \over \partial\lambda}\left( (1+h)u \right) - {1 \over \cos\phi}
{\partial \over \partial\phi}\left((1+h)v \cos\phi\right),
\quad\eqno(6) $$

$${\partial a \over \partial t}= {\partial \over \partial\phi}
(u\thinspace b - v\thinspace a)
+ {a \over \cos\phi}\left({\partial u \over \partial\lambda}
+ {\partial \over \partial\phi}(v \thinspace \cos\phi)\right)
- {u \over \cos\phi}\left({\partial a \over \partial\lambda}
+ {\partial \over \partial\phi}(b \thinspace \cos\phi)\right),
\quad\eqno(7) $$

$${\partial b \over \partial t}= -{1 \over \cos\phi}
{\partial \over \partial\lambda} (u\thinspace b - v\thinspace a)
+ {b \over \cos\phi}\left({\partial u \over \partial\lambda}
+ {\partial \over \partial\phi}(v \thinspace \cos\phi)\right)
- {v \over \cos\phi}\left({\partial a \over \partial\lambda}
+ {\partial \over \partial\phi}(b \thinspace \cos\phi)\right),
\quad\eqno(8) $$

$${\partial \over \partial\lambda}\left((1+h)a\right) +
{\partial \over \partial\phi}\left((1+h)b\cos\phi\right)=0.
\quad\eqno(9) $$

To obtain these dimensionless equations (4-9), $r_0$, the radius of the fluid shell is used as unit dimensionless length, and $1/\omega_c$ (where $\omega_c$ is the core-rotation rate) as the unit dimensionless time. This gives rise to a nondimensional parameter ($G$), called the ``effective gravity", so that $G=gH/{r_0}^2 {\omega_c}^2$. $G$ is a measure of subadiabaticity of the shallow fluid layer, and relates to the actual gravity $g$ at the radius ($r_0$) of the shell, thickness ($H$) of the shell and the core-rotation rate ($\omega_0$).

These equations are solved along with mass conservation (equation 9) and divergence-free conditions for magnetic fields, which is $\nabla.\left((1+h)B\right)=0$. To avoid nonphysical singularities at the poles, global 3D-thin-shell and shallow-water-type systems are solved in latitude-longitude space using spherical harmonics basis functions.

\subsection{Development of PINNBARDS as an inverse framework for deriving MHD state vectors from observations}

We build here a numerical code for deriving initial state-vectors from  observed AR distribution. We implement an algorithm based on PINN, and for the purpose of creating a convenient unique identifier, we name the code as PINNBARDS, which stands for PINN-Based Active Regions Distribution Simulator. 

The physical model employed in PINNBARDS is a global MHD shallow-water model, for which the formulation and detailed derivation of full-set of nonlinear equations already exist \citep[see, e.g.,][]{dikpati2018role}, and have already been briefly presented in \S2.2. Note that equations (4-8) contain time-derivatives on the left hand side. The state-vectors ($u, v, a, b, h$) have, in general, two components each, namely the reference state (unperturbed time-independent component) and a perturbation part (i.e. the time-dependent component). The reference states are the toroidal magnetic field and the azimuthal flow, i.e. the differential rotation. Both the field and flow being in the longitude direction, they will not interact, because magnetic forces only act on the component of motion that is perpendicular to the magnetic field direction. Therefore, to initiate the interaction between the global flows and fields, the system must be perturbed by a component in the latitude direction. Thus the nonlinear interactions among the state-variables in the model-system will not occur until it is perturbed. The unperturbed reference states correspond to $m=0$ components only, and all $m>0$ modes are perturbations to the reference states. The observed warped toroid patterns contain $m=0$  as well as $m>0$ modes. Thus they already contain finite perturbations, and hence the model-system can be initialized by them to study their nonlinear evolution. So our goal is to derive the state-vectors that best-fit the observed warped toroid patterns. To reach that goal we train the PINN so that it yields the spatial distribution of state-vectors that has minimal loss. This means, in turn, we need to minimize the right hand sides of equations (4-8), along with the constraints mass-conservation and divergence-free magnetic field conditions.

Thus, to derive the state vectors at $t = 0$ for initializing the MHD simulation, we must minimize the errors in the PINN solution. These errors arise from the mismatch between the PINN-derived state vectors and the ones that satisfy Equations (4)–(8) in time-independent regime. We reduce these errors by adjusting the parameters of the PINN model during training. At the same time, the PINN solution must also satisfy mass-conservation and divergence-free magnetic field conditions, and the contraint from observationally derived toroids discussed in \S2.1.

\begin{figure}[t!]
\hspace{-0.0\textwidth}

\includegraphics[width=1.0\linewidth]
{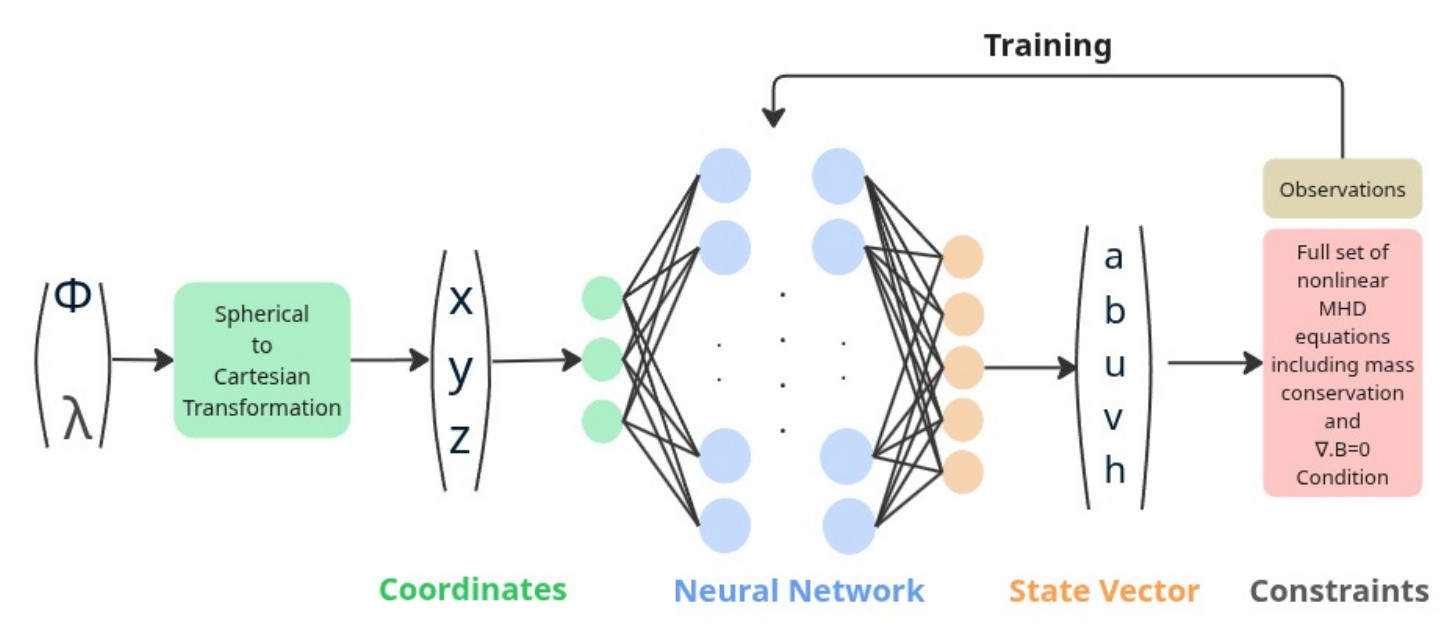}
\caption{
Workflow of the Physics Informed Neural Network (PINN) for an MHD shallow-water model. Latitude–longitude coordinates ($\phi$-$\lambda$) are converted to Cartesian coordinates ($x,y,z$) and fed into a neural network that predicts the physical state variables ($a, b, u, v, h$), namely longitude and latitude components of magnetic fields ($a,b$) and velocity fields ($u,v$) and $h$ is related to thickness of the thin-shell as $1+h$. The model is trained using observational data and constrained by the full set of nonlinear magnetohydrodynamic equations, including mass conservation and divergence-free magnetic field conditions.}
\label{fig:flow}
\end{figure}

We describe the detailed workflow (Figure~\ref{fig:flow}) consisting of the PINN architecture, loss function and the training setup in \S2.3.1, \S2.3.2, and \S2.3.3, respectively. We describe the integration of various packages at the end of the paper in \textit{Softwares} .

\newpage
\subsubsection{PINN Architecture}
We build the PINN as a fully connected neural network with $N_l$ hidden layers, each containing $N_d$ nodes that can be represented as a vector-valued non-linear function ($f^{\theta}$) performing the mapping:
$$(a, b, u, v, h) = f^{\theta}(\cos\phi\cos\lambda, \cos\phi\sin\lambda, \sin\phi) \quad\eqno(10)$$
In the input, we transform latitude ($\phi$) and longitude ($\lambda$) to Cartesian coordinates for hard-coding the periodic boundary conditions. We use $\tanh$ non-linearity as the activation function for all the neural-network layers. $\theta$ defines the model parameters that are tuned to minimize the loss function during the training phase. The process of selecting optimal values of $N_d$ and $N_l$ is described in \S 3.1. 

\subsubsection{Loss Function}
We construct the loss function of the PINN by combining the observational and physical constraints defined previously. To inform the model further with a solar-like configuration, we provide additional constraints, namely the differential rotation constraint on $u$ as
$$\bar u = \omega_d\cos{\phi} \quad\eqno(11)$$ and mass-conservation constraint on $h$ as
$$
 G\partial \bar h/\partial \mu = -2\omega_d\mu - \omega_d^2\mu + \alpha_0^2
\quad\eqno(12)$$
$$\bar h_{\phi = 0} = 0 \quad\eqno(13)$$
Where, $\mu = \sin\phi$, $\omega_d = 0.047 - 0.15\mu^2 - 0.06\mu^ 4$, and $\alpha_0 = A(e^{-\beta(\mu-\mu^0_N)^2} - e^{-\beta(\mu+\mu^0_S)^2})$. $\bar u, \bar h$ stand for longitudinal means of $u$ and $h$ respectively.
$A$ defines the magnetic toroid amplitude, $\mu^0_N$ and $\mu^0_S$ represent the sine of mean-latitudes derived from North and South toroid fits on observational data, respectively. 

We also construct the band loss term using the following constraints-
$$a = sAe^{-(\frac{\phi-P_c(\lambda)}{\sigma})^2}\cos{\psi} \quad\eqno(14)$$
$$b = sAe^{-(\frac{\phi-P_c(\lambda)}{\sigma})^2}\sin{\psi} \quad\eqno(15)$$
We derive the toroid fits $P_c(\lambda)$ for Northern and Southern hemispheres separately and then add them to constrain $a, b$. We define $\psi$ as $\tan^{-1}\frac{dP_c}{d\lambda}$. Assuming a Gaussian latitudinal profile of the toroids as described in equations (14) and (15), we derive $\sigma$ from the prescribed Full Width Half Maxima (FWHM) of the north and south toroids as $\sigma = \frac{FWHM}{\sqrt{8ln(2)}}$. $s$, defined as 1 for north and -1 for South, captures the direction of the magnetic field vector along the toroids. 

We construct the boundary loss by constraining the magnetic and velocity fields to be 0 at the poles and $h$ to be matched for $\lambda$ intervals $[0, \pi]$ and $[\pi, 2\pi]$ for each pole.

We combine the residuals of the time-dependent unperturbed MHD equations, the differential rotation equation, the prolateness equations, band loss, and boundary loss in the form of mean squared errors, assigning equal weight to each, to construct the loss function.
The tuple $(G, A, FWHM)$ is kept constant for each training run.

\subsubsection{Training}
\begin{figure}[htbp!]
\centering
\hspace{0.0\textwidth}
\includegraphics[width=1.0\linewidth]{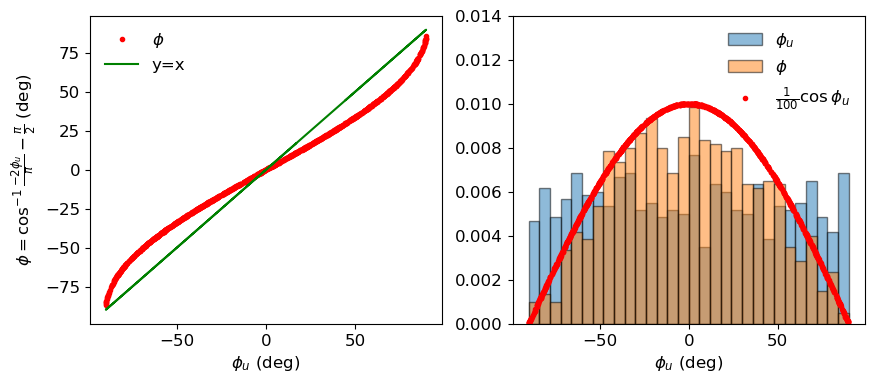}
\caption{Sampling of latitudes for PINN training. The transformation on the left converts $\phi_u$ to $\phi$ (red points), lowering the latitudes away from the poles (`y=x' is added for reference). This converts, as shown on the right panel, the uniform distribution (histogram density in light blue) of $\phi_u$ in $[-\pi/2, \pi/2]$ to a cosine-modulated distribution (histogram density in light orange) that follows the distribution of latitudes when points are uniformly sampled from the surface of a sphere. 
}
\label{fig:lat_sample}
\end{figure}

To train the PINN, we need to minimize the loss function on a set of collocation points apart from the boundaries. We randomly select a set of collocation points from the $(\phi, \lambda)$ space $[0, 2\pi]\times[-\pi/2+\delta, \pi/2-\delta]$ such that the poles are avoided ($\delta = 10^{-8}$) and points are uniformly distributed on a sphere with number points reducing gradually from equator to pole. To realize this we first uniformly sample $(\phi_u, \lambda)$ and then perform the transformation-
$$\phi = \cos^{-1}(-\frac{2\phi_u}{\pi}) - \frac{\pi}{2} \quad\eqno(16)$$
This transformation can also be rewritten as $\sin^{-1}{2\phi_u/\pi}$ and maps a value of $\pi/2$ to $\pi/2$ and $-\pi/2$ to $-\pi/2$. As shown in the Figure~\ref{fig:lat_sample}, apart from the poles and equator, all other latitudes move closer to the equator as a result of this transformation. This causes a cosine-modulated reduction in the number of samples selected while moving to higher latitudes from the equator, as shown in the histograms in Figure~\ref{fig:lat_sample}. This conforms to a uniform distribution of points on the surface of a sphere, and adds more weightage to points lying between low to mid latitudes because of the cosine distribution.

We use a total of 5,000 collocation points in the $(\phi, \lambda)$ grid to enforce the physical constraints, consisting of the residuals of the MHD equations (4)--(9) in time-independent, unperturbed, and the band constraints defined in equations (14)-(15). To use the additional loss function constraints defined in equations (11)-(13), we select 300 latitude ($\phi$) points using the sampling technique defined in equation (16). We then uniformly sample 32 longitude ($\lambda$) points from $[0, 2\pi]$ and calculate the longitudinal means ($\bar u, \bar h$) for each of 300 $\phi$ points. To satisfy the polar boundary conditions, we use 200 longitude ($\lambda$) points at each of the North and South poles. We minimize the combined loss using an Adam optimizer with a learning rate of $10^{-3}$ to train the model ($f_{\theta}$) for 20000 epochs. We find the best model by choosing the epoch at which the loss reaches the minimum. 

\section{Results}

As already mentioned in \S1, to relate the modeled toroidal bands to the observed warped toroid patterns at the surface, we need to consider two issues. One is the time delay associated with flux emergence and the underlying strength of the toroidal field at depth. Because strong magnetic fields rise more rapidly than weak ones, only the largest active regions can reliably be traced back to a common subsurface toroidal structure a week or two earlier; therefore, by deriving surface toroids using a high magnetic threshold can corroborate focusing on the strong-field emergences with a fixed one or two weeks time-delay. 

The second issue is that the surface magnetograms do not directly reveal the mean toroidal field strength, which is expected to be significantly greater in the tachocline where the bands originate. By combining constraints from flux-emergence theory and shallow-water MHD instability properties, we select a plausible range of deep toroidal field strengths and central latitudes. 

The above logic is the basis for the cases, for which Physics-Informed Neural Network (PINN) inversion has been used to obtain physically consistent state variables that reproduce the observed warped toroid patterns. The basis includes a plausible set of parameters at the tachocline. From theoretical studies of MHD shallow-water model system \citep{dikpati2003}, we select four different solar-like peak field strengths, namely 2kG, 20kG, 30kG and 100kG. Because the surface observations indicate that the ARs, tightly strung in warped toroid-patterns, are most likely coming from a narrow toroidal band at the tachocline, we consider four different latitudinal-widths of those band, namely, $5^{\circ}$, $10^{\circ}$, $15^{\circ}$ and $20^{\circ}$, again guided by theoretical MHD tachocline shallow-water models. The other two parameters are the tachocline differential rotation amplitude and the effective gravity (a measure of subadiabaticity). The differential rotation is a much less sensitive parameter; we consider 21\% pole-to-equator differential rotation throughout our calculations. For the effective gravity value ($G$), we consider $G=0.5$, which applies to the overshoot part of the tachocline \citep[see, e.g.][for the relation between $G$ and subadiabaticity]{dikpati2001, dikpati2003}. Certain latitude-longitude locations can form, where the magnetic flux coinciding with the positive pressure-departure region (or bulging of the top surface) of the global thin-shell MHD tachocline can be pushed-up towards the convection zone, leading to eventual emergence of flux to the surface. However, we will also consider a case of $G=10$, which would imply the radiative part of the tachocline.

In the subsection below (\S3.1), we first select a case for the tachocline toroidal band of $10^{\circ}$ latitudinal width (FWHM) and 30 kG peak field strength, placed at $\pm15^{\circ}$ latitudes, a pole-to-equator differential rotation amplitude of 21\%, $G=0.5$, and show how the converged solutions come up with the iterations of PINN. We also describe how we choose the optimal architecture by studying the loss for different numbers of layers ($N_l$) and number of nodes/layer ($N_d$). 

\begin{figure}[t!]
\centering
\hspace{0.0\textwidth}
\includegraphics[width=1.0\linewidth]{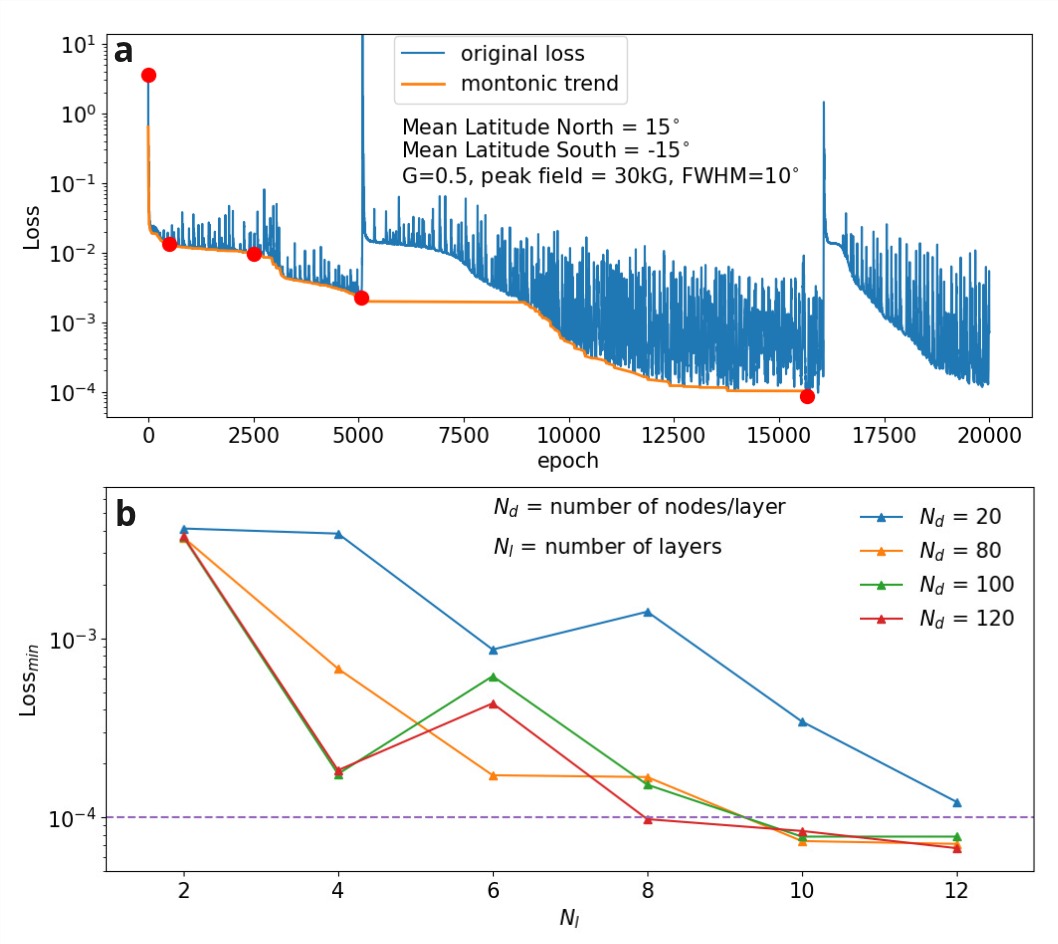}
\caption{Convergence of PINN and identification of optimal architecture. Plot in panel (a) shows the evolution of PINN loss with epoch for the architecture with 10 layers and 100 nodes/layer. The loss is a combination of equation loss and data loss.  The blue curve shows the loss for all the epochs, whereas the orange curve shows the monotonic version with epochs ($e$) defined by the set $S = \{\, e \mid Loss(e) \le \min_{k = 0}^{e - 1} Loss(k) \,\}$. We highlight the PINN convergence for a set of four epochs, marked in filled red circles, by visualizing the state vector outcomes in Figure~\ref{fig:pinn_cnvrg}. Panel (b) shows how the minimum loss (Loss$_{min}$) changes when the number of layers ($N_l$) and the number of nodes/layer ($N_d$) are changed from 10 and 100, respectively. It can be seen that for a much smaller $N_d$, such as 20, the loss is larger for all $N_l$, causing less accurate results. However for $N_d\geq80$ and $N_l\geq10$ the Loss$_{min}$ goes below $10^{-4}$ in an asymptotic manner, and this difference is likely due to statistical fluctuation from model initialization. We thus stick to an $N_d=100$ and $N_l=10$ as an optimal PINN architecture in terms of speed and accuracy. We perform all further experiments with the same architecture.}
\label{fig:pinn_loss}
\end{figure}


\subsection{PINN convergence and selection of optimal architecture}

\begin{figure}[ht]
\epsscale{0.57}
\hspace{-0.014\textwidth}
\plotone{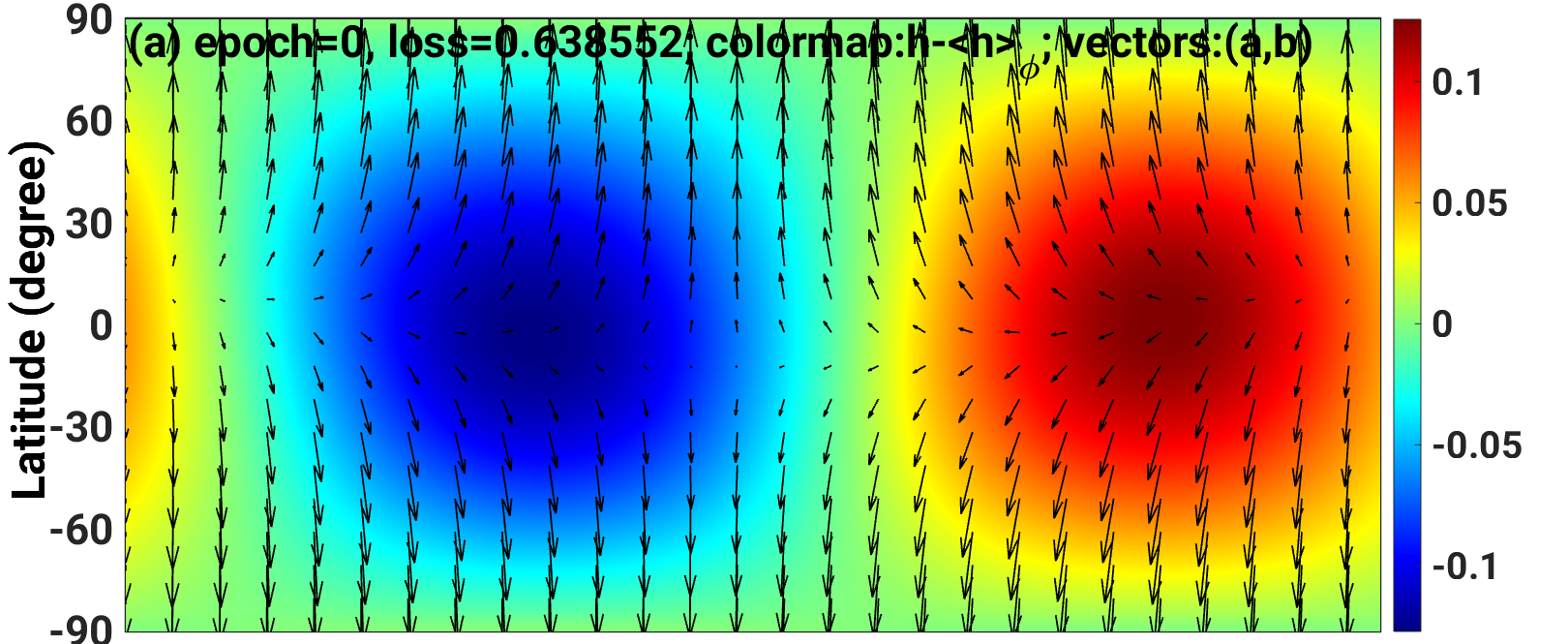}
\plotone{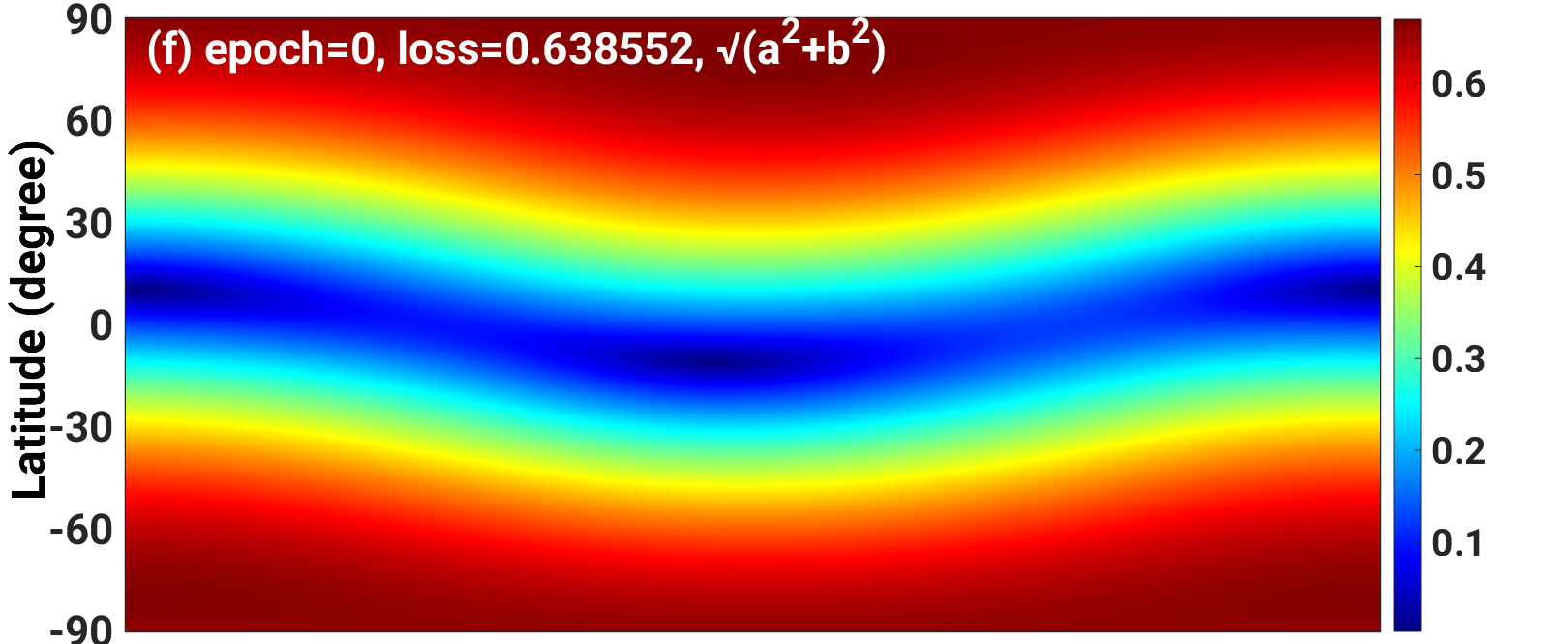}
\plotone{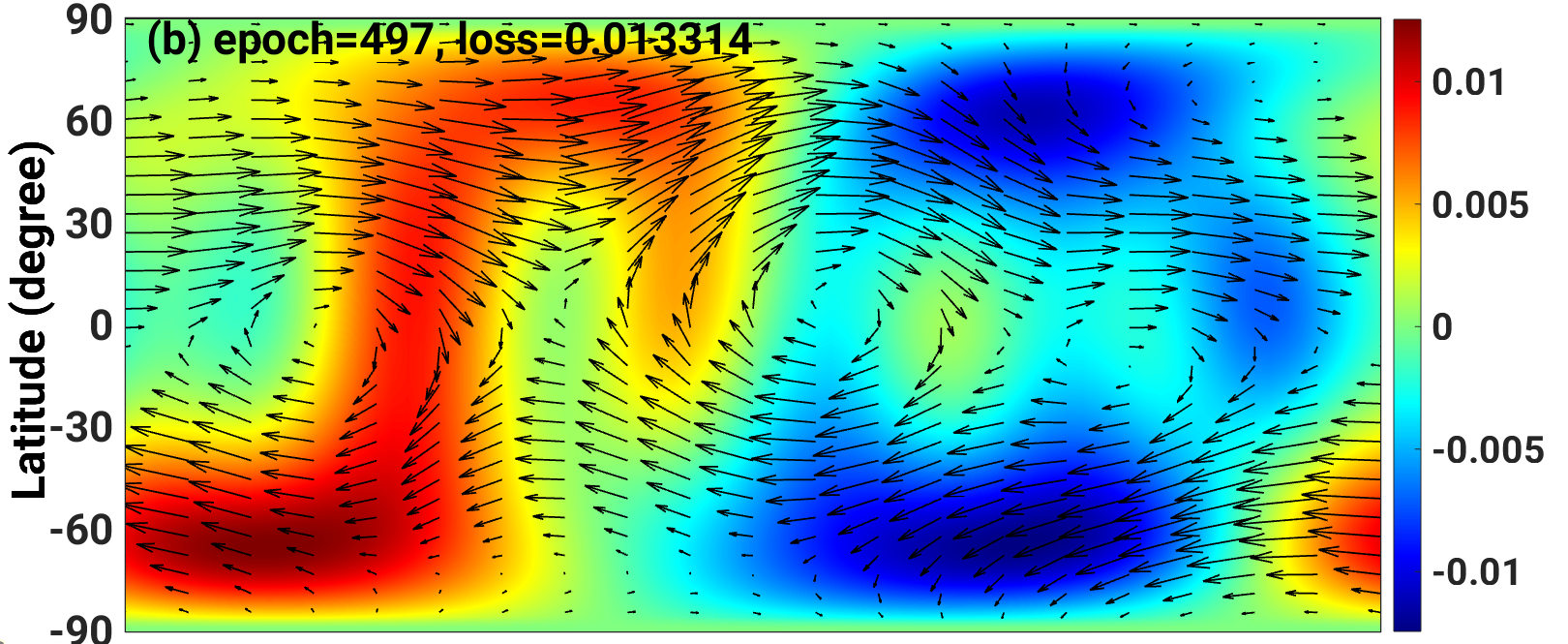}
\plotone{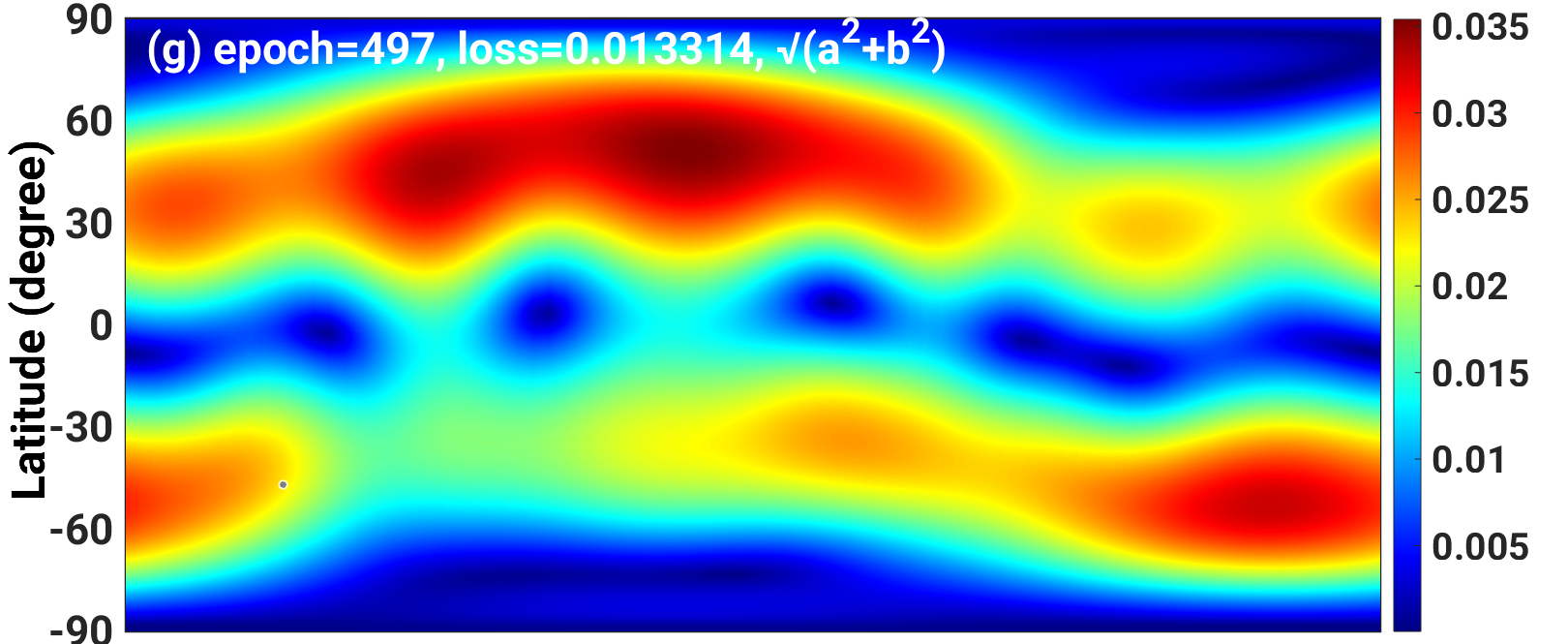}
\plotone{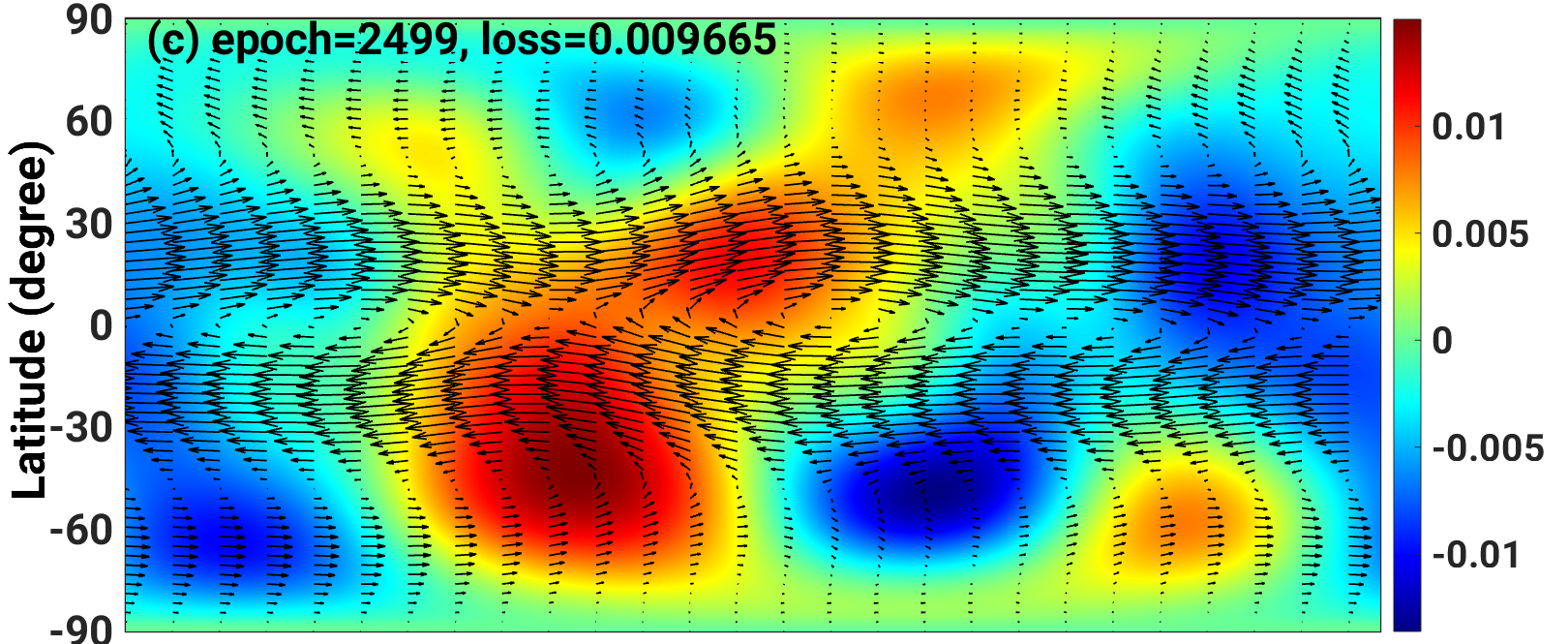}
\plotone{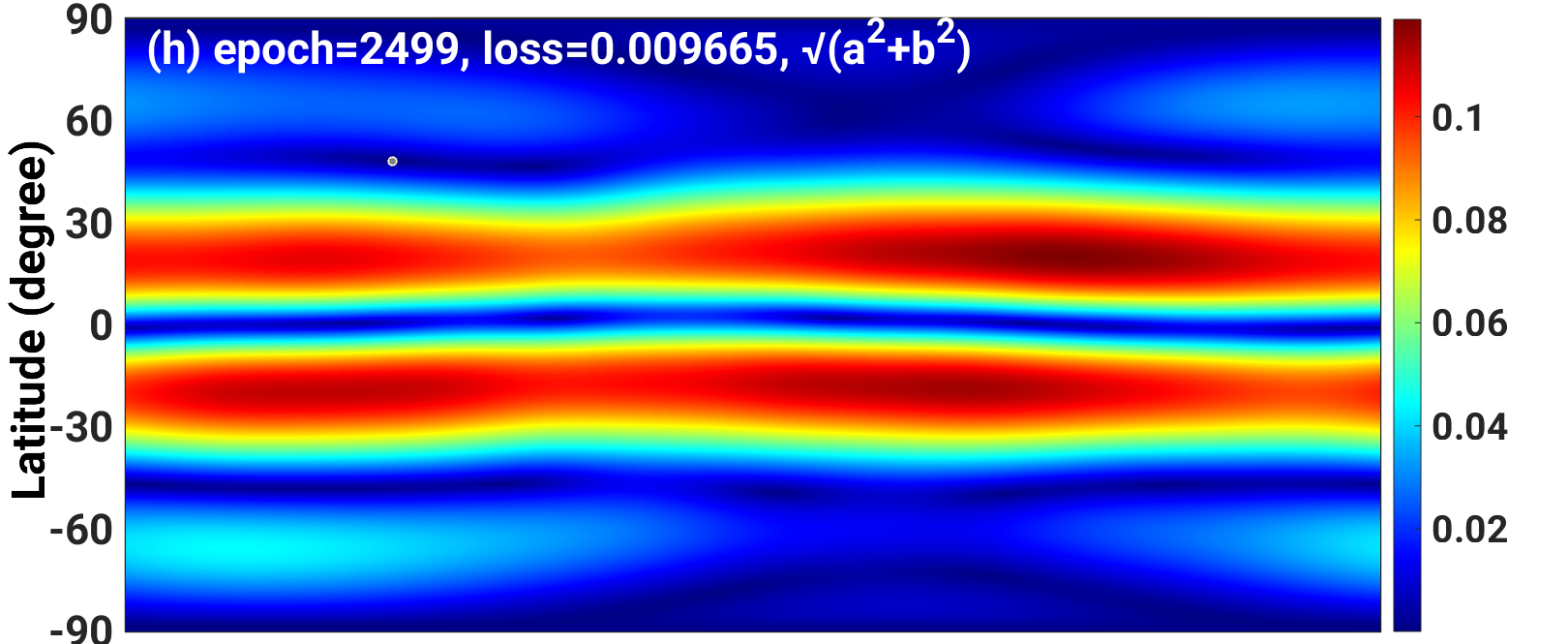}
\plotone{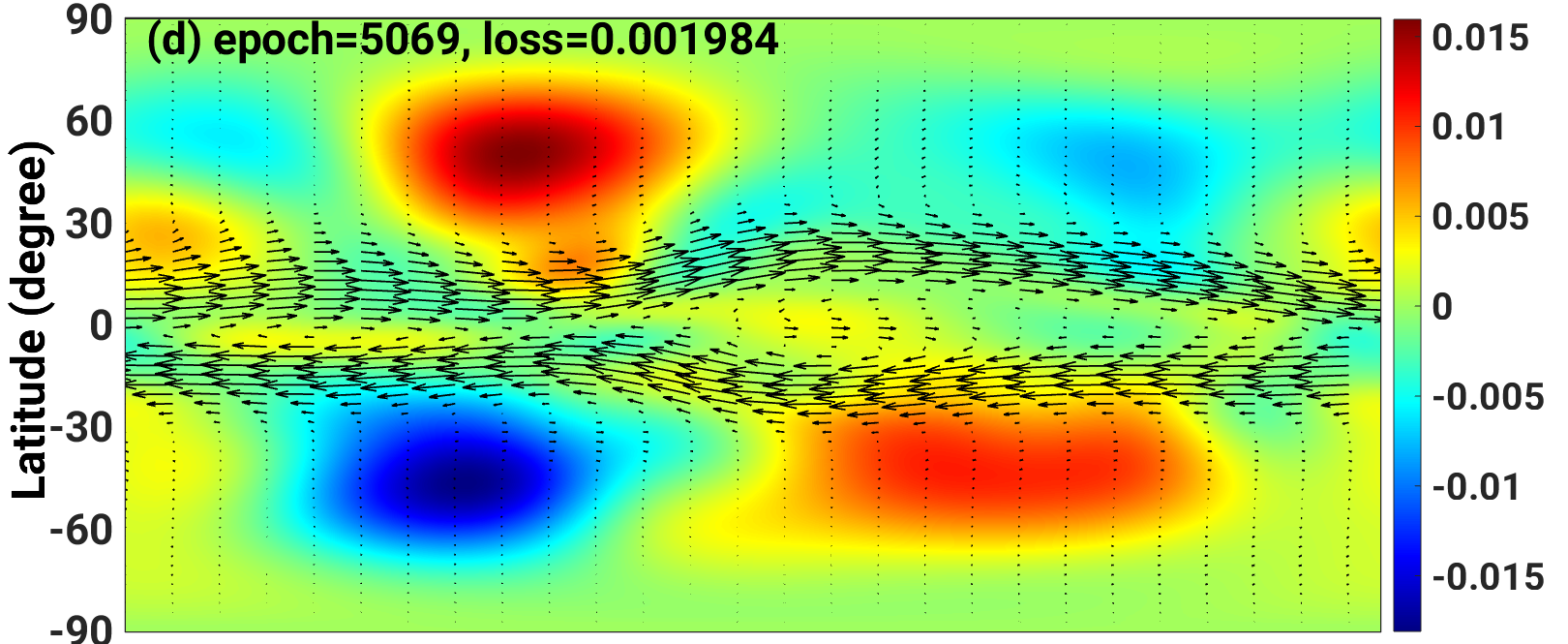}
\plotone{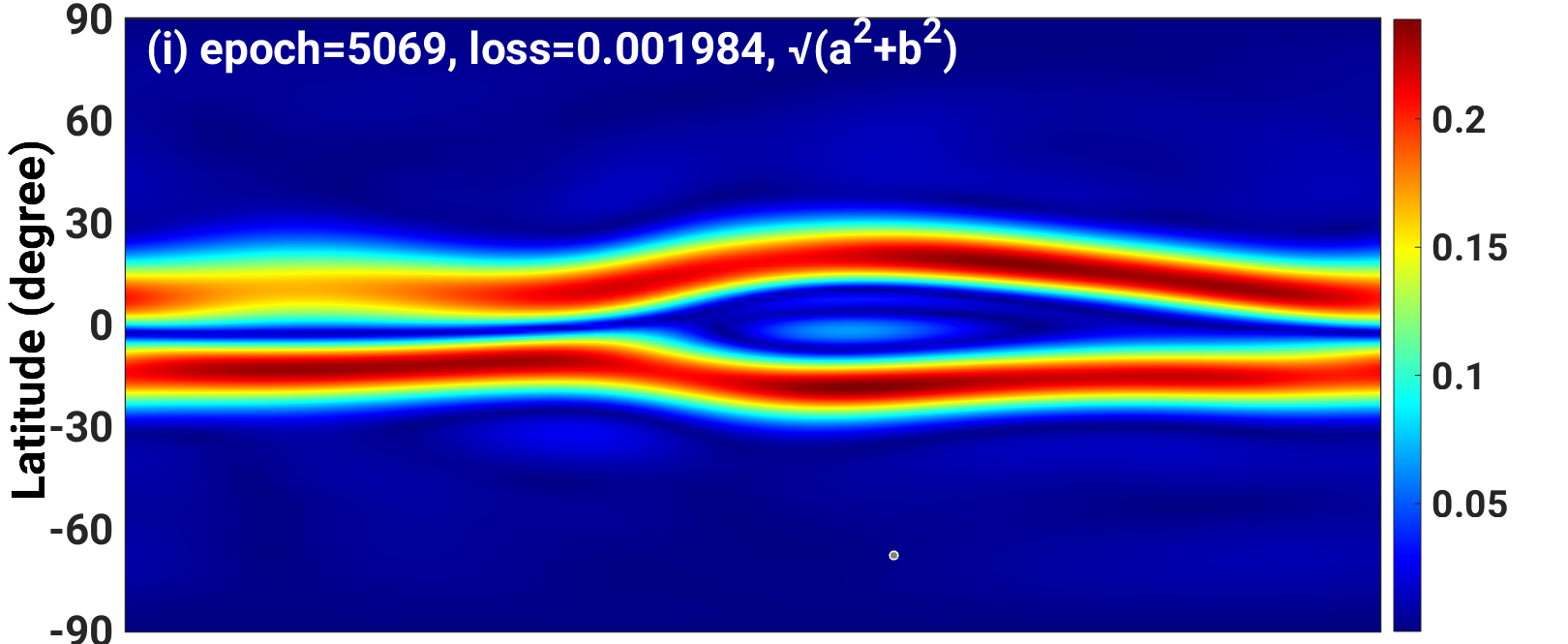}
\plotone{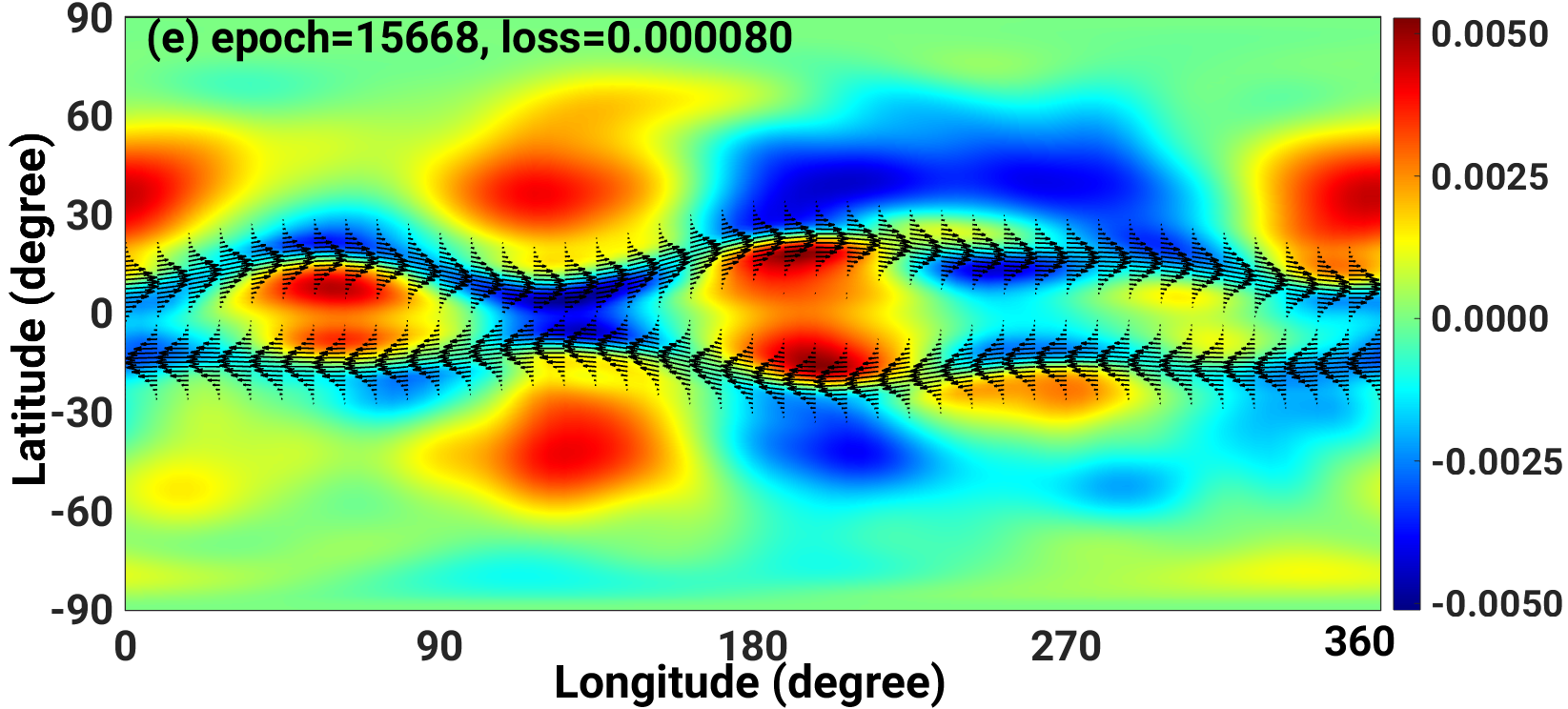}
\plotone{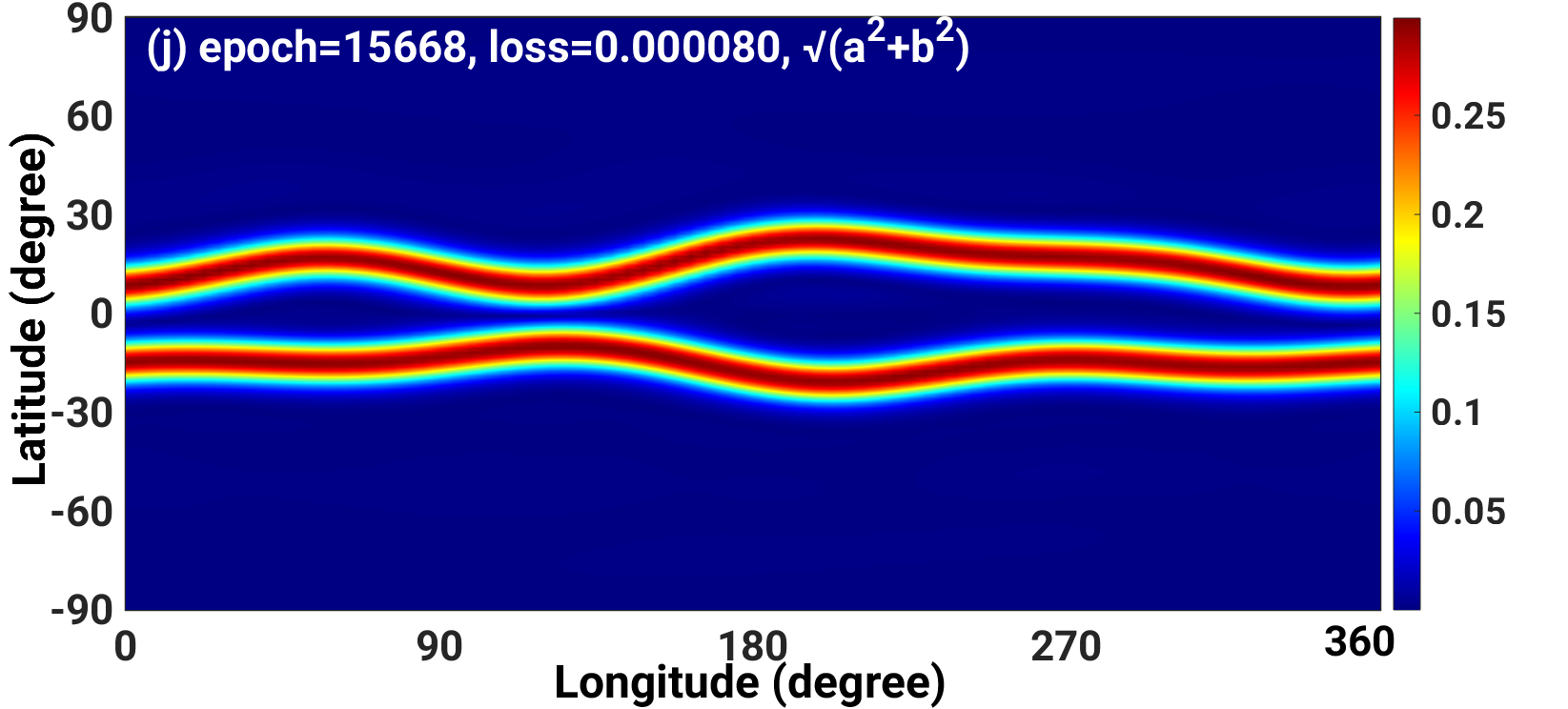}
\caption{PINN-simulated MHD shallow-water solutions to replicate the observed warped toroidal magnetic bands of February 14, 2024. Left panels: magnetic field vectors (black arrows) overlaid on height-deformation contours ($h-<h_\phi>$) in rainbow color-map, where red (blue) indicates bulging (depression); right panels: magnitude of toroidal bands ($\sqrt{a^2+b^2}$). These two columns from top to the bottom display how the PINN progressively converges from broad, unstructured fields (high loss) to well-defined, antisymmetric toroidal bands that closely reproduce the observed longitudinal warping, demonstrating convergence of solution for global magnetic topology consistent with solar observations.}
\label{fig:pinn_cnvrg}
\end{figure}

We train the PINN for a range of $N_d\in[20, 80, 100, 120]$ for each $N_l\in[2, 4, 6, 8, 10, 12]$ with the same value of prescribed physical parameters. The blue curve in Figure~\ref{fig:pinn_loss}a shows the PINN loss as a function of epoch for $N_d=100$ and $N_l=10$. The monotonic version of the same is depicted in orange. We calculate the minimum value of Loss (Loss$_{min}$) as a function of $N_l$ for different values of $N_d$ as shown in Figure~\ref{fig:pinn_loss}b. We find that for a small $N_d$, such as 20, the loss is larger for all $N_l$, causing less accurate results. However, for $N_d\geq80$ and $N_l\geq10$, the Loss$_{min}$ goes below $10^{-4}$ in an asymptotic manner, and at this point the difference is attributed to statistical fluctuation from model initialization. We stick to an $N_d=100$ and $N_l=10$ as an optimal PINN architecture in terms of speed and accuracy, as the performance gain is not significant for a higher value of $N_d$ (such as 120) or $N_l$ (such as 12). We perform all further experiments with the same architecture.

`We identify 4 points from the loss vs epoch curve, as shown by red circles in Figure \ref{fig:pinn_loss}a, and demonstrate graphically in a latitude-longitude grid (displayed in Figure \ref{fig:pinn_cnvrg}) how the magnetic state-vectors and bulging patterns are changing with the epoch to ultimately converge to the solutions that have the lowest loss.
Five left panels of Figure \ref{fig:pinn_cnvrg} show the PINN-derived magnetic field vectors ($a, b$), overlaid on the top-surface deformation ($h-<h>_{\phi}$, shown in the rainbow color map) at five different training epochs, while the right panels display the magnitude of the toroidal field ($\sqrt{a^2 + b^2}$) at the respective epochs. The panels are arranged from top to bottom in order of decreasing loss, representing progressive convergence toward the observationally constrained AR distribution pattern. The PINN model is trained to satisfy the full set of MHD shallow-water equations, enforcing mass conservation and the divergence-free magnetic field condition, while simultaneously matching the observed warped toroid pattern of February 14, 2024 (Figure \ref{fig:AR_toroid}). Through this physics-informed training, the network identifies initial state vectors at $t=0$ that are physically consistent due to the minimization of the residuals of the governing Equations (4–9), and reproduction the observed large-scale warped magnetic bands.

At high loss (top row), the PINN solution exhibits broad, low-latitude magnetic field structures and large-scale dipolar asymmetry—indicating an early-stage approximation of the global field without clear toroidal band confinement. As training progresses (second, third and fourth rows), warped toroidal band emerges, and the magnetic pattern becomes concentrated in two antisymmetric belts about the equator, qualitatively matching the toroidal bands inferred from observations.

By the lowest-loss stage (bottom row), PINN reproduces well-defined, warped toroidal bands that align closely with the observed toroidal structure seen in the synoptic map Figure~\ref{fig:AR_toroid}. The southern toroid, in particular, displays longitudinal undulation near 180°–200°, consistent with the observed deformation linked to active-region clustering and flux emergence at those longitudes. Other locations of active-region clustering, such as near zero-degree in the north, poleward side of the band coincide with the bulging (yellow-orange-red), whereas for $0-60^{\circ}$ locations equatorward side of both bands in the north and south coincide with bulging. Around $130^{\circ}$ longitude in PINN reference frame (150$^{\circ}$ absolute longitude), the two bands come closer to each other; there exists a depression (blue color) across the equator at that longitude, but for both bands the poleward sides coincide with the bulging, revealing the plausibility of AR emergence there. Comparing with Figure \ref{fig:AR_toroid}, it can be seen that he ARs indeed emerged there. The phase and amplitude of the simulated warps correspond closely to those observed, indicating that the PINN has successfully captured the warping of the toroidal field.

\begin{figure}[t!]
\centering
\hspace{0.0\textwidth}
\includegraphics[width=1.0\linewidth]{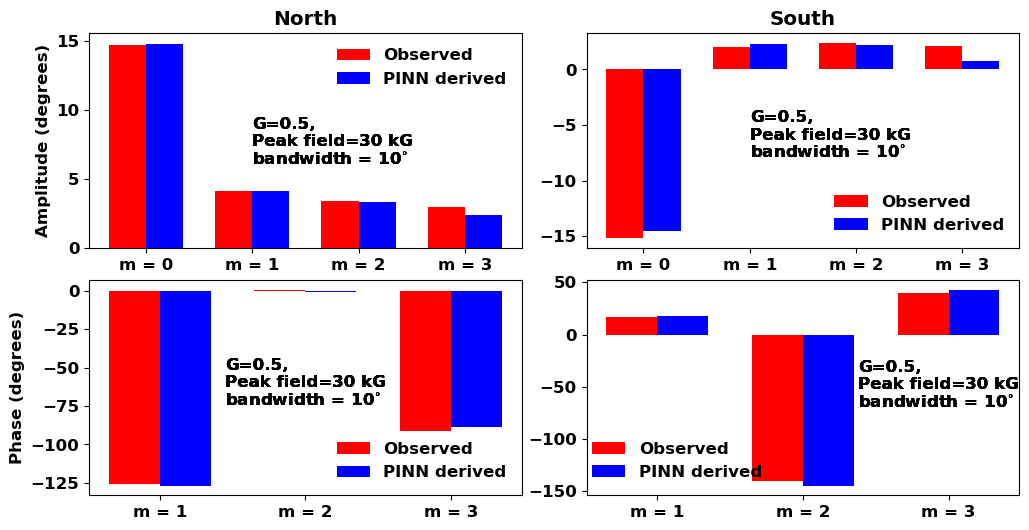}
\caption{Quantitative correspondence between observed vs. PINN-derived toroidal structure for G=0.5, 30 kG peak field and 10$^{\circ}$ bandwidth. We derive the amplitude (top row) and phase (bottom) of different modes ($m = 0, 1, 2, 3$) in North (left column) and South (right column) for both observed (red bars) and PINN-derived toroids (blue bars) through correlation with sine and cosine basis functions. Apart from the amplitude of $m=3$ in South, we find a near-perfect match for all the modes in amplitude and phase between observation and PINN solution for both hemispheres.}
\label{fig:toroid_modes}
\end{figure}

In essence, the final PINN solution replicates the key morphological and dynamical features of the observed toroidal structure, namely for two antisymmetric toroidal bands of Gaussian profile with $10^{\circ}$ latitudinal width and 30 kG peak field strength, centered at/near $\pm 15^{\circ}$ latitudes, longitude-dependent warping, consistent with modulation created through nonlinear interactions among magneto-Rossby waves, toroidal magnetic field and tachocline differential rotation, is obtained, mirroring the hemispheric tipped-away patterns seen in the synoptic magnetogram. To quantify the match between the observed toroid and PINN solution, we first derive from the field strength ($\sqrt{a^2+b^2}$) map (Figure~\ref{fig:pinn_cnvrg}j) the peak strength latitudes for North and South toroids at all longitudes. This provides a latitude vs longitude curve ($Pc(\lambda)$) for the PINN derived solution. We then correlate it with cosine and sine basis functions to derive respective $m>1$ amplitudes as $A_m = \frac{1}{\pi}\int P_c(\lambda)\cos m\lambda d\lambda, B_m = \frac{1}{\pi}\int P_c(\lambda)\sin m\lambda d\lambda$ and the $m=0$ amplitude as $A_0 = \frac{1}{2\pi}\int P_c(\lambda) d\lambda$. We derive the phases ($\zeta_m$) by satisfying the equations $A_m = \sin \zeta_m, B_m = \cos \zeta_m$. As depicted in Figure~\ref{fig:toroid_modes}, we see that the converged PINN solution produces mean north and south toroid latitudes (i.e., $m=0$ amplitude) close to the observed target of $\pm15^{\circ}$. For higher modes ($m>0$), we achieve a great match between the observed and PINN-derived toroids (except for $m=3$ amplitude in South) with a relative error ($100|\frac{observed - PINN}{observed}|$) of $<19\%$ in both amplitude and phase for both hemispheres. 

This close agreement demonstrates that the PINN-MHD-shallow-water approach can self-consistently recover large-scale, physically plausible magnetic field configurations that reproduce observed global toroidal deformations. It thereby offers a powerful data-constrained framework for initializing the solar tachocline MHD shallow-water model-system for forward integration to simulate evolution of global toroid patterns for weeks ahead.

\subsection{PINN-solutions across solar-like parameter regimes}

The warped toroid patterns of emerged active regions provide important
observational constraints on the latitude, width, and geometry of the
underlying toroidal magnetic bands in each hemisphere. Guided by these
patterns, we train the PINN to derive self-consistent initial state
vectors for multiple solar-like conditions. Following the observational
guidance of Figure \ref{fig:AR_toroid} and arguments provided in the
introduction \S1, we can consider bands at $\pm 15^{\circ}$ latitudes,
and bandwidths covering from narrow to somewhat broader bands of Gaussian profile, namely 
within the range of $5^{\circ}-15^{\circ}$ latitudinal full width half maxima (FWHM, marked as bandwidth in Figures 5-8) in steps of $5^{\circ}$. Because 
the correspondence between the AR field-strengths and the strength of
the tachocline toroidal bands, from which the ARs likely originate and
manifest at the surface, is not directly obtainable yet through 
observations and/or models, we explore a broad range of plausible peak
toroidal field strengths, from 2 to 100 kG, consistent with both thin 
and wide flux-tube emergence simulations. Global MHD shallow-water 
models also include the effective gravity parameter $G$, representing 
the degree of subadiabaticity; we consider $G=0.5$ for the overshoot 
region and $G=10$ for the radiative tachocline. Since the tachocline
dynamics are relatively insensitive to the precise amplitude of 
differential rotation, the pole-to-equator differential rotation is 
fixed at 21\% throughout.

\subsubsection{PINN-derived state-vectors for various peak field strength}

\begin{figure}[ht]
\epsscale{0.57}
\hspace{-0.014\textwidth}
\plotone{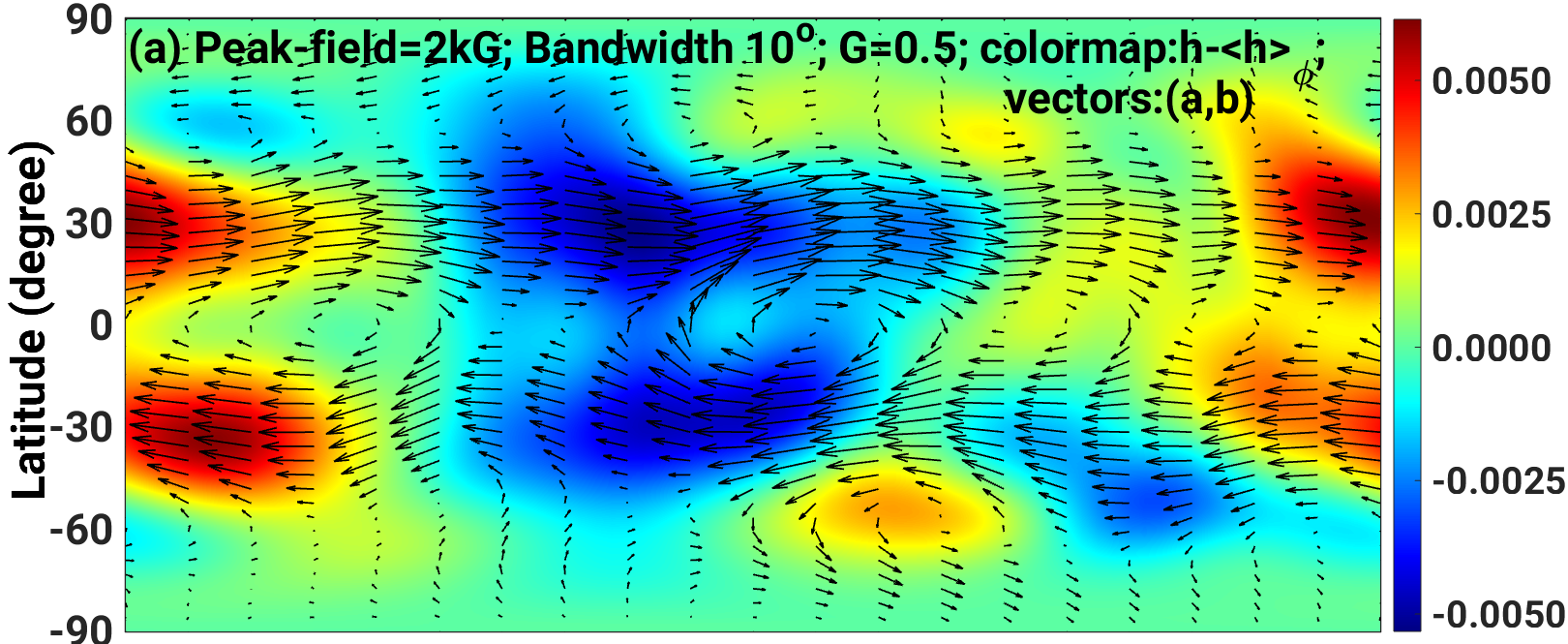}
\plotone{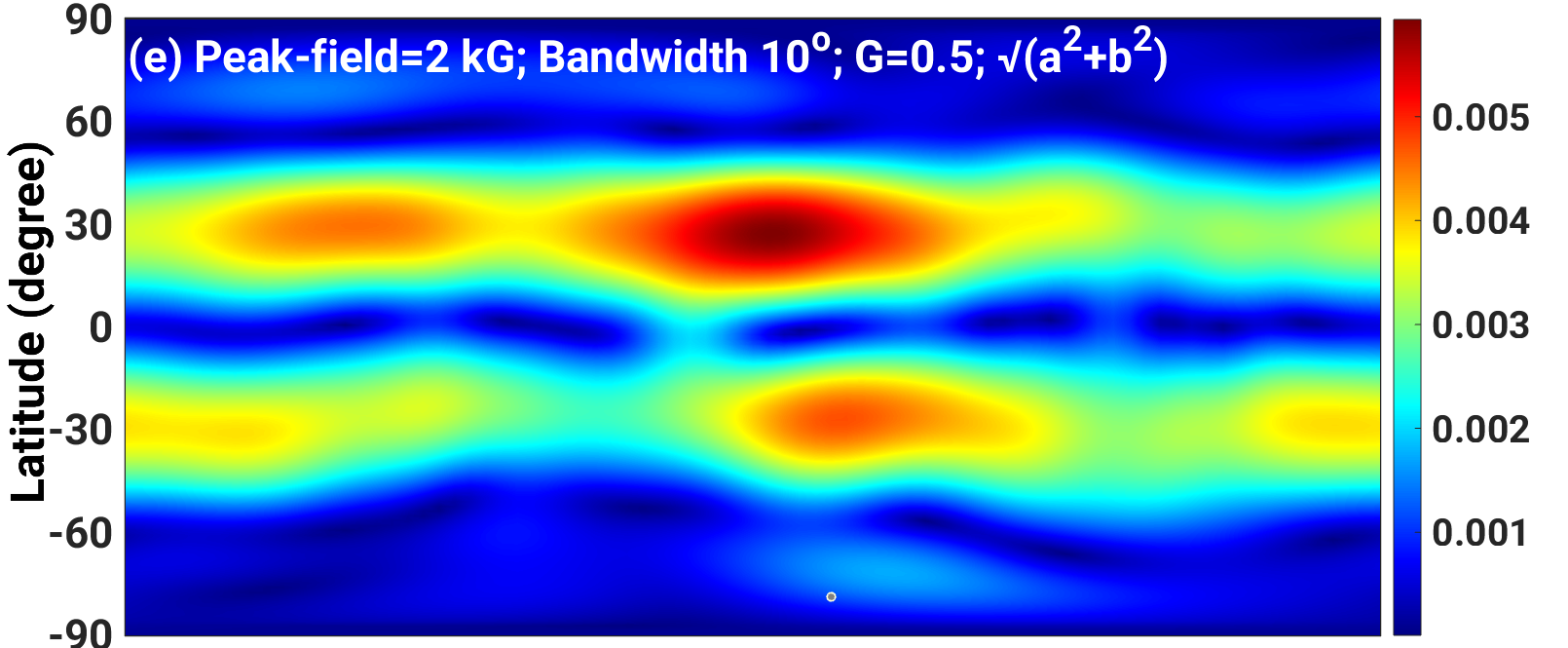}
\plotone{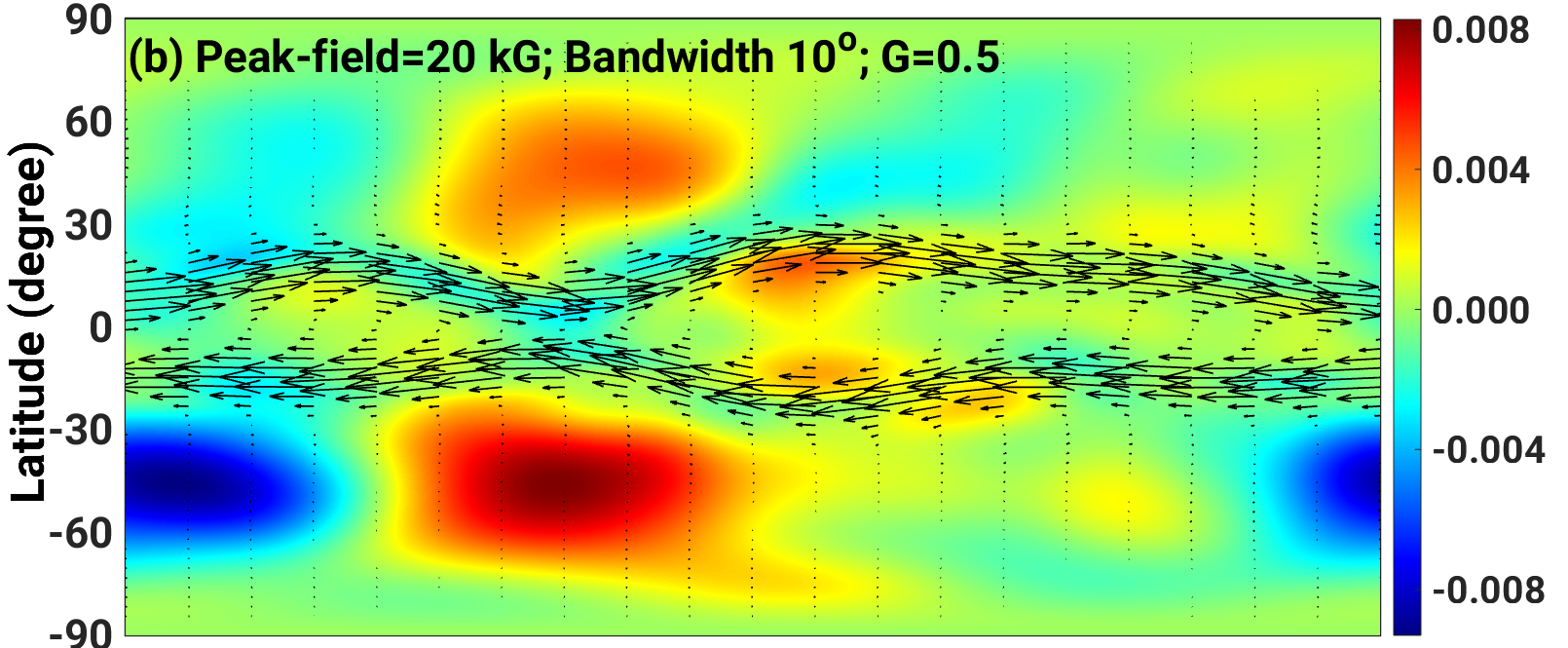}
\plotone{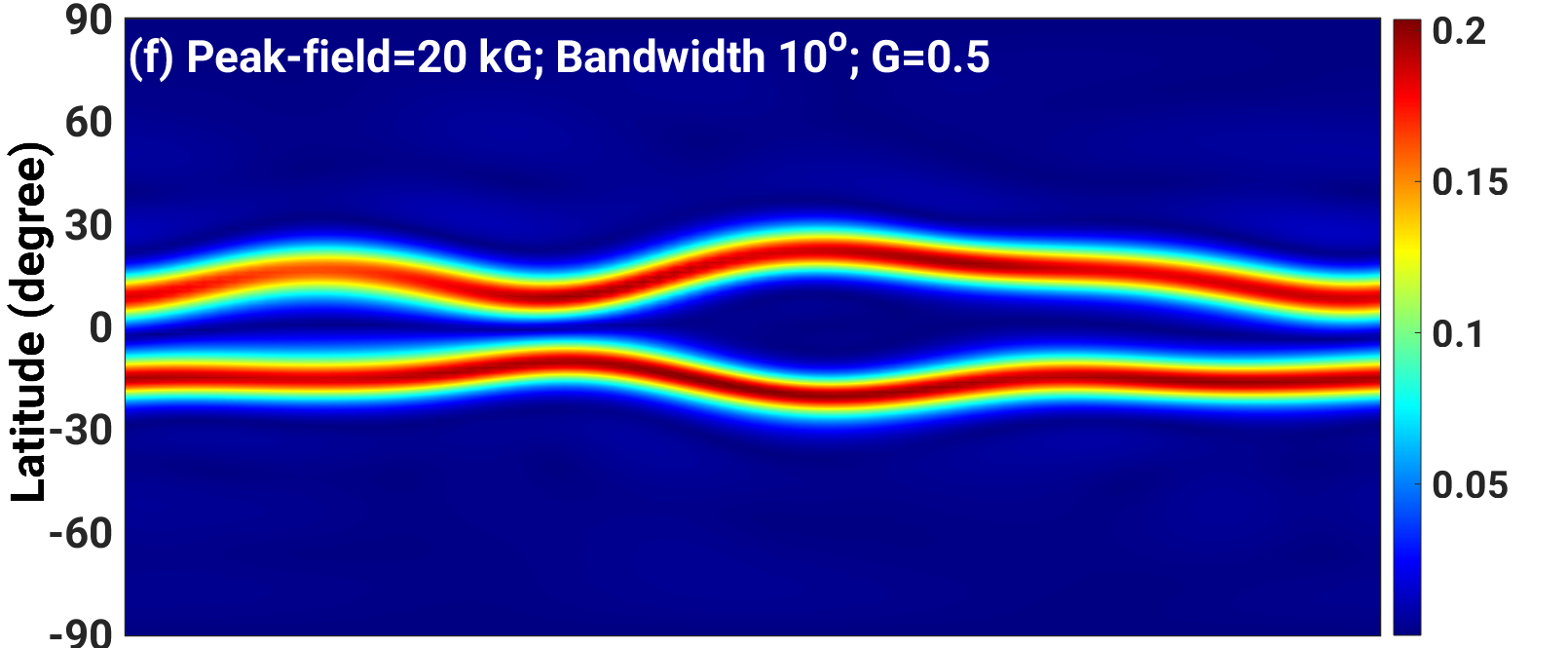}
\plotone{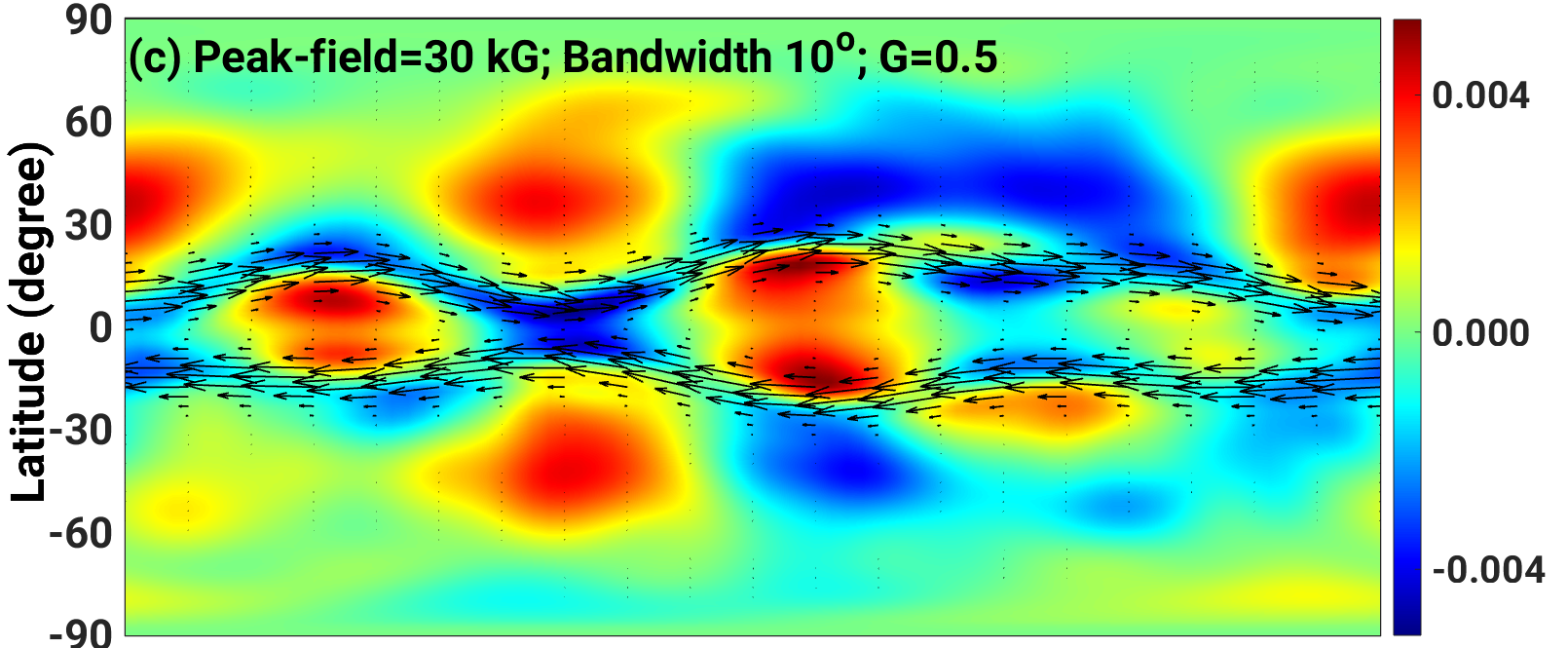}
\plotone{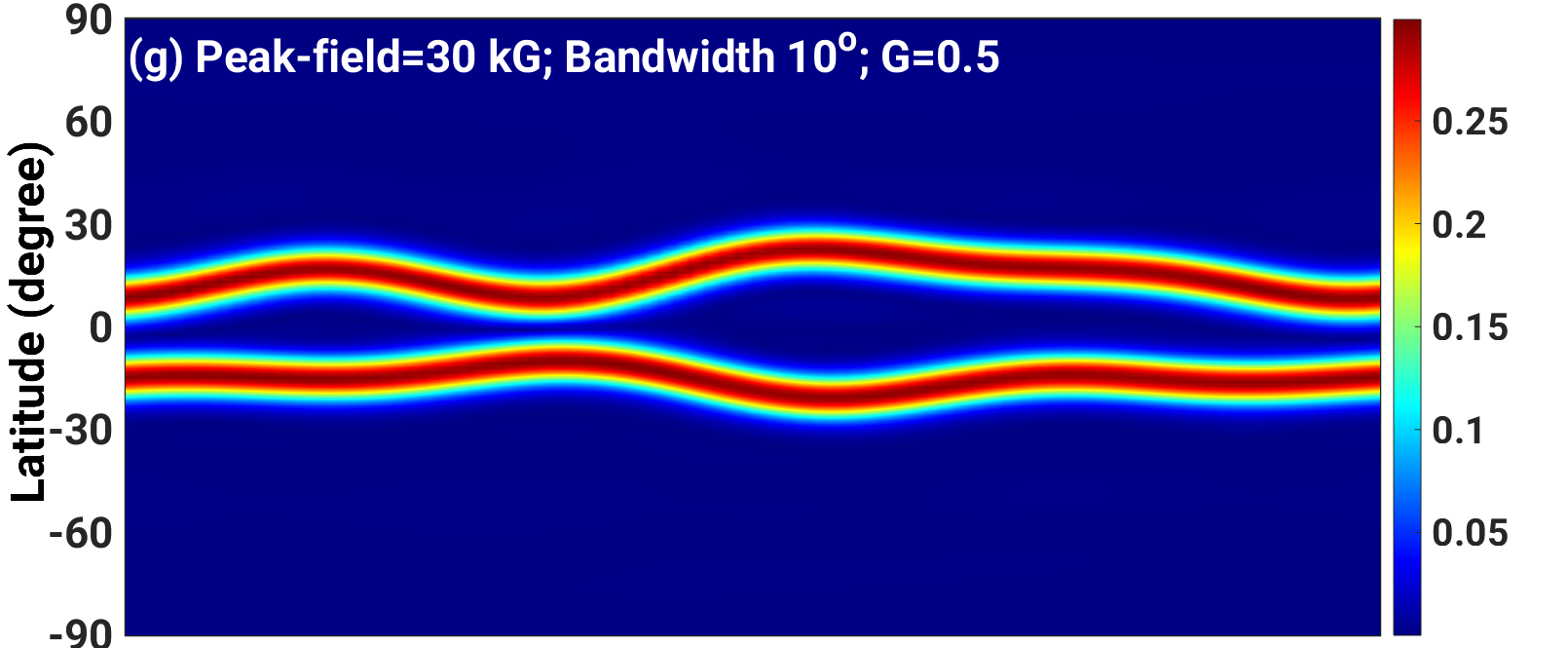}
\plotone{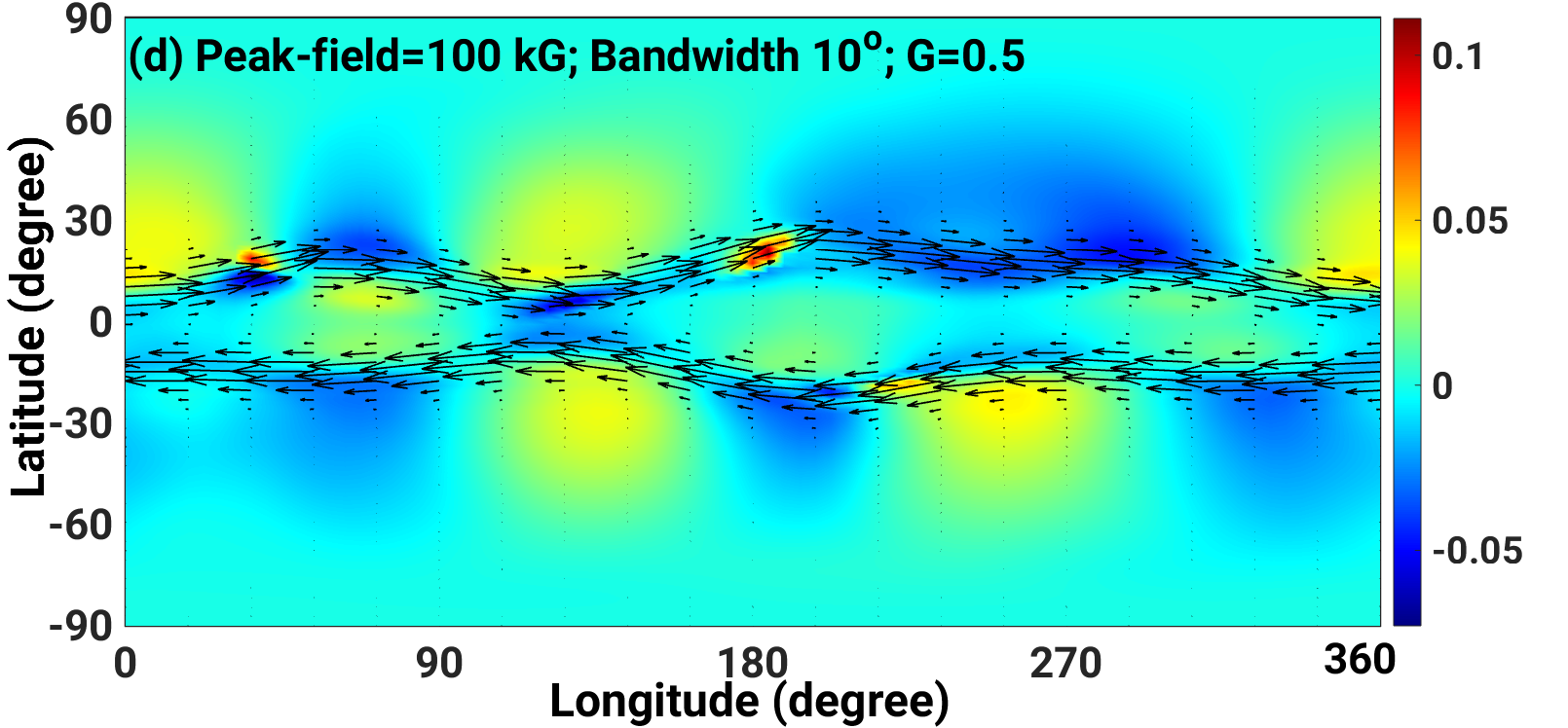}
\plotone{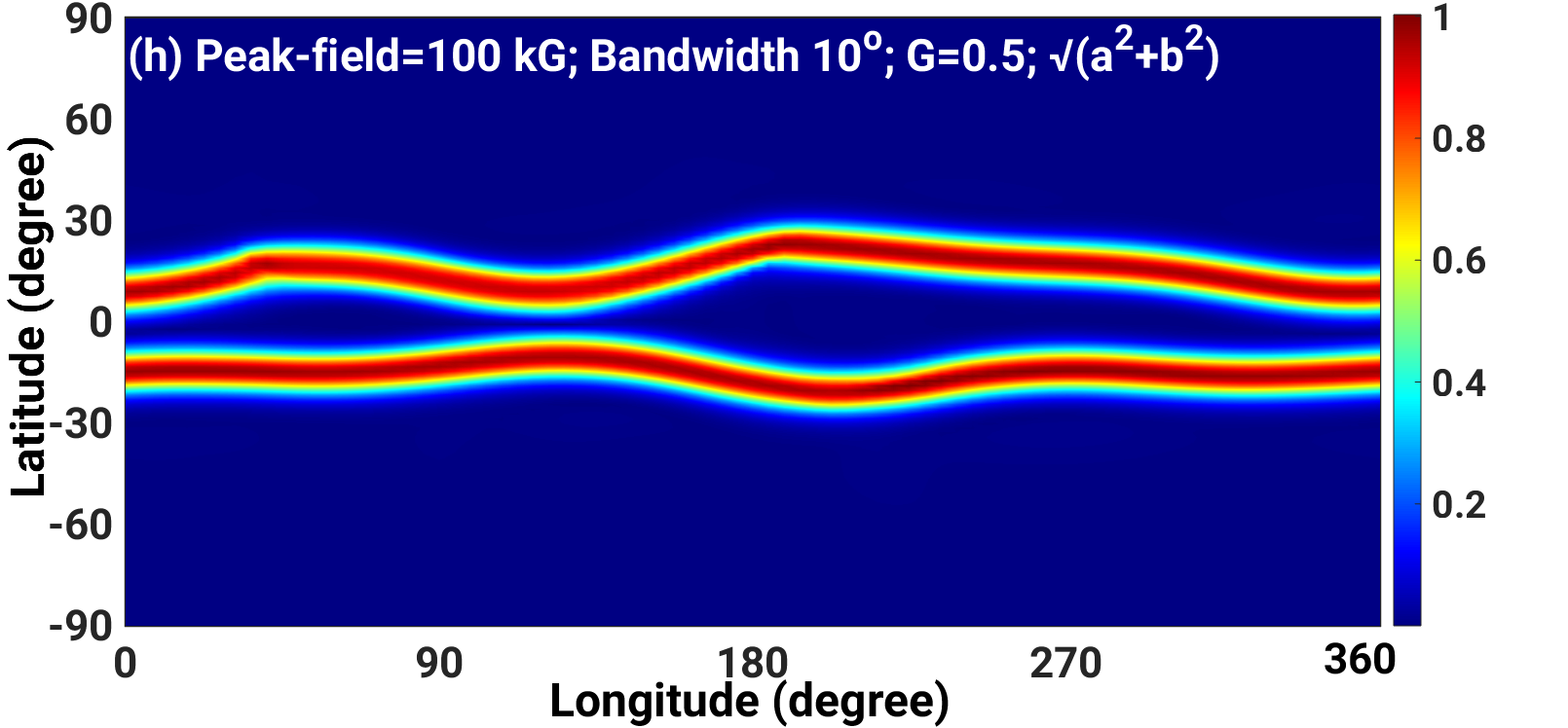}
\caption{Visualization of PINN-derived initial state vector for multiple peak field strengths given G=0.5 and FWHM=10 deg. The four rows from top to bottom show results for peak field strengths of 2 kG, 20 kG, 30 kG, and 100 kG, respectively. The left column shows derived bulging (in red) and depression (in blue) regions from derived longitudinal mean subtracted height profiles ($h-<h>_{\phi}$). The velocity field, shown with white arrows, is overlaid on the height profile. The right column shows the derived field strengths ($\sqrt{a^2+b^2}$) at all latitude and longitude locations. We achieve the best results for 30 kG, both in terms of matching mode configuration with the observed toroids and enforced peak field strength. 
}
\label{fig:g_0.5_fwhm_10}
\end{figure}

Starting with $10^{\circ}$ bands at $\pm15^{\circ}$ latitudes we implement
PINN, along with MHD shallow-water model equations and observations of
Figure \ref{fig:AR_toroid}, and derive the state-vectors for four different peak field
strength, namely 2, 20, 30 and 100 kG.

Figure \ref{fig:g_0.5_fwhm_10} presents the PINN-derived state vectors for four
representative peak field strengths (2, 20, 30, and 100 kG) for
oppositely-directed $10^{\circ}$ toroidal bands at $\pm 15^{\circ}$
latitudes. The four rows show cases with progressively stronger
tachocline toroidal fields from the top to the bottom. Left panels display the height-deformation/pressure-departures ($h - <h>_{\phi}$), on which are overlaid magnetic field vectors ($a,b$), while the right
panels map the derived toroidal field magnitudes $\sqrt{a^2+b^2}$.
The panels (a,e) reveal that, for the weakest field case (2 kG),
the $(a,b)$ state-vector solutions (panel a) as well as toroidal
field magnitudes (panel e) converge to wider bands. It is not
surprising that the PINN-derived features reflect the regime, more
hydrodynamically- than magnetically-dominated, in which magnetic
energy is smaller compared to the kinetic and potential energies
of the system. The PINN solution corresponds to the excitation of
low-order hydrodynamic modes, namely $m=1$ symmetric and $m=2$
antisymmetric modes. Despite inclusion of magnetic fields, these
modes can certainly be identified as hydrodynamically-dominated
modes, because their phase speeds correspond to higher latitudes
(see Figure 5 of  \citet{Dikpati_2018}) than
the local differential rotation speed at the toroidal band-location.
The PINN gets informed about this physics from the model, and hence
finds the solution as displayed in Figure \ref{fig:g_0.5_fwhm_10}(a,e). The loss value
in this case is $\sim 0.000098$.

For 20 kG field, the PINN derives stronger, more confined bands and a
clearer latitudinal banding in the field-magnitude map, displayed respectively in panels (b) and (f).
Comparison with synoptic observations, warped patterns and latitudinal
bandwidths are reasonable match. Pretty significant north-south
asymmetry is revealed from both the panels (b) and (f), respectively
in height-deformation map (panel b) and toroidal field amplitudes
(panel f). And the loss obtained in this case is $\sim 0.000043$, which implies, not surprisingly, a pretty good convergence of PINNBARDS.

However, the results for the 30 kG case provides a better overall
agreement with the observed toroids of Figure \ref{fig:AR_toroid}, compared to
that for 20 kG case. Not only are the warping patterns and latitudinal
band-widths a reasonable match, north-south asymmetry is optimal.
Here the north-south asymmetry is seen in the height-deformation
(color-map in panel c), but not in toroidal field amplitude
displayed in panel (g). The loss in this case is again very good ($\sim 0.000078$).

For the cases of 20 kG and 30 kG fields, the reasons of good matches
between the PINN-derived features of the state-vectors and that of the
observed toroid-patterns of Figure \ref{fig:AR_toroid} are that the physical regime
for these two cases change from hydrodynamically-dominated to
magnetically-dominated regimes; $m=1$ symmetric and $m=2$ antisymmetric
modes are excited and can provide suitable frameworks to the PINN for
successful derivation of the state-vectors. The PINN loss is lowest for
these two cases, indicating a close fit to both the MHD constraints and
the target surface pattern.

For the strongest specified field (100 kG) the PINN-solution for the
magnetic bands fit the observational constraint except for the kinks
in the north toroidal band, which are clearly visible at $\sim
45^{\circ}$ and $\sim 210^{\circ}$ longitudes (see panel h). To
derive this solution through the training of observations and
physical model, the PINN has to adjust at those two longitudes the
height-deformations also in such a way as to create unusually sharp
bulges. PINN provides the solution with a loss value of $\sim 0.000092$, pushing itself in an over-driven regime where the dominant modes
differ from those inferred observationally. This over-driven situation
arises because of the physics of the system. Global MHD tachocline
model's eigen-system points out that the only mode excited for such a
strong field at $\pm 15^{\circ}$ latitude is $m=1$ symmetric mode
(see figure 5c of \citet{Dikpati_2018}). It is
nearly impossible to derive a PINN-solution matching our target surface
pattern with only one $m=1$ mode, because a 100 kG toroidal band behaves
like a "steel ring", which only tips, but cannot deform to produce the
required warping. pattern. Such an extreme field strengths also conflict
with expectations from some flux-emergence models, particularly those that
consider non-thin tubes of finite extent, and thus are less plausible for
reproducing real surface toroid warps.

In brief, we find the following: (i) there is a clear sensitivity of the
warped-toroid morphology to subsurface peak field strength; (ii) weak
fields fail to produce coherent low-order warping, overly strong fields excite nonphysical kinky structure in the derived warped bands, and (iii)
an intermediate strength (20-30 kG) yields the best match for toroidal
bands of each $10^{\circ}$ width at $\pm 15^{\circ}$ latitudes in this
PINN-MHD shallow-water framework.

\subsubsection{PINN-derived state-vectors for narrower and wider than 10-degree bands}

\begin{figure}[ht]
\epsscale{0.57}
\hspace{-0.014\textwidth}
\plotone{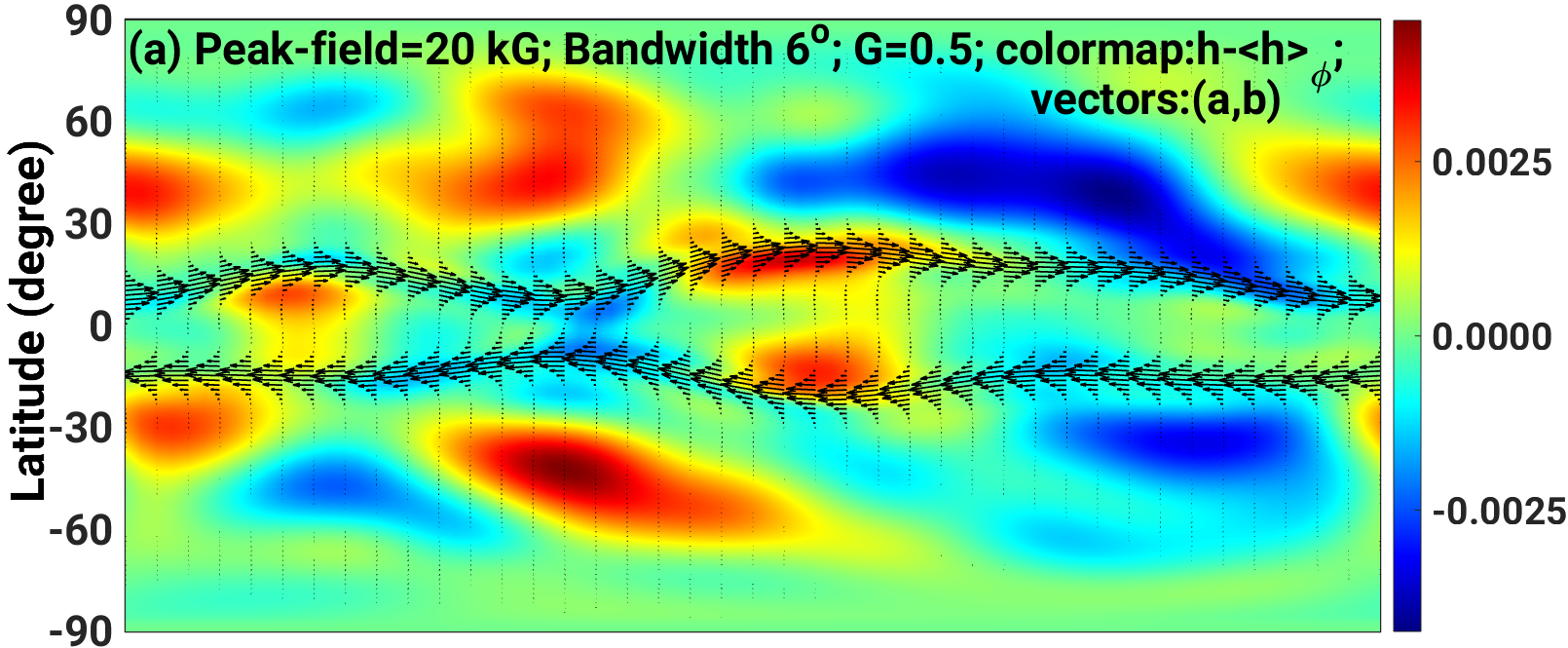}
\plotone{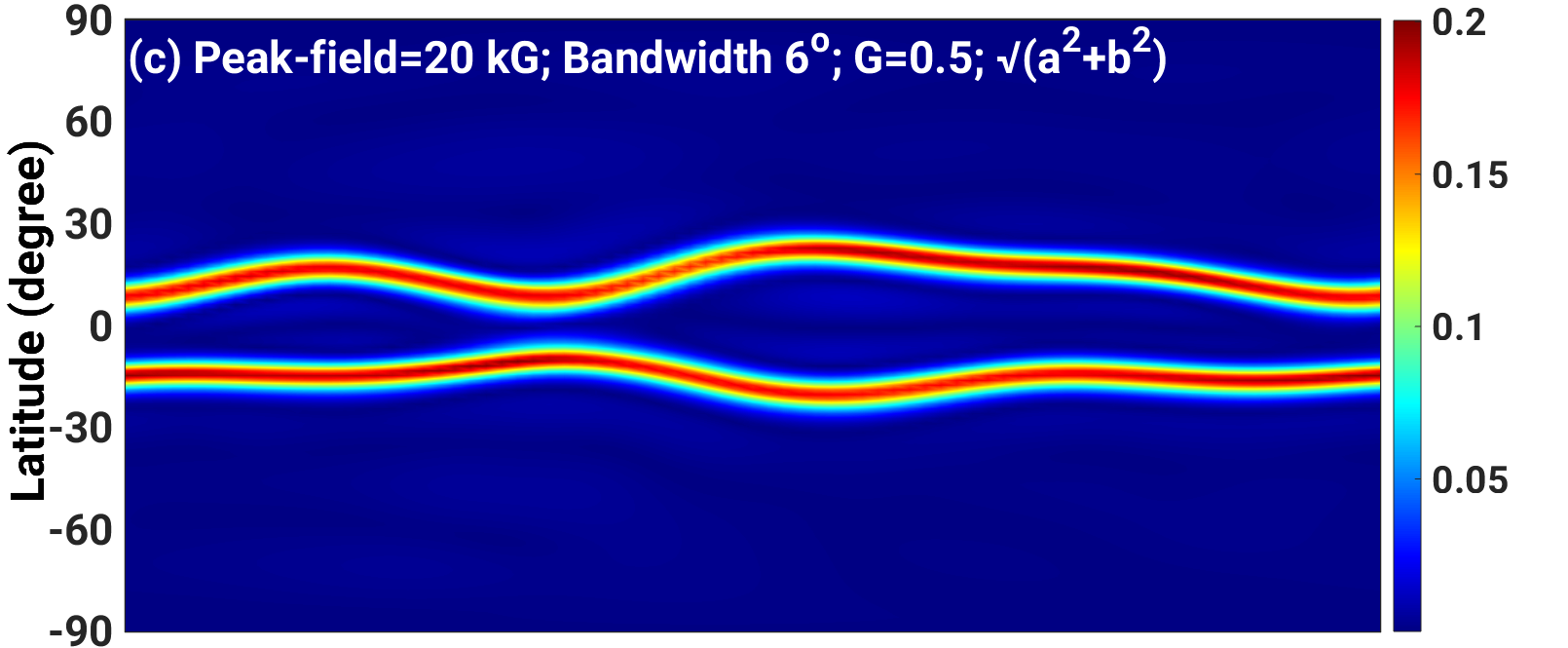}
\plotone{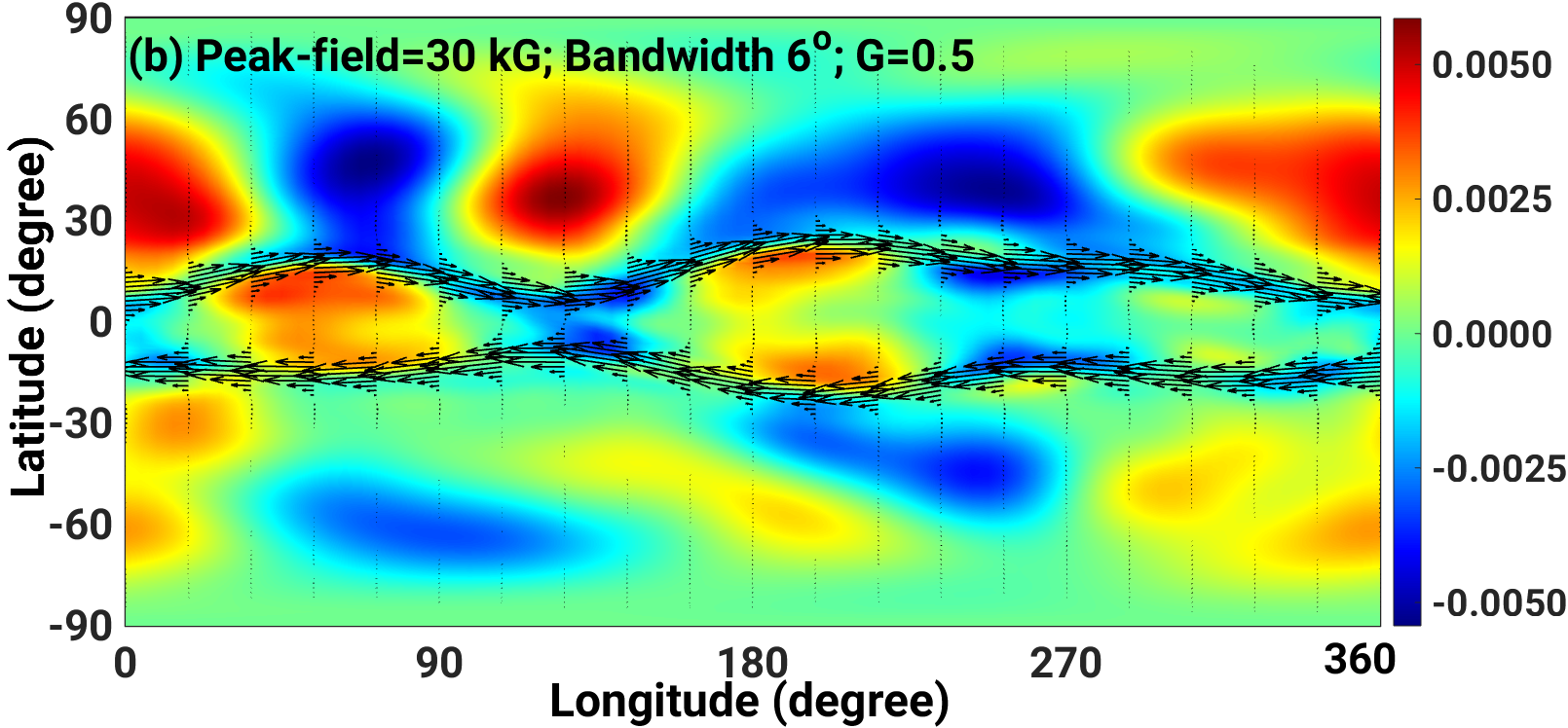}
\plotone{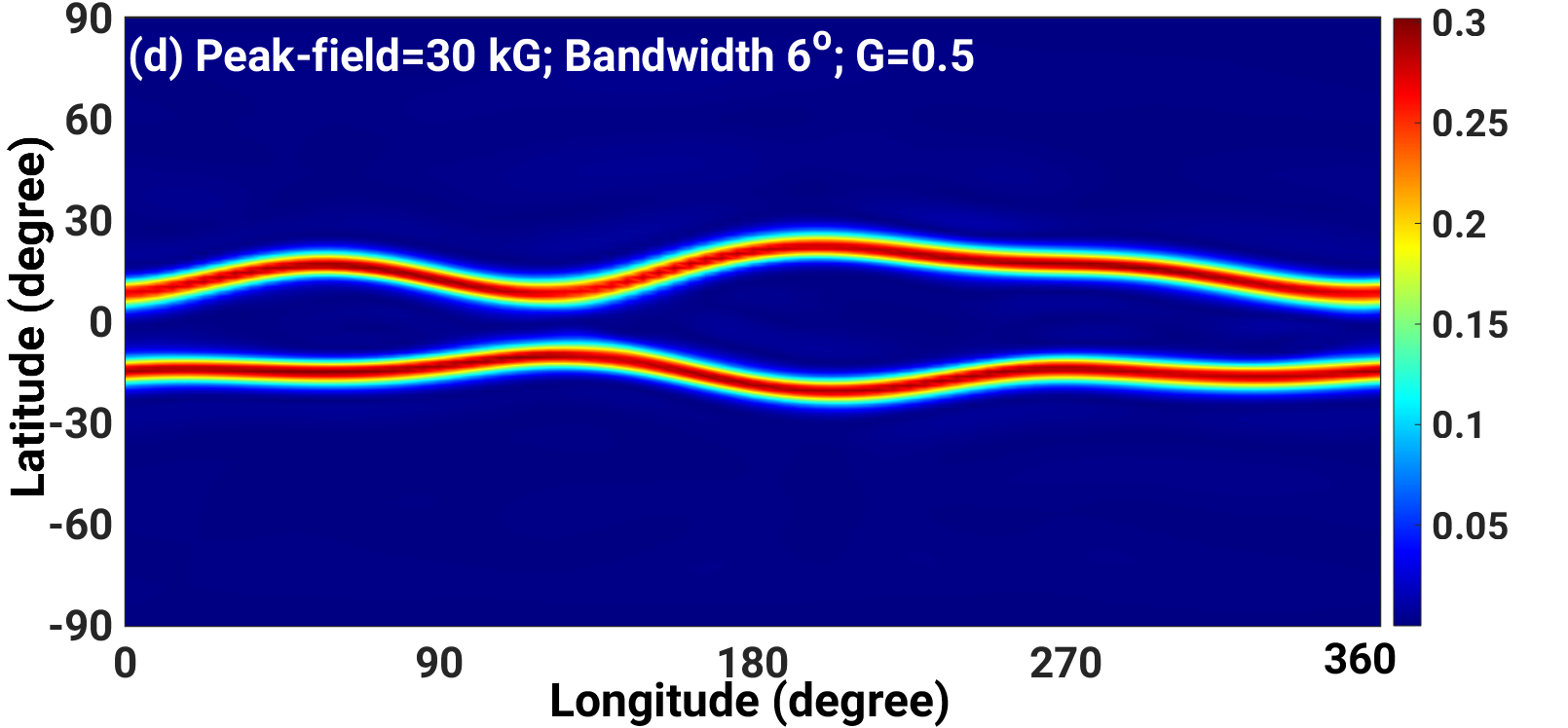}
\caption{PINN-derived initial state vector for $6^{\circ}$ bands with peak field strengths of 20 kG (top) and 30 kG (bottom) respectively. Again left column shows derived height-deformations ($h-<h>_{\phi}$, i.e. departure from mean shell-thickness) in the form of bulging (red-orange) and depression (blue-sky blue) regions. Magnetic field vectors are overlaid on height-deformation in black arrows. Right column shows amplitude of magnetic fields ($\sqrt{a^2+b^2}$) at all latitude and longitude locations. Losses are $\sim 0.000119$ and $\sim 0.000071$ respectively for 20 and 30 kG field strengths.} 
\label{fig:g_0.5_fwhm_6}
\end{figure}

In \S3.2.1 for $10^{\circ}$ bandwidth, we find that experiments with
wide range of field-strengths using the PINN–MHD shallow-water formalism
reveal certain sensitivity of warped-toroid morphology to the underlying
tachocline field strength. Weak fields ($<10$ kG) fail to sustain coherent
low-order deformations to produce tight warped toroid patterns, while
very strong fields ($>50$ kG) produce overly rigid morphologies containing
some unphysical features, like kinky structures in toroids. Intermediate
field strengths of 20–30 kG yield the best correspondence with
observations, both in the geometry of the toroidal bands and in the
balance between magnetic and hydrodynamic effects. The obvious next question
is: how does the PINN-MHD-shallow-water framework perform in state-vector
derivation for narrower and wider bands than $10^{\circ}$ latitudinal width?

In Figure \ref{fig:g_0.5_fwhm_6} we present results for $6^{\circ}$ toroidal bands with
peak field strengths of 20 kG and 30 kG. Again, like Figure \ref{fig:g_0.5_fwhm_10},
left panels display the height-deformation/pressure-departures
($h - <h>_{\phi}$) in rainbow colormap, and black arrows overlaid
on that denote magnetic field vectors ($a,b$); the right panels display
the derived toroidal field magnitudes $\sqrt{a^2+b^2}$. 

The PINN-derived
state vectors reveal several key features. First, the solutions for the
state-vectors for both the 20 kG and 30 kG field strengths appear as
combinations of low-order longitudinal modes. This feature can be seen
in the arrow-vectors in panels (a,b) as well as in toroidal field
amplitudes in panels (c,d). Warped toroid patterns reveal good match with the
observed north and south toroids of Figure \ref{fig:AR_toroid}.

Second, north-south asymmetry is seen in both cases, and is prominent
the height-deformation colormaps. Such asymmetry has also been seen
for $10^{\circ}$ bandwidth cases in Figure 5 (panels b,c),
but by contrast, the asymmetry in panels (a,b) of Figure \ref{fig:g_0.5_fwhm_6} is modest
for 20 kG field strength, and a little more enhanced for 30 kG case.

Third, due to narrower bandwidth, magnetic fields at $\sim 130^{\circ}$
longitudes (in PINN reference frame i.e. $\sim150^{\circ}$ absolute longitude) do not coincide with the bulges. Although the poleward side
of both north and south bands are in neutral thickness region at
$\sim 130^{\circ}$ longitudes (in PINN reference frame), this reduces the likelihood of AR emergence
there to some extent compared to corresponding $10^{\circ}$ band cases.
Thus this deteriorates the match with observations of February 14, 2024
toroid patterns, which contain emerged ARs at $130^{\circ}$ (in PINN reference frame) also. However, this $15^{\circ}$ band case (wider than $10^{\circ}$ is important in its own right, because the band being wider, a portion can coincide with the bulges, creating possibilities of AR emergence. 

\begin{figure}[ht]
\epsscale{0.57}
\hspace{-0.014\textwidth}
\plotone{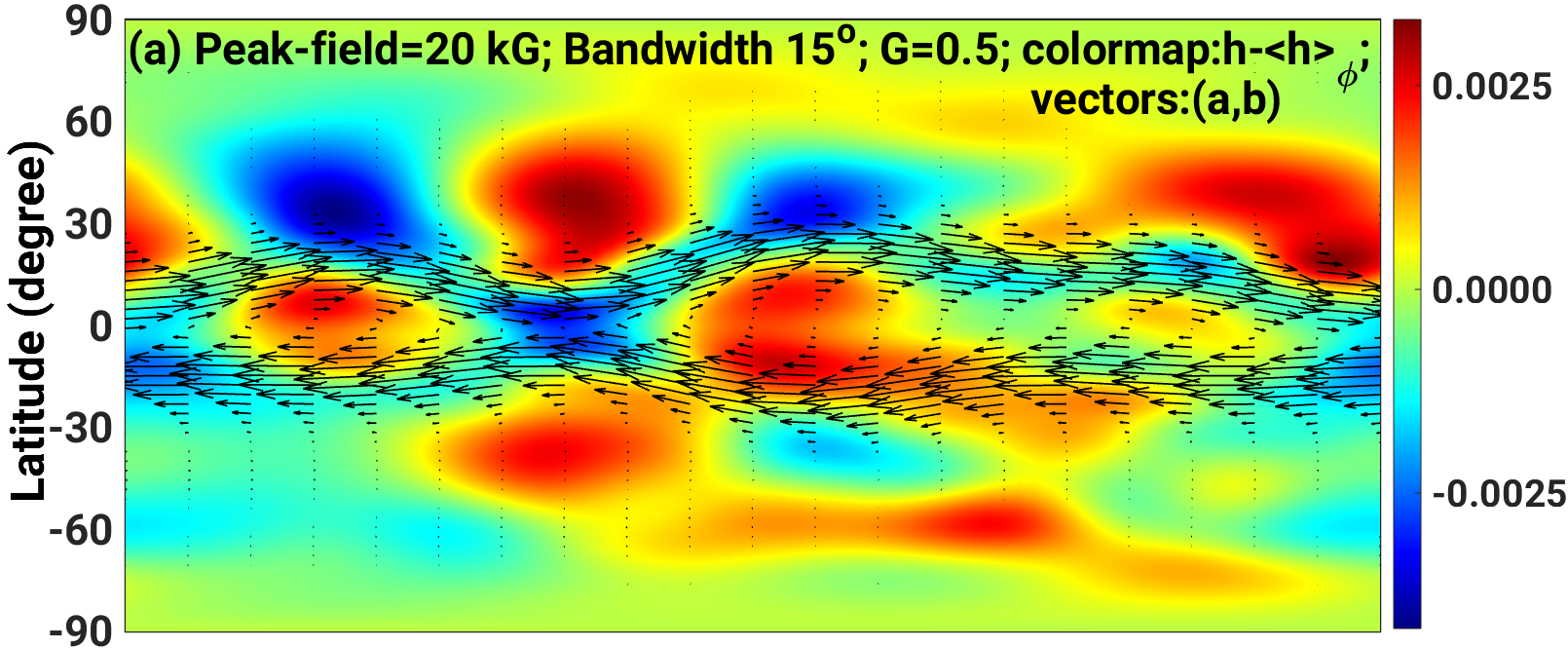}
\plotone{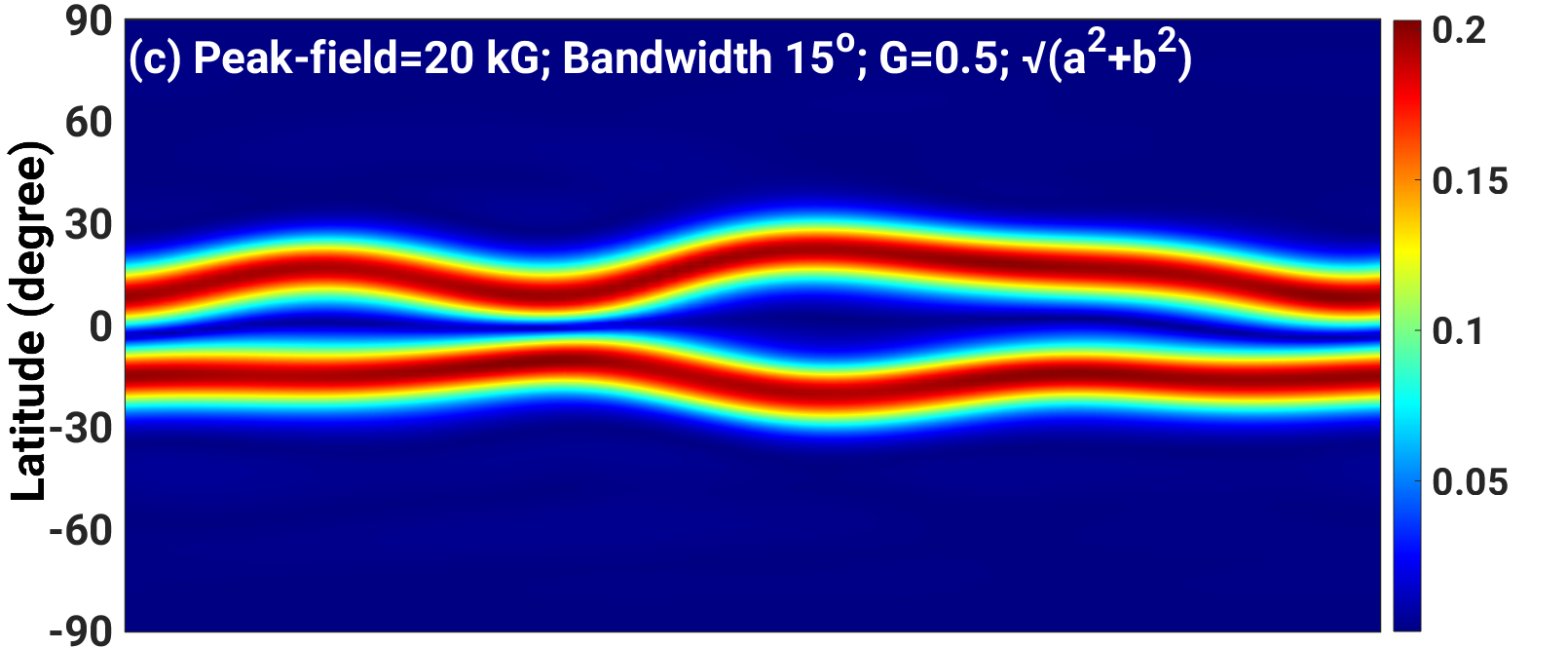}
\plotone{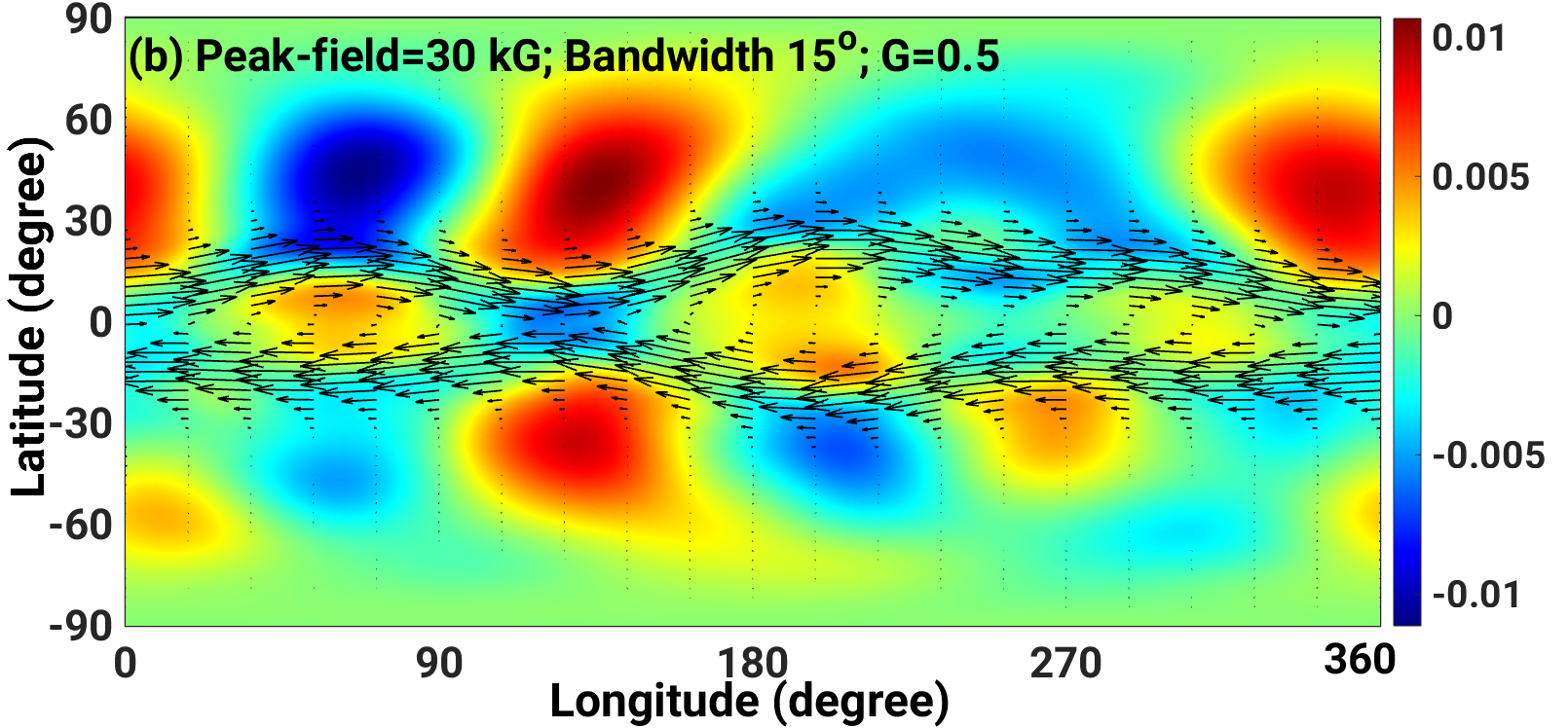}
\plotone{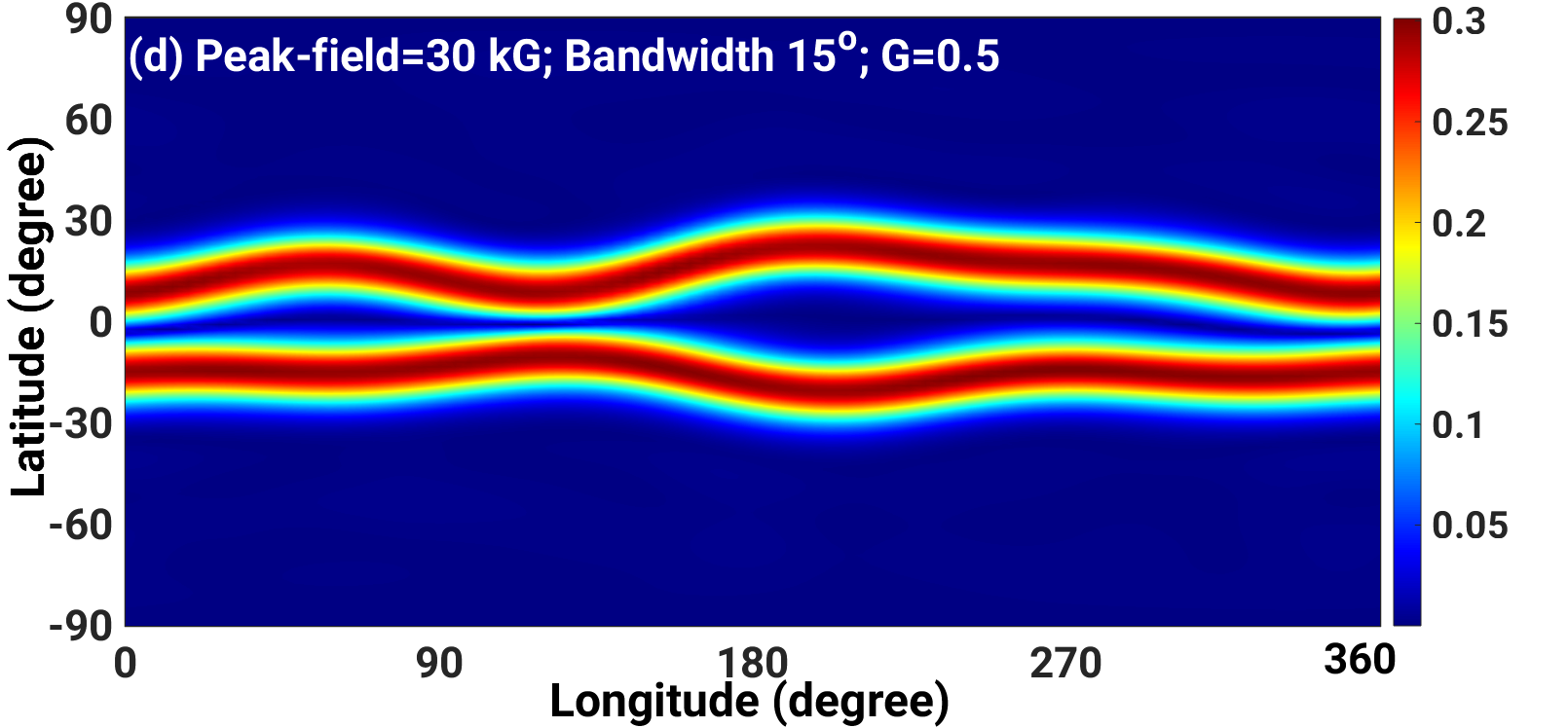}
\caption{Visualization of PINN-derived initial state vector for multiple peak field strengths given G=0.5 and FWHM=10 deg. The four rows from top to bottom show results for peak field strengths of 2 kG, 20 kG, 30 kG, and 100 kG, respectively. The left column shows derived bulging (in red) and depression (in blue) regions from derived longitudinal mean subtracted height profiles ($h-<h>_{\phi}$). The velocity field, shown with white arrows, is overlaid on the height profile. The right column shows the derived field strengths ($\sqrt{a^2+b^2}$) at all latitude and longitude locations. We achieve the best results for 30 kG, both in terms of matching mode configuration with the observed toroids and enforced peak field strength. The losses for 20 and 30 kG cases are respectively $\sim 0.000044$ and $\sim 0.000108$.} 
\label{fig:g_0.5_fwhm_15}
\end{figure}

To examine how the PINN derives the state-vectors for wider bands, we
consider toroidal bands with $15^{\circ}$ latitudinal width. Figure
\ref{fig:g_0.5_fwhm_15} shows the PINN solutions for magnetic field vectors ($a,b$)
inblack arrows, overlaid on height-deformation/pressure-departures
($h - <h>_{\phi}$) in rainbow colomap in the left panels (a,b), whereas
the right panels (c,d) display the magnitude of toroidal fields,
$\sqrt{a^2+b^2}$. We find that the warped toroid patterns in Figure \ref{fig:g_0.5_fwhm_15}
are broader and smoother, covering a larger latitudinal extent as expected
for a wider bands, compared to $6^{\circ}$ band case. Furthermore,
the north-south asymmetry is modest.

Comparison with observed toroids of Figure \ref{fig:AR_toroid} reveals that the
morphological match is qualitatively well, but it is degraded around
$0^{\circ}$ and $360^{\circ}$ longitudes. At these two ends, the
PINN-derived north and south toroids come closer to each other
compared to the observations in Figure \ref{fig:AR_toroid}. However, due to broader
band-structure in latitude, the overlap of the
north and south toroids, particularly their poleward sides, with the
bulges at $\sim 130^{\circ}$ (in PINN reference frame i.e. $\sim150^{\circ}$ in absolute longitude) longitudes, thereby indicating the
likelihood of AR emergence there.

Band-width alters which low-order modes are most easily excited.
For broader bands, $m=1$ modes get excited. As the bands become narrower,
more $m>1$ modes get excited along with $m=1$ modes. The PINN gets
trained with this physics of the MHD shallow-water model, and can more
flexibly combine various modes to derive the warped toroid patterns with
good match with the observed pattern. For broader bands, this becomes
a little difficult to derive the required warping with $m=1$ modes only.
The degradation at $0^{\circ}$ and $360^{\circ}$ longitudes for the
toroidal bands of $15^{\circ}$ latitudinal widths is the reflection
of this physics.
\begin{figure}[!htbp]
\epsscale{0.57}
\hspace{-0.014\textwidth}
\plotone{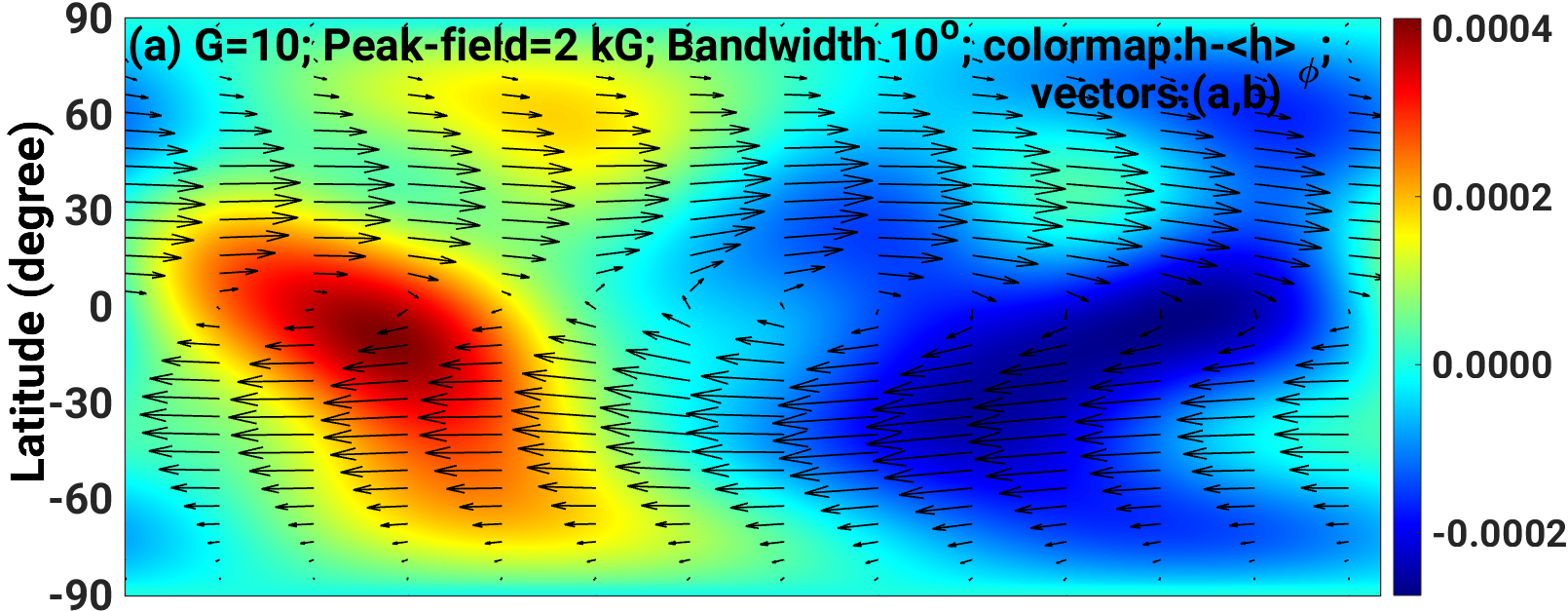}
\plotone{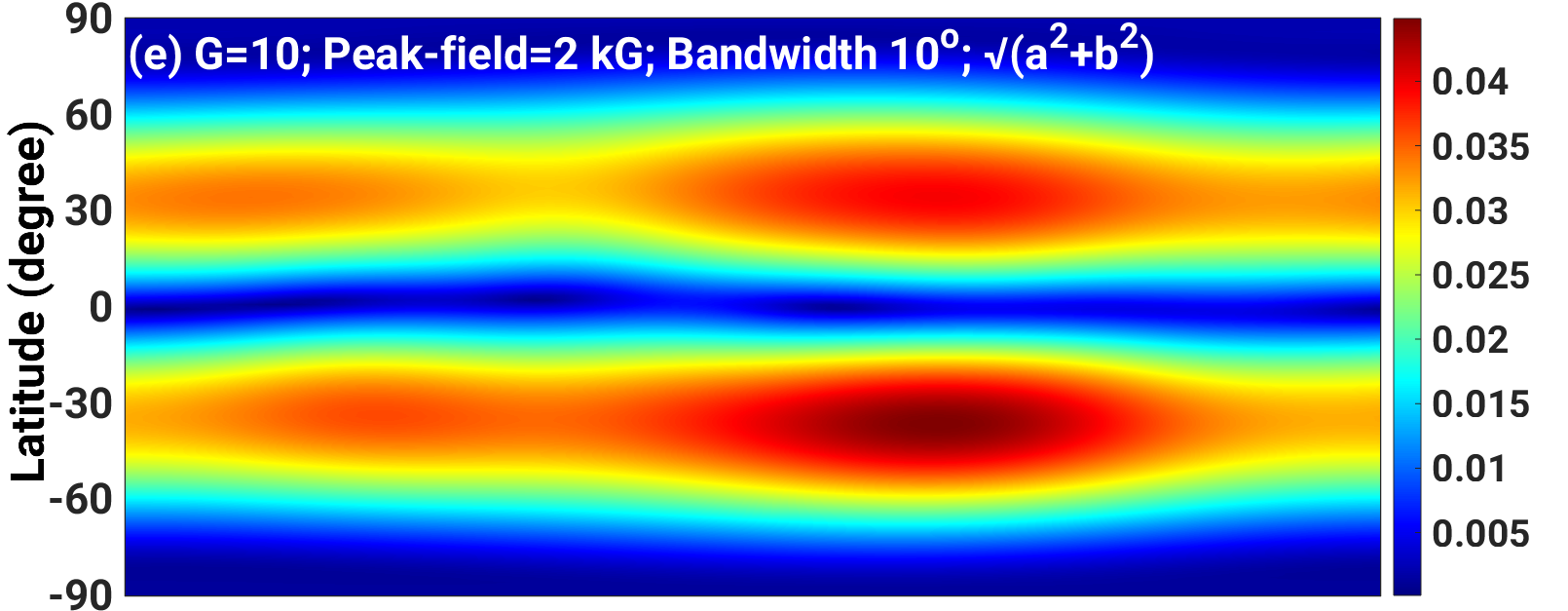}
\plotone{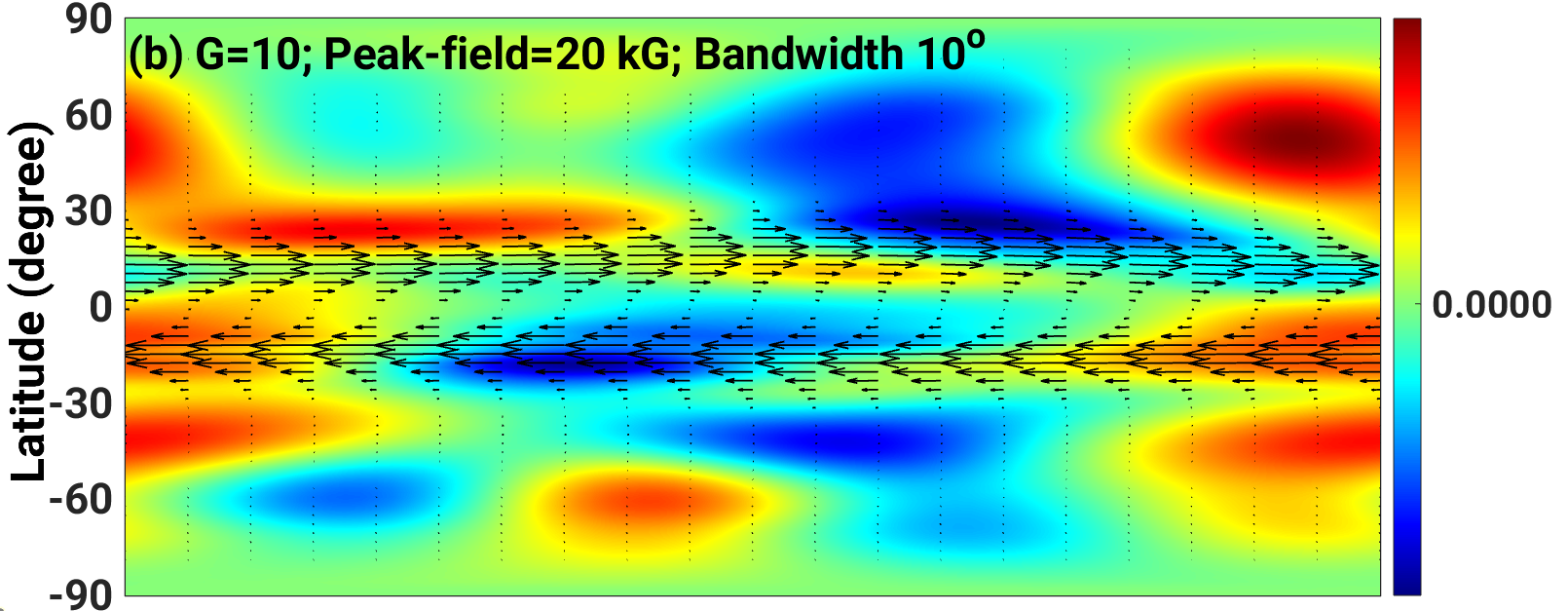}
\plotone{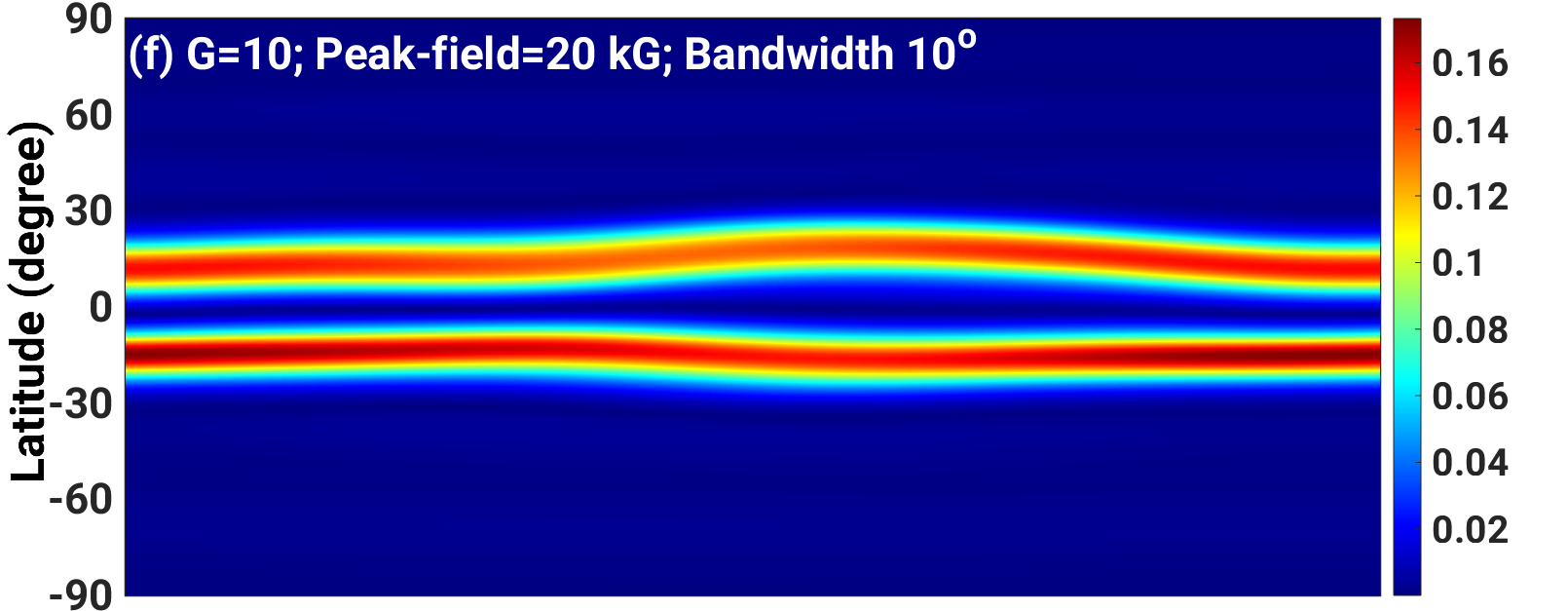}
\plotone{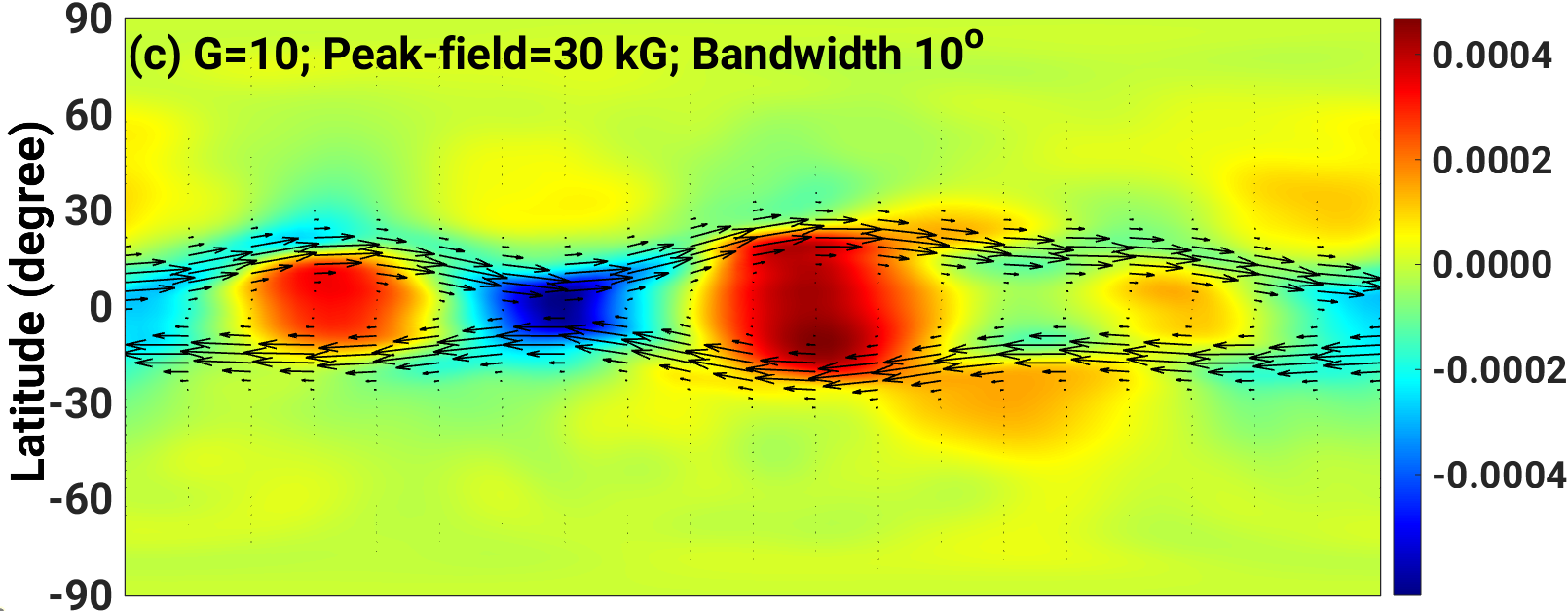}
\plotone{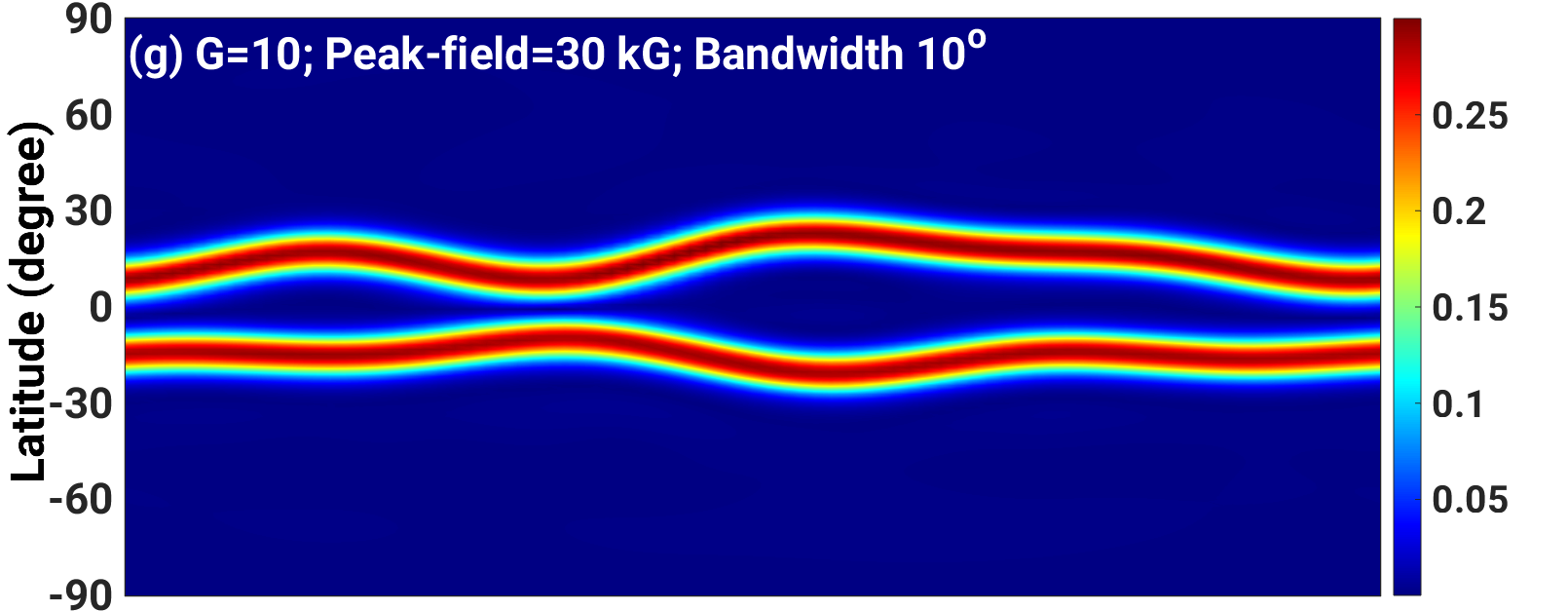}
\plotone{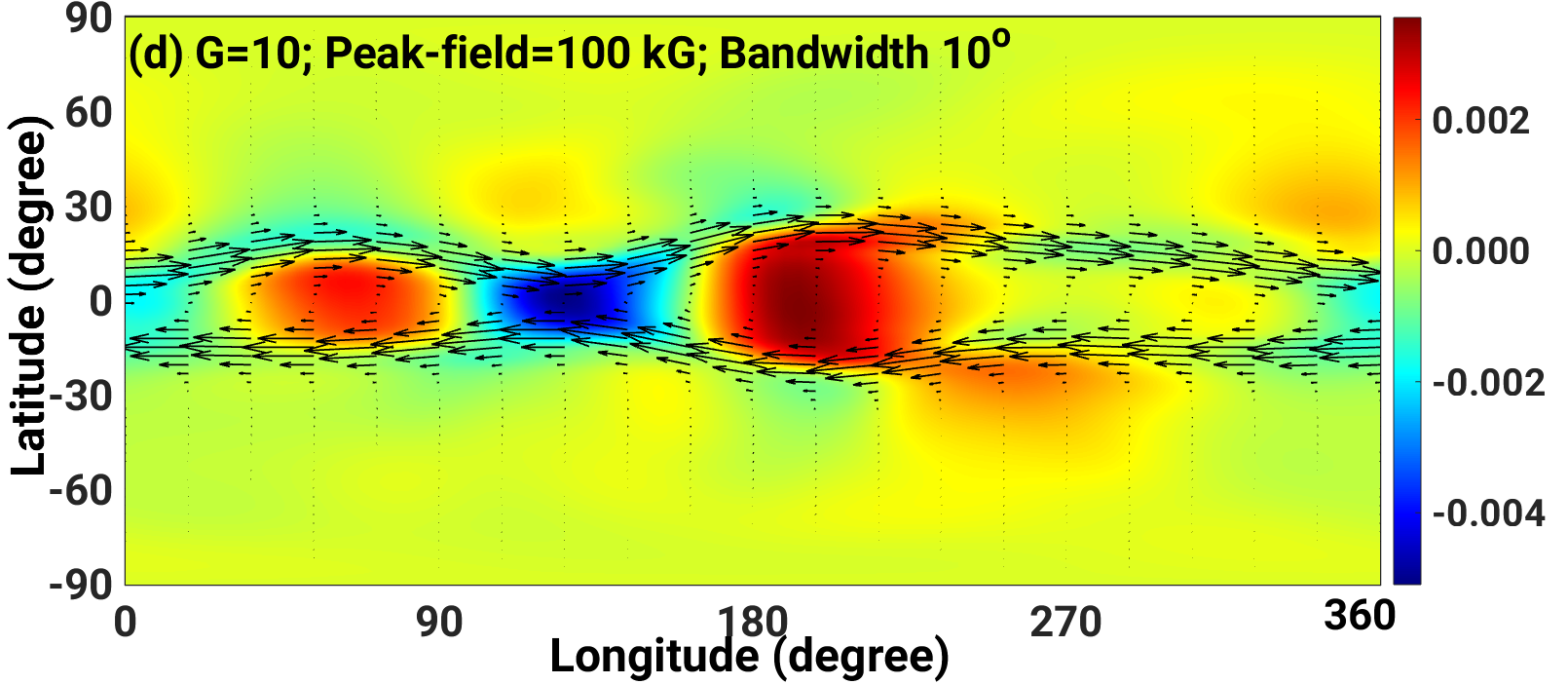}
\plotone{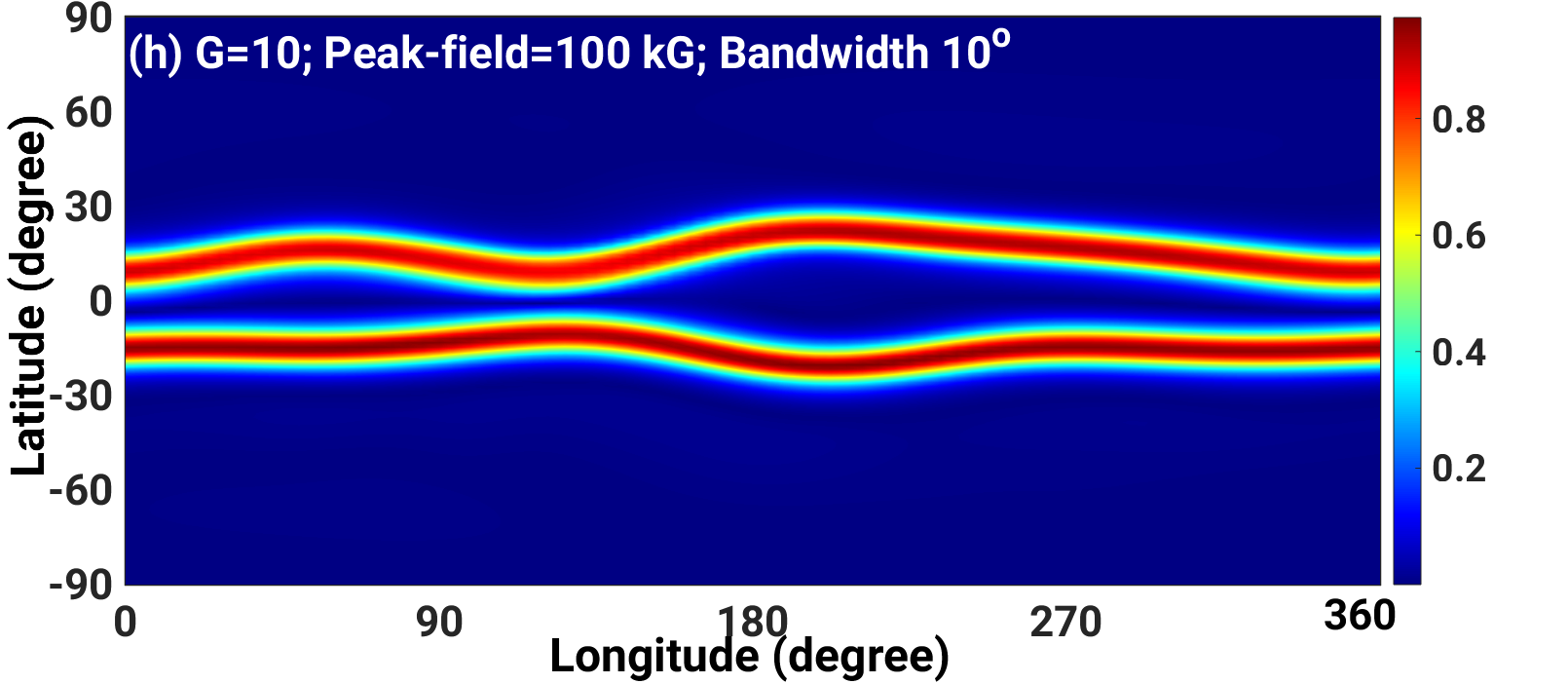}
\caption{PINN-derived initial state vectors for G=10 case; bandwidth is $10^{\circ}$. Four rows from the top to the bottom show results for peak field strengths of 2 kG, 20 kG, 30 kG, and 100 kG, respectively. Left column shows derived bulging (in red) and depression (in blue) regions from derived longitudinal mean subtracted height profiles ($h-<h>_{\phi}$). Overlaid on colormaps are the magnetic field vectors in black arrows. Right column shows the amplitudes of warped toroidal fields ($\sqrt{a^2+b^2}$) at all latitude and longitude locations. Loss values are $\sim 0.000483$, $\sim 0.000076$, $\sim 0.000097$ and $\sim 0.000403$ respectively for 2, 20, 30 and 100 kG fields.
}
\label{fig:g_10_fwhm_10}
\end{figure}
\subsubsection{PINN-derived state-vectors for high $G$ case}

So far, in the previous three subsections we presented results for
the effective gravity $G=0.5$, which characterizes slightly subadiabatic overshoot part
of the tachocline. We present in this section the cases for the
strongly subadiabatic region, the radiative part of the tachocline,
characterizing that by $G=10$. To compare the solutions with Figure \ref{fig:g_10_fwhm_10}
for $G=0.5$, we consider $10^{\circ}$ toroidal bands at $\pm 15^{\circ}$
latitudes and peak field strengths of 2, 20, 30 and 100 kG.
Figure \ref{fig:g_10_fwhm_10} shows that the PINN solutions exhibit a clear,
field-strength–dependent progression in morphology that can be directly
compared with the observed warped toroid of Figure \ref{fig:AR_toroid}, as well as
with Figure \ref{fig:g_0.5_fwhm_10} derived with $G=0.5$. Again like other figures,
PINN solutions for magnetic field vectors ($a,b$) are displayed
in black arrows, overlaid on height-deformation/pressure-departures
($h - <h>_{\phi}$) in rainbow colomap in left column, and magnetic
field amplitudes in right column.

Looking at the panels from the top to the bottom rows we find that the
top row (panels a,e) reveal the height-deformation is smoothly-varying
and pretty broad, even broader than the corresponding $G=0.5$ case for
2 kG field. The magnetic vectors are weak and widely spread, do not
confine into $10^{\circ}$ band. The solutions are hydrodynamically
dominated. This PINN-derived state does not match the observed toroid.

Second row (panels b,f), although show band structure, the warping
is poor. This is because the excited modes have poor growth rates, and
so they are close to neutral. They cannot extract enough energy to
develop their tip-patterns. Actually for high $G$, 20 kG field strength
case is also hydrodynamically dominated. Strongly subadiabatic case
behaves like 2D, and it is well-known that in pure 2D case strong
magnetic fields are needed to excite unstable modes. The threshold
field-strength for this case is about 25 kG. Thus the PINN-derived
warped toroids do not provide a good agreement with observations.
A modest hemispheric asymmetry is visible in the height-deformation.

For 30 kG field, shown in the third row from the top (panels c,g), solutions
are qualitatively better. This case reveals a predominantly antisymmetric
pattern about the equator. This is because for high $G$ case, the excited
modes are dominant $m=1$ antisymmetric modes. Overall morphology and
symmetry resemble the magnetogram observations of February 14, 2024
toroids.

In bottom-most row (panels d,h), for 100 kG field, the solution is
very similar to that of 30 kG case. This is not surprising, because the excited modes
and their properties do not change, instead  the growth rates of the
$m=1$ modes reach asymptotic value with the increase of field strength
(see, e.g., Figure 3 of \citet{Cally_Dikpati_Gilman2003}). Note that a larger
$G$ implies a more subadiabatic tachocline, thus increasing the restoring
buoyancy for radial displacements and favoring more horizontal deformations
than strong vertical deformation. Under these conditions the 
toroidal band’s
ability to support $m>1$ longitudinal modes becomes difficult.
PINN-derived solutions for high $G$ cases can provide best possible match
for our target warped toroids for $\ge$30 kG fields.


\subsubsection{Robustness of PINNBARDS}

We implement PINNBARDS with the same hyperparameter setting for another case study, namely for the observed, warped toroid pattern of August 2, 2011, a week before the second biggest X-flare of cycle 24 occurred. On August 9, 2011, AR 11263 erupted into X9.96 flare from Carrington longitude of $\sim 320^{\circ}$ \citep{raphaldini2023deciphering}.

\begin{figure}[ht]
\epsscale{0.97}
\hspace{-0.028\textwidth}
\plotone{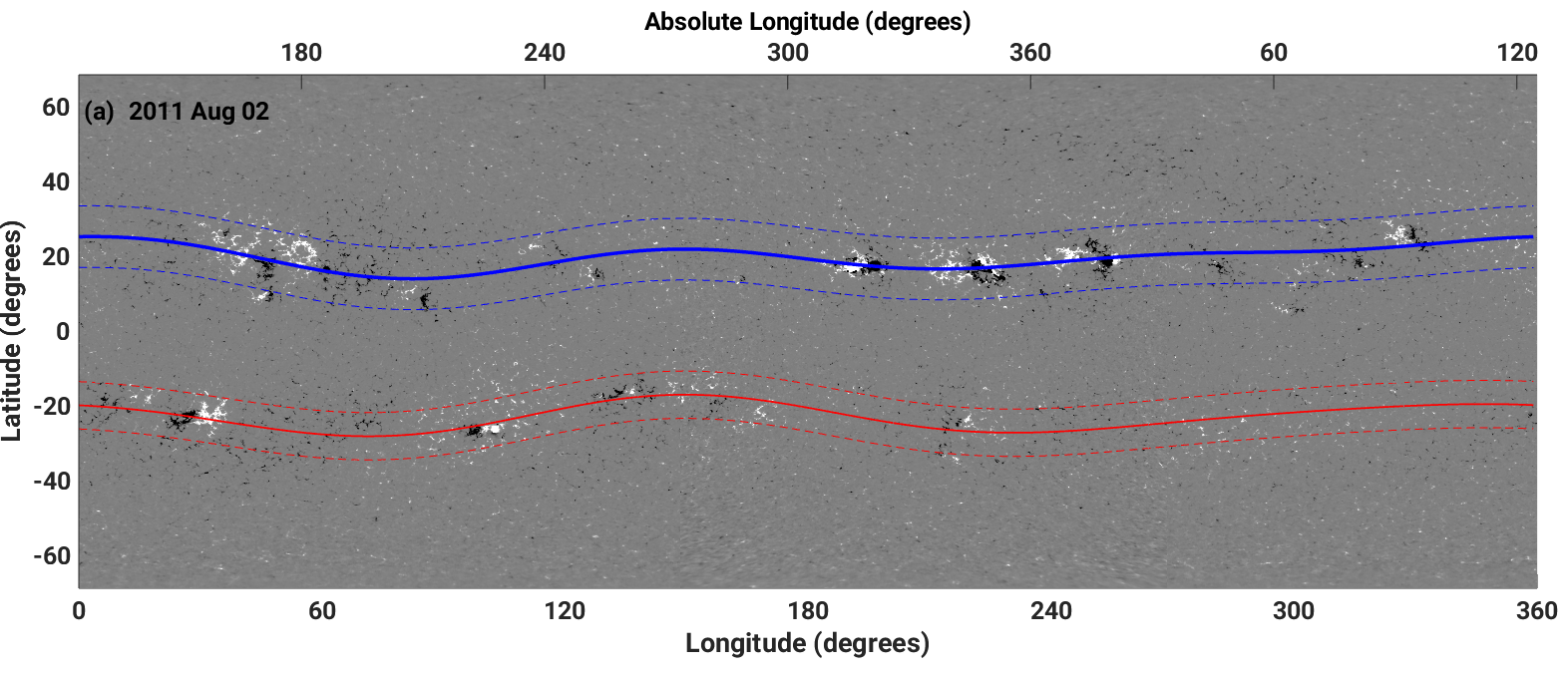}
\epsscale{1.0}
\plotone{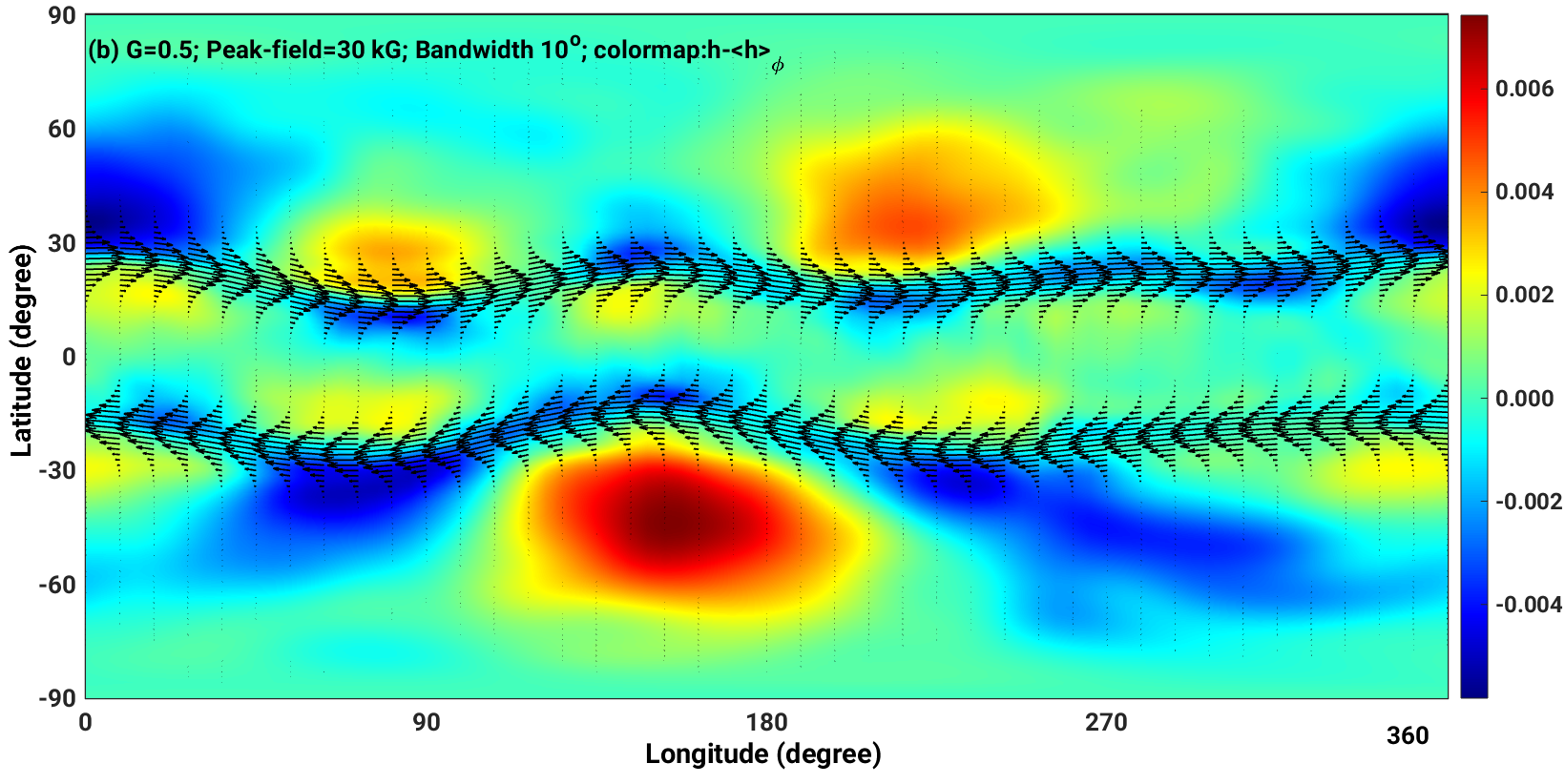}
\plotone{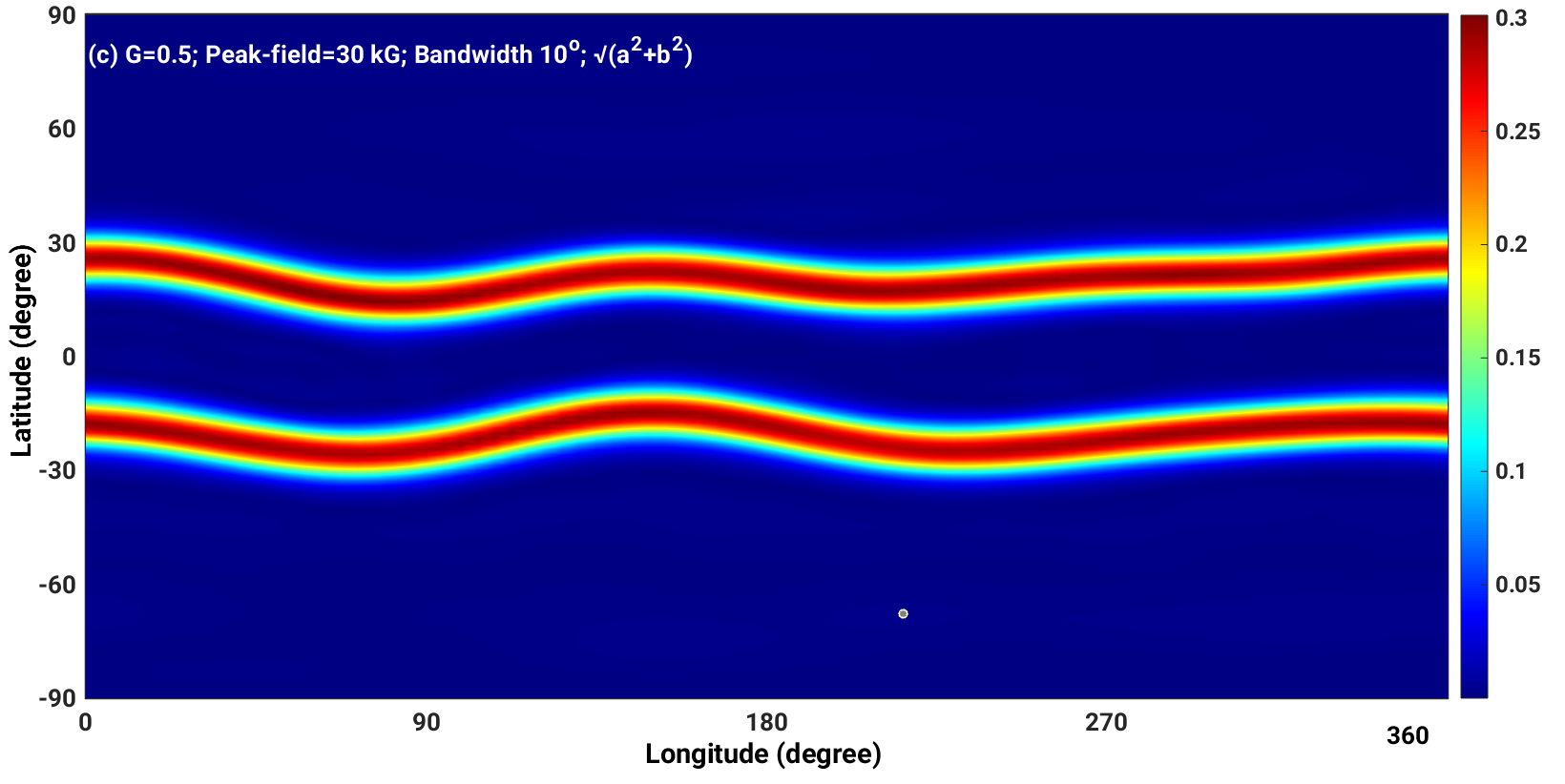}
\caption{Three panels from the top to the bottom display (a) observed warped toroid pattern of August 2, 2011; solid blue (red) lines indicate warped band's central latitude location in north (south), and dashed blue (red) curves indicate north (south) toroid's width; (b) PINN-derived magnetic field vectors (black arrows) overlaid on height-deformation contours ($h-<h_\phi>$) in rainbow colormap, where red (blue) indicates bulging (depression); (c) magnitude of toroidal bands ($\sqrt{a^2 + b^2}$). PINN-derivation is performed for 30 kG peak-field strength, $10^{\circ}$ bandwidth and for $G=0.5$. Loss value is $\sim 0.000091$ for the converged solution.
}
\label{fig:another_case}
\end{figure}

We present the results in three panels of Figure \ref{fig:another_case}. The top panel shows the latitude–longitude distribution of active regions on 2 August 2011 from an HMI synoptic map. The pattern exhibits the characteristic warped toroidal bands in both hemispheres. In the north, the central latitude (blue solid line) varies with longitude, producing clear wave-like warps; dashed curves indicate the band width. In the south, the toroidal band (red curves) shows a similar but not identical modulation, introducing a degree of hemispheric asymmetry. Active regions cluster at preferred longitudes—particularly near $\sim 240^{\circ}$ longitudes of our reference axis at the bottom ($\sim 320^{\circ}$ absolute longitude marked at the top horizontal axis). These clusters tend to fall near the poleward crests or equatorward troughs of the warped bands.

The pattern reflects a superposition of low-order longitudinal modes ($m = 1–3$), forming a smoothly varying global toroid. The central latitudes ($\pm 20.5^{\circ}$) are higher than those of the February 2024 case because the August 2011 pattern occurred earlier in cycle 24's rising phase, whereas the February 2024 case lies near cycle 25's maximum. The warped toroids here display a largely symmetric longitudinal undulations, namely when the north band moves poleward at a certain longitude-range, the south band moves equatorward at that longitude-range. This is in contrast to the  pattern of the February 2024 toroids, which tip away from one another in two longitude sectors.

This observed distribution provides a complementary geometric constraint to the first case study. Guided by earlier results, showing optimal PINN performance for 20–30 kG fields and $10^{\circ}$ widths, and noting that narrower bands excite higher $m$-modes, we adopt 30 kG peak field strength, $10^{\circ}$ bands at central latitudes $\pm 20.5^{\circ}$. We again consider the weakly subadiabatic overshoot tachocline ($G = 0.5$) as the flux-launching layer and use the same 21\% differential rotation as used in the first case study. With these inputs, PINNBARDS is run to construct the model's state-vectors that correspond to the observations of Figure \ref{fig:another_case}a.

The middle panel shows the PINN-derived latitude-longitude structure of the shallow-water MHD state variables, namely the magnetic fields (arrow vectors) overlaid on height-deformation contours $h-<h>_{\phi}$ in rainbow colormap. Red–yellow regions correspond to positive bulging of the tachocline top surface; blue regions correspond to depressions. These bulges mark locations where magnetic pressure or mode interference favors upward displacement of flux, meaning they are potential launch-locations for future active-region emergence. 

The vectors form two latitude-belts, tightly confined near $\pm 20.5^{\circ}$ latitude, matching the observed band positions. Warped geometry of the toroids is clearly visible; however, the longitudinal undulations are pretty much symmetric, maintaining approximately a fixed distance between the north and south toroids, in contrast to that seen in the first case study of cycle 25 toroids of February 14, 2024. The warping displays the expected low-order magneto-Rossby mode signatures (m $\approx$ 1–3), producing alternating bulges and depressions; either poleward or equatorward sides of the toroids align well with longitudes of observed active-region clustering in panel (a).

Overall, the same PINNBARDS reproduces both the latitudinal confinement and longitudinal modulation of the observed toroids and provides the associated dynamical fields needed for model initialization for eventually predicting the eruption into X-flare in a week.

The bottom panel shows the magnitude of the magnetic field within the toroidal bands. This mimics the patterns shown in panel b. The north and south bands appear as smooth, continuous ribbons, each $\sim 10^{\circ}$ wide, centered at approximately $20.5^{\circ}$ latitude, consistent with the observational fits. The field strength peaks along the central axis of each band and decreases gradually toward the edges. The overall symmetry between hemispheres is strong.

As in the bottom panel of Figure \ref{fig:g_0.5_fwhm_10} for the first case study, the PINN solution displayed in Figure \ref{fig:another_case}(c) converged to a physically stable configuration with a low loss value of 0.000091. Thus this panel also confirms the robustness of PINNBARDS, revealing that the PINN successfully constructs a physically consistent, smoothly varying toroidal band morphology whose structure aligns with the warped toroid inferred from observations.
\newpage
\section{Summary and Conclusions}

In this study we have developed a Physics-Informed Neural Network–based 
inversion framework, PINNBARDS (PINN-Based Active Regions Distribution
Simulator), for deriving dynamically consistent MHD state vectors from
the observed global organization of active regions (ARs). The primary
motivation is that forward modeling of the evolution of deep-seated
toroidal magnetic fields, required for multi-week prediction of
active regions emergence and associated space weather hazards, demands
physically consistent initial conditions. However, synoptic magnetograms
provide only the geometric pattern of AR distributions in warped toroids,
not the full set of dynamical variables ($a, b, u, v, h$) needed to
initialize global MHD shallow-water simulations. This has been one of
the biggest challenges until now. In this work we have built a plausible
framework to address that challenge and demonstrate that a PINN,
trained simultaneously on observations and the fully nonlinear MHD
shallow-water equations at $t=0$, can derive the state-vectors from
the observed geometric patterns for warped toroid.

The framework incorporates three essential classes of constraints:
(1) observational constraints, enforcing that the latitude–longitude
structure of the derived magnetic fields reproduces the warped toroidal
bands extracted from HMI synoptic maps; (2) physical constraints,
including the complete nonlinear MHD shallow-water system, mass
conservation and divergence-free magnetic field condition; (3) boundary
constraints, ensuring regularity at the poles and periodicity in longitude.
Through these combined requirements, the PINN self-consistently derives
the magnetic and velocity fields that are compatible with both the physics
of the system and the warped-toroid morphology observed at the surface.

We have first demonstrated that the PINN converges reliably from broad,
unstructured initial patterns to tight-fit warped toroids, which are
predominantly antisymmetric, and whose longitudinal phase, amplitude,
and latitudinal widths closely match the observed toroid pattern of 14
February 2024. The convergence analysis confirms that the PINN
simultaneously satisfies the governing MHD equations and observational
geometry, thereby yielding a physically meaningful set of initial state
vectors suitable for forward integration.

We have then explored the sensitivity of the derived state vectors to
key tachocline parameters, namely the peak toroidal field strength,
band width, and effective gravity $G$, all of which are constrained but
not uniquely determined by surface magnetograms. For $10^{\circ}$ wide
toroidal bands centered at $\pm 15^{\circ}$ latitude, we examined four
representative field strengths (2, 20, 30, and 100 kG).

Weak fields ($\sim 2$ kG) produce solutions dominated by hydrodynamic
modes, yielding overly broad structures inconsistent with the observed
warping. Very strong fields ($\sim 100$ kG) behave as overly rigid "steel
rings" that cannot sustain multi-mode longitudinal deformation, leading
to unrealistic kinks. Intermediate fields (20–30 kG) produce the closest
match to the observed toroids, with appropriate latitudinal confinement,
longitudinal warping, and hemispheric morphology. Thus, in this framework,
the observed warped toroidal pattern is most naturally reproduced by
tachocline toroidal fields in the 20–30 kG range, consistent with
flux-emergence theory for strong ARs.

We have further examined narrower (6$^{\circ}$) and wider (15$^{\circ}$) band
widths. Narrow bands can match some features of the observed warping
but fail to reproduce AR emergence around certain longitudes (e.g.,
$130^{\circ}$ in PINN reference frame or $\sim150^{\circ}$ in absolute longitude), while wider bands tend to generate overly smooth
structures and degrade the latitudinal gap between North and South toroids at longitude boundaries $0^{\circ}$ and $360^{\circ}$.

The best overall correspondence with observations arises for bandwidths
of roughly $10^{\circ}$, which also allow simultaneous excitation of
low-order longitudinal modes ($m = 1$ and $m > 1$).

Finally, we investigated how the effective gravity ($G$) influences
the solution.$G$ represents the degree of subadiabaticity in the
tachocline. For $G = 0.5$, which is an appropriate value for the
overshoot layer, intermediate fields (20–30 kG) yield the best
match. For $G = 10$, characteristics of the strongly subadiabatic
radiative tachocline, fields must exceed $\sim 25$ kG to excite
unstable modes capable of producing observationally consistent
warping. The best matches for this high $G$ regime occur for peak
fields of $\ge 30$ kG, with the morphology dominated by
antisymmetric $m = 1$ modes, as expected from thin-shell theory.

The second case-study of deriving state-vectors for warped toroids during August 2, 2011, demonstrates that PINNBARDS can successfully reconstruct the complete state-vectors corresponding to the underlying geometric pattern of an observed warped toroid at an earlier phase of the solar cycle 24, in addition to the peak-phase toroids of cycle 25. The HMI-derived geometry shows primarily a symmetric warping of toroidal bands centered at $\pm 20.5^{\circ}$, with active regions clustering near longitudinal bulges and depressions. Using this geometry together with physically motivated parameters, i.e., $10^{\circ}$ wide bands with 30 kG peak field strength in weakly subadiabatic overshoot tachocline ($G = 0.5$), and 21\% pole-to-equator solar differential rotation, the PINN produces magnetic fields that preserve the observed latitudinal confinement and longitudinal modulation while generating physically consistent height deformations and toroidal-field magnitudes. The smooth band profiles, and low loss value indicate a stable, self-consistent solution. This case, complementary to the February 14, 2024 case study, confirms that the same PINNBARDS framework can robustly infer tachocline state vectors across different phases of the solar cycle, providing coherent initial conditions for forward modeling of flux emergence and space-weather-relevant activity.

In summary, the PINNBARDS inversion framework resolves a longstanding
challenge by demonstrating that the MHD-model-consistent state vectors
can be derived directly from observed warped toroidal patterns using a
PINN, constrained by a fully nonlinear MHD shallow-water physical model.

The observed morphology of AR-bearing toroids strongly constrains the
plausible tachocline field strength, bandwidth, and subadiabaticity.
Intermediate toroidal field strengths (20–30 kG), band widths near 10°,
and moderate-to-weak subadiabaticity ($G \approx 0.5$) produce the
best agreement with the observed February 2024 warped toroids.

Both overly weak and overly strong fields are inconsistent with the
observed multi-mode warping, the former due to hydrodynamic dominance
and the latter due to excessive magnetic rigidity.

The derived state vectors provide strong physical foundations for
the initial conditions required to integrate forward the tachocline
MHD shallow-water model-system for simulating magnetic evolution
and forecasting of big, complex, flare-producing active-region's
emergence with a few weeks lead-time.

Together, these results establish a new path for combining
observations with physics-based modeling to reconstruct subsurface
magnetic structures. The inversion capability demonstrated here
represents an essential step toward predictive simulations of
global toroidal-band evolution and improved lead times for
intermediate scale space-weather forecasting.

\newpage
\section*{Acknowledgments}
\begin{acknowledgments}
We thank an anonymous reviewer for his/her thorough review and many helpful comments, taking care of which has significantly improved our paper. This work is supported by the NSF National Center for Atmospheric Research, which is a major facility sponsored by the National Science Foundation under cooperative agreement 1852977. SC acknowledges support by the NASA HGIO award with grant number 80NSSC23K0416 and MD acknowledges support from NASA-DRIVE Center award 80NSSC22M0162 to Stanford. This work utilizes data from HMI onboard NASA's SDO spacecraft, courtesy of NASA/SDO and the HMI Science Teams. In conducting all PINN simulations, including code development, testing and production-runs, $\sim$5k NVIDIA A100 gpu hours have been used in NWSC-3 (NCAR Wyoming Supercomputer Center), Derecho, under High Altitude Observatory, NSF-NCAR's project allocation number P22100000.
\end{acknowledgments}

\facilities{SDO/HMI}

\software{To develop PINNBARDS, we used different \texttt{Python} version 3.10.12 from installation at NCAR-Wyoming-SuperComputer-3 packages namely,(1)\texttt{TensorFlow} version 2.16.1+nv24.7 from NGC container version 24.07 (https://docs.nvidia.com/deeplearning/frameworks/tensorflow-release-notes/rel-24-07.html) to develop the neural network architecture, physics-informed loss function, and training routine, (2)\texttt{Numpy} version 1.24.4 for generating training data, (3) \texttt{Pickle} protocol version 4 for saving the loss, (4) \texttt{Matplotlib} version 3.10.3 for producing plots after model training. All latitude-longitude planform figures are produced using NSF-NCAR's \texttt{MATLAB} version R2024b Update 3.
}
\newpage
\appendix
\section{Extension to other datasets}
\begin{figure}[htbp!]
\hspace{0.0\textwidth}
\includegraphics[width=1.0\linewidth]%
{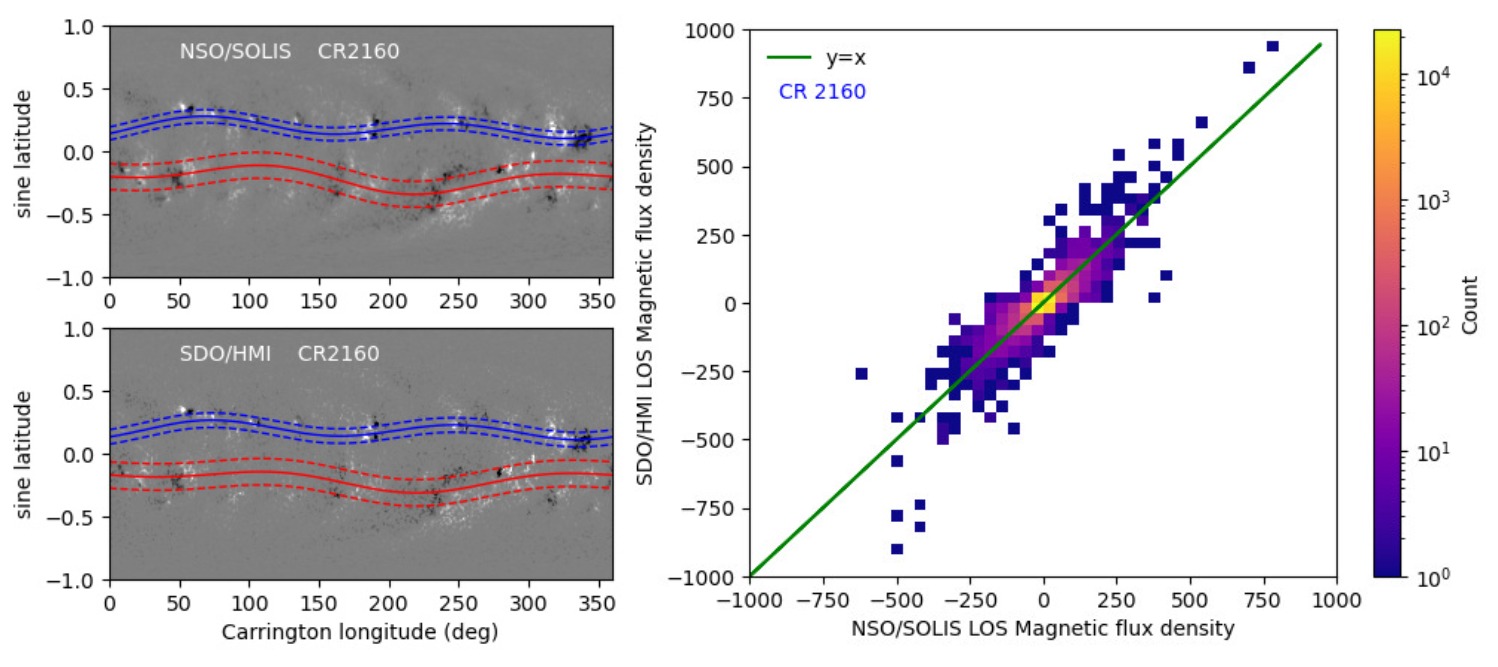}
\caption{Match between fitted toroids on NSO/SOLIS and SDO/HMI data. The North (blue) and South (red) toroids are fitted on the NSO/SOLIS (top left) and SDO/HMI (bottom left) synoptic maps from rotation CR2160 with the same threshold of magnetic flux density. The pixel-wise match of the flux density is depicted by the 2D histogram on the right panel. The histogram aligns well with the `y=x' line shown in green.  
}
\label{fig:other_data}
\end{figure}

Although the present study uses only SDO/HMI synoptic maps of the magnetic flux density, it can be easily extended to other magnetogram datasets such as NSO/SOLIS, NSO/GONG, and SoHO/HMI. Figure~\ref{fig:other_data} shows that fitted toroids on a representative Carrington rotation CR2160 from SOLIS and HMI with the same threshold reveal a great match for both northern and southern hemispheres. Comparing the maps pixel to pixel (2D histogram in Figure~\ref{fig:other_data}) does not show a significant deviation from `y=x' line, ruling out the need for a scaling factor for homogenization between SOLIS and HMI. However, it should be noted that such a scaling factor, different from unity, is necessary for other datasets such as SoHO/MDI and NSO/GONG, as has been described in previous studies \citep{Muñoz-Jaramillo_2024}. To perform a more reliable homogenization of these data surveys, a more sophisticated deep-learning-based homogenization \citep{Muñoz-Jaramillo_2024} with uncertainty quantification \citep{Chatterjee_2023} should be considered. In that way, the same toroid fitting approach can be used across multiple magnetogram surveys.

\bibliography{ms.bib}{}

\begin{thebibliography}{}
\expandafter\ifx\csname natexlab\endcsname\relax\def\natexlab#1{#1}\fi
\providecommand{\url}[1]{\href{#1}{#1}}
\providecommand{\dodoi}[1]{doi:~\href{http://doi.org/#1}{\nolinkurl{#1}}}
\providecommand{\doeprint}[1]{\href{http://ascl.net/#1}{\nolinkurl{http://ascl.net/#1}}}
\providecommand{\doarXiv}[1]{\href{https://arxiv.org/abs/#1}{\nolinkurl{https://arxiv.org/abs/#1}}}

\bibitem[{Bihlo \& Popovych(2022)}]{bihlo2022}
Bihlo, A., \& Popovych, R.~O. 2022, Journal of Computational Physics, 456, 111024, \dodoi{https://doi.org/10.1016/j.jcp.2022.111024}

\bibitem[{Branch {et~al.}(1999)Branch, Coleman, \& Li}]{branch1999subspace}
Branch, M.~A., Coleman, T.~F., \& Li, Y. 1999, SIAM Journal on Scientific Computing, 21, 1, \dodoi{10.1137/S1064827595289108}

\bibitem[{{Cally} {et~al.}(2003){Cally}, {Dikpati}, \& {Gilman}}]{Cally_Dikpati_Gilman2003}
{Cally}, P.~S., {Dikpati}, M., \& {Gilman}, P.~A. 2003, \apj, 582, 1190, \dodoi{10.1086/344746}

\bibitem[{{Chatterjee} {et~al.}(2022){Chatterjee}, {Mu{\~n}oz-Jaramillo}, \& {Lamb}}]{chatterjee2022}
{Chatterjee}, S., {Mu{\~n}oz-Jaramillo}, A., \& {Lamb}, D.~A. 2022, Nature Astronomy, 6, 796, \dodoi{10.1038/s41550-022-01701-3}

\bibitem[{Chatterjee {et~al.}(2023)Chatterjee, Muñoz-Jaramillo, Dayeh, Bain, \& Moreland}]{Chatterjee_2023}
Chatterjee, S., Muñoz-Jaramillo, A., Dayeh, M.~A., Bain, H.~M., \& Moreland, K. 2023, The Astrophysical Journal Supplement Series, 268, 33, \dodoi{10.3847/1538-4365/ace9d7}

\bibitem[{Dikpati {et~al.}(2018)Dikpati, Belucz, Gilman, \& McIntosh}]{Dikpati_2018}
Dikpati, M., Belucz, B., Gilman, P.~A., \& McIntosh, S.~W. 2018, The Astrophysical Journal, 862, 159, \dodoi{10.3847/1538-4357/aacefa}

\bibitem[{{Dikpati} {et~al.}(2017){Dikpati}, {Cally}, {McIntosh}, \& {Heifetz}}]{dikpati2017origin}
{Dikpati}, M., {Cally}, P.~S., {McIntosh}, S.~W., \& {Heifetz}, E. 2017, Scientific Reports, 7, 14750, \dodoi{10.1038/s41598-017-14957-x}

\bibitem[{{Dikpati} \& {Gilman}(2001)}]{dikpati2001}
{Dikpati}, M., \& {Gilman}, P.~A. 2001, \apj, 551, 536, \dodoi{10.1086/320080}

\bibitem[{{Dikpati} {et~al.}(2003){Dikpati}, {Gilman}, \& {Rempel}}]{dikpati2003}
{Dikpati}, M., {Gilman}, P.~A., \& {Rempel}, M. 2003, \apj, 596, 680, \dodoi{10.1086/377708}

\bibitem[{{Dikpati} \& {McIntosh}(2020)}]{dikpati2020spaceweather}
{Dikpati}, M., \& {McIntosh}, S.~W. 2020, Space Weather, 18, e02109, \dodoi{10.1029/2019SW002109}

\bibitem[{{Dikpati} {et~al.}(2018){Dikpati}, {McIntosh}, {Bothun}, {Cally}, {Ghosh}, {Gilman}, \& {Umurhan}}]{dikpati2018role}
{Dikpati}, M., {McIntosh}, S.~W., {Bothun}, G., {et~al.} 2018, \apj, 853, 144, \dodoi{10.3847/1538-4357/aaa70d}

\bibitem[{{Dikpati} {et~al.}(2021){Dikpati}, {McIntosh}, {Chatterjee}, {Norton}, {Ambroz}, {Gilman}, {Jain}, \& {Munoz-Jaramillo}}]{dikpati2021deciphering}
{Dikpati}, M., {McIntosh}, S.~W., {Chatterjee}, S., {et~al.} 2021, \apj, 910, 91, \dodoi{10.3847/1538-4357/abe043}

\bibitem[{{Dikpati} {et~al.}(2023){Dikpati}, {Anderson}, {Belucz}, {Biesecker}, {Bothun}, {Chatterjee}, {Fan}, {Gibson}, {Gilbert}, {Gilman}, {Guerrero}, {Hoeksema}, {Jain}, {Kitiashvili}, {Korsos}, {Kosovichev}, {Leamon}, {Linkmann}, {McIntosh}, {Norton}, {Raoafi}, {Raphaldini}, {Rempel}, {Tripathy}, {Upton}, {Wang}, {Wing}, \& {Zaqarashvili}}]{dikpati2022space}
{Dikpati}, M., {Anderson}, J.~L., {Belucz}, B., {et~al.} 2023, in Bulletin of the American Astronomical Society, Vol.~55, 097, \dodoi{10.3847/25c2cfeb.eb8f0fdc}

\bibitem[{{Dikpati} {et~al.}(2025){Dikpati}, {Kors{\'o}s}, {Norton}, {Raphaldini}, {Jain}, {McIntosh}, {Gilman}, {Teruya}, \& {Raouafi}}]{dikpati2025}
{Dikpati}, M., {Kors{\'o}s}, M.~B., {Norton}, A.~A., {et~al.} 2025, \apj, 988, 108, \dodoi{10.3847/1538-4357/addd09}

\bibitem[{{Fournier} {et~al.}(2017){Fournier}, {Arlt}, {Ziegler}, \& {Strassmeier}}]{fournier2017}
{Fournier}, Y., {Arlt}, R., {Ziegler}, U., \& {Strassmeier}, K.~G. 2017, \aap, 607, A1, \dodoi{10.1051/0004-6361/201629989}

\bibitem[{Gilman(2000)}]{Gilman_2000}
Gilman, P.~A. 2000, The Astrophysical Journal, 544, L79, \dodoi{10.1086/317291}

\bibitem[{{Hughes} {et~al.}(1997){Hughes}, {Wissink}, {Matthews}, \& {Proctor}}]{hughes1997}
{Hughes}, D.~W., {Wissink}, J.~G., {Matthews}, P.~C., \& {Proctor}, M.~R.~E. 1997, in Astronomical Society of the Pacific Conference Series, Vol. 118, 1st Advances in Solar Physics Euroconference. Advances in Physics of Sunspots, ed. B.~{Schmieder}, J.~C. {del Toro Iniesta}, \& M.~{Vazquez}, 66

\bibitem[{{Inceoglu} {et~al.}(2018){Inceoglu}, {Jeppesen}, {Kongstad}, {Hern{\'a}ndez Marcano}, {Jacobsen}, \& {Karoff}}]{inceoglu2018}
{Inceoglu}, F., {Jeppesen}, J.~H., {Kongstad}, P., {et~al.} 2018, \apj, 861, 128, \dodoi{10.3847/1538-4357/aac81e}

\bibitem[{{Jouve} \& {Brun}(2009)}]{jouve2009}
{Jouve}, L., \& {Brun}, A.~S. 2009, \apj, 701, 1300, \dodoi{10.1088/0004-637X/701/2/1300}

\bibitem[{{Kors{\'o}s} {et~al.}(2021){Kors{\'o}s}, {Erd{\'e}lyi}, {Liu}, \& {Morgan}}]{korsos2021}
{Kors{\'o}s}, M.~B., {Erd{\'e}lyi}, R., {Liu}, J., \& {Morgan}, H. 2021, Frontiers in Astronomy and Space Sciences, 7, 113, \dodoi{10.3389/fspas.2020.571186}

\bibitem[{{Liu} {et~al.}(2021){Liu}, {Wang}, {Huang}, {Kors{\'o}s}, {Jiang}, {Wang}, \& {Erd{\'e}lyi}}]{liu2021}
{Liu}, J., {Wang}, Y., {Huang}, X., {et~al.} 2021, Nature Astronomy, 5, 108, \dodoi{10.1038/s41550-021-01310-6}

\bibitem[{{Manek} \& {Brummell}(2024)}]{manek2024}
{Manek}, B., \& {Brummell}, N. 2024, \apj, 971, 7, \dodoi{10.3847/1538-4357/ad5993}

\bibitem[{{Manek} {et~al.}(2018){Manek}, {Brummell}, \& {Lee}}]{manek2018}
{Manek}, B., {Brummell}, N., \& {Lee}, D. 2018, \apjl, 859, L27, \dodoi{10.3847/2041-8213/aac723}

\bibitem[{{Miesch} {et~al.}(2007){Miesch}, {Gilman}, \& {Dikpati}}]{Miesch_Gilman_Dikpati2007}
{Miesch}, M.~S., {Gilman}, P.~A., \& {Dikpati}, M. 2007, \apjs, 168, 337, \dodoi{10.1086/509880}

\bibitem[{Muñoz-Jaramillo {et~al.}(2024)Muñoz-Jaramillo, Jungbluth, Gitiaux, Wright, Shneider, Maloney, Baydin, Gal, Deudon, \& Kalaitzis}]{Muñoz-Jaramillo_2024}
Muñoz-Jaramillo, A., Jungbluth, A., Gitiaux, X., {et~al.} 2024, The Astrophysical Journal Supplement Series, 271, 46, \dodoi{10.3847/1538-4365/ad12c2}

\bibitem[{{Nelson} {et~al.}(2013){Nelson}, {Brown}, {Brun}, {Miesch}, \& {Toomre}}]{nelson2013}
{Nelson}, N.~J., {Brown}, B.~P., {Brun}, A.~S., {Miesch}, M.~S., \& {Toomre}, J. 2013, \apj, 762, 73, \dodoi{10.1088/0004-637X/762/2/73}

\bibitem[{Norton \& Gilman(2005)}]{norton2005recovering}
Norton, A.~A., \& Gilman, P.~A. 2005, The Astrophysical Journal, 630, 1194, \dodoi{10.1086/431961}

\bibitem[{{Raissi} {et~al.}(2019){Raissi}, {Perdikaris}, \& {Karniadakis}}]{Raissi2019}
{Raissi}, M., {Perdikaris}, P., \& {Karniadakis}, G.~E. 2019, Journal of Computational Physics, 378, 686, \dodoi{10.1016/j.jcp.2018.10.045}

\bibitem[{Raissi {et~al.}(2020)Raissi, Yazdani, \& Karniadakis}]{Raissi2020}
Raissi, M., Yazdani, A., \& Karniadakis, G.~E. 2020, Science, 367, 1026, \dodoi{10.1126/science.aaw4741}

\bibitem[{{Raphaldini} {et~al.}(2023){Raphaldini}, {Dikpati}, {Norton}, {Teruya}, {McIntosh}, {Prior}, \& {MacTaggart}}]{raphaldini2023deciphering}
{Raphaldini}, B., {Dikpati}, M., {Norton}, A.~A., {et~al.} 2023, \apj, 958, 175, \dodoi{10.3847/1538-4357/acfef0}

\bibitem[{{Raphaldini} {et~al.}(2024){Raphaldini}, {Dikpati}, {Teruya}, {Jain}, {Norton}, \& {McIntosh}}]{Raphaldini2024deciphering}
{Raphaldini}, B., {Dikpati}, M., {Teruya}, A.~S.~W., {et~al.} 2024, \aap, 691, A3, \dodoi{10.1051/0004-6361/202451428}

\bibitem[{{Schrijver} \& {Zwaan}(2000)}]{schrijver2000}
{Schrijver}, C.~J., \& {Zwaan}, C. 2000, {Solar and Stellar Magnetic Activity} (New York: Cambridge University Press)

\bibitem[{{Weber}(2014)}]{weber2014}
{Weber}, M.~A. 2014, PhD thesis, Colorado State University

\bibitem[{Whitman {et~al.}(2023)Whitman, Egeland, Richardson, Allison, Quinn, Barzilla, Kitiashvili, Sadykov, Bain, Dierckxsens, Mays, Tadesse, Lee, Semones, Luhmann, Núñez, White, Kahler, Ling, Smart, Shea, Tenishev, Boubrahimi, Aydin, Martens, Angryk, Marsh, Dalla, Crosby, Schwadron, Kozarev, Gorby, Young, Laurenza, Cliver, Alberti, Stumpo, Benella, Papaioannou, Anastasiadis, Sandberg, Georgoulis, Ji, Kempton, Pandey, Li, Hu, Zank, Lavasa, Giannopoulos, Falconer, Kadadi, Fernandes, Dayeh, Muñoz-Jaramillo, Chatterjee, Moreland, Sokolov, Roussev, Taktakishvili, Effenberger, Gombosi, Huang, Zhao, Wijsen, Aran, Poedts, Kouloumvakos, Paassilta, Vainio, Belov, Eroshenko, Abunina, Abunin, Balch, Malandraki, Karavolos, Heber, Labrenz, Kühl, Kosovichev, Oria, Nita, Illarionov, O’Keefe, Jiang, Fereira, Ali, Paouris, Aminalragia-Giamini, Jiggens, Jin, Lee, Palmerio, Bruno, Kasapis, Wang, Chen, Sanahuja, Lario, Jacobs, Strauss, Steyn, {van den Berg}, Swalwell, Waterfall, Nedal, Miteva, Dechev, Zucca, Engell,
  Maze, Farmer, Kerber, Barnett, Loomis, Grey, Thompson, Linker, Caplan, Downs, Török, Lionello, Titov, Zhang, \& Hosseinzadeh}]{Whitman2023}
Whitman, K., Egeland, R., Richardson, I.~G., {et~al.} 2023, Advances in Space Research, 72, 5161, \dodoi{https://doi.org/10.1016/j.asr.2022.08.006}

\bibitem[{Zaqarashvili {et~al.}(2007)Zaqarashvili, Oliver, Ballester, \& Shergelashvili}]{Zaq2007}
Zaqarashvili, T., Oliver, R., Ballester, J., \& Shergelashvili, B. 2007, Astronomy \& Astrophysics, 470, 815, \dodoi{10.1051/0004-6361:20077382}

\end{thebibliography}
\bibliographystyle{aasjournal}
\end{document}